%% file: main.tex
\documentclass{ieeeaccess}
\usepackage{cite}
\usepackage{amsmath,amssymb,amsfonts}
\usepackage{algorithmic}
\usepackage{graphicx}
\usepackage{textcomp}
\usepackage{lipsum}

\usepackage{hyperref}
\usepackage{overpic,graphics}
\newcommand{\ds}{\displaystyle}

\def\BibTeX{{\rm B\kern-.05em{\sc i\kern-.025em b}\kern-.08em
    T\kern-.1667em\lower.7ex\hbox{E}\kern-.125emX}}
\begin{document}
\history{Date of publication xxxx 00, 0000, date of current version xxxx 00, 0000.}
\doi{10.1109/ACCESS.2017.DOI}

\bstctlcite{IEEEexample:BSTcontrol} 

\title{Scoring the Terabit/s Goal: \\ Broadband Connectivity in 6G}

\author{
\uppercase{Nandana~Rajatheva}\authorrefmark{1}, \uppercase{Italo~Atzeni}\authorrefmark{1}, 
\uppercase{Simon~Bicaïs} \authorrefmark{2}, \uppercase{Emil~Björnson}\authorrefmark{3}, \uppercase{André~Bourdoux}\authorrefmark{4}, \uppercase{Stefano~Buzzi}\authorrefmark{5}, \uppercase{Carmen~D'Andrea}\authorrefmark{5}, \uppercase{Jean-Baptiste~Dor{\'e}}\authorrefmark{2}, \uppercase{Serhat~Erkucuk}\authorrefmark{6}, \uppercase{Manuel~Fuentes}\authorrefmark{7}, \uppercase{Ke~Guan}\authorrefmark{8}, \uppercase{Yuzhou~Hu}\authorrefmark{9}, \uppercase{Xiaojing~Huang}\authorrefmark{10}, \uppercase{Jari~Hulkkonen}\authorrefmark{11}, \uppercase{Josep~Miquel~Jornet}\authorrefmark{12}, 
\uppercase{Marcos Katz}\authorrefmark{1}, 
\uppercase{Behrooz Makki}\authorrefmark{13}, 
\uppercase{Rickard Nilsson}\authorrefmark{14}, \uppercase{Erdal~Panayirci}\authorrefmark{6}, \uppercase{Khaled~Rabie}\authorrefmark{15}, \uppercase{Nuwanthika~Rajapaksha}\authorrefmark{1}, \uppercase{MohammadJavad~Salehi}\authorrefmark{1}, \uppercase{Hadi~Sarieddeen}\authorrefmark{16}, \uppercase{Shahriar~Shahabuddin}\authorrefmark{17}, \uppercase{Tommy~Svensson}\authorrefmark{18}, \uppercase{Oskari~Tervo}\authorrefmark{11}, \uppercase{Antti~Tölli}\authorrefmark{1}, \uppercase{Qingqing~Wu}\authorrefmark{19}, 
\uppercase{and Wen~Xu}\authorrefmark{20}}

\address[1]{Centre for Wireless Communications, University of Oulu, Finland (email: \{nandana.rajatheva, italo.atzeni, marcos.katz, nuwanthika.rajapaksha, mohammadjavad.salehi, antti.tolli\}@oulu.fi).} 

\address[2]{CEA-Leti, France (email: jean-baptiste.dore@cea.fr).} 

\address[3]{KTH Royal Institute of Technology, Sweden, and  Linköping University, Sweden (email: emilbjo@kth.se).}

\address[4]{IMEC, Belgium (email: andre.bourdoux@imec.be).}

\address[5]{University of Cassino and Southern Latium, Italy, and Consorzio Nazionale Interuniversitario per le Telecomunicazioni (CNIT), Italy (email: buzzi@unicas.it, carmen.dandrea@unicas.it).}

\address[6]{Department of Electrical and Electronics Engineering, Kadir Has University, Turkey (email:\{serkucuk, eepanay\}@khas.edu.tr).}

\address[7]{Fivecomm, Valencia, Spain (email: manuel.fuentes@fivecomm.eu).}

\address[8]{State Key Laboratory of Rail Traffic Control and Safety, Beijing Jiaotong University, China, and with the Beijing Engineering Research Center of High-Speed Railway Broadband Mobile Communications, China (email: kguan@bjtu.edu.cn).} 

\address[9]{Algorithm Department, ZTE Corporation, China (email: hu.yuzhou@zte.com.cn).} 

\address[10]{School of Electrical and Data Engineering, Faculty of Engineering and Information Technology, University of Technology Sydney, Australia (email: xiaojing.huang@uts.edu.au).} 

\address[11]{Nokia Bell Labs, Finland (email: \{jari.hulkkonen, oskari.tervo\}@nokia-bell-labs.com).}

\address[12]{Institute for the Wireless Internet of Things, Department of Electrical and Computer Engineering, Northeastern University, USA (email: j.jornet@northeastern.edu).}

\address[13]{Ericsson Research, Gothenburg, Sweden (email: behrooz.makki@ericsson.com).}

\address[14]{Department of Computer Science, Electrical and Space Engineering, Lule{\aa} University of Technology, Sweden (email: rickard.o.nilsson@ltu.se).} 

\address[15]{Department of Engineering, Manchester Metropolitan University, UK (email: k.rabie@mmu.ac.uk).} 

\address[16]{Division of Computer, Electrical and Mathematical Sciences and Engineering, King Abdullah University of Science and Technology, Saudi Arabia (email: hadi.sarieddeen@kaust.edu.sa).}

\address[17]{Nokia, Finland (email: shahriar.shahabuddin@nokia.com).}

\address[18]{Department of Electrical Engineering, Chalmers University of Technology, Sweden (email: tommy.svensson@chalmers.se).} 

\address[19]{State Key Laboratory of Internet of Things for Smart City and Department of Electrical and Computer Engineering, University of Macau, Macau, China (email: qingqingwu@um.edu.mo).}

\address[20]{Huawei Technologies, Germany (email: wen.dr.xu@huawei.com).}

\tfootnote{The work of N.~Rajatheva and N. Rajapaksha was supported by the Academy of Finland 6Genesis Flagship (grant 318927). The work of I.~Atzeni was supported by the Marie Sk\l{}odowska-Curie Actions (MSCA-IF 897938 DELIGHT). The work of S.~Buzzi and C.~D'Andrea was supported by the MIUR Project ``Dipartimenti di Eccellenza 2018-2022'' and by the MIUR PRIN 2017 Project ``LiquidEdge''. The work of E.~Panayirci was supported by the Scientific and Technical Research Council of Turkey (TUBITAK) under the 1003-Priority Areas R\&D Projects Support Program No. 218E034.
}

\markboth
{N. Rajatheva \headeretal: Scoring the Terabit/s Goal: Broadband Connectivity in 6G}
{N. Rajatheva \headeretal: Scoring the Terabit/s Goal: Broadband Connectivity in 6G}

\corresp{Corresponding author: Nandana Rajatheva (e-mail: nandana.rajatheva@oulu.fi).}



\begin{abstract}
This paper explores the road to vastly improving the broadband connectivity in future 6G wireless systems. Different categories of use cases are considered,  with peak data rates up to 1~Tbps.
Several categories of enablers at the infrastructure, spectrum, and protocol/algorithmic levels are required to realize the intended broadband connectivity goals in 6G. At the infrastructure level, we consider ultra-massive MIMO technology (possibly implemented using holographic radio), intelligent reflecting surfaces, user-centric cell-free networking, integrated access and backhaul, and integrated space and terrestrial networks. At the spectrum level, the network must seamlessly utilize sub-6 GHz bands for coverage and spatial multiplexing of many devices, while higher bands will be mainly used for pushing the peak rates of point-to-point links. 
Finally, at the protocol/algorithmic level, the enablers include improved coding, modulation, and waveforms to achieve lower latency, higher reliability, and reduced complexity. 
\end{abstract}



\maketitle

\section{Introduction}

\input{1_introduction}



\section{From 5G to 6G} \label{sec:use_cases}

This section describes the state-of-the-art in 5G and presents our vision for 6G use cases and performance indicators.

\subsection{5G Today}

\input{2_5G_today}

\subsection{6G Use Cases}

\input{2_use_cases}

\subsection{6G Key Performance Indicators}

\input{2_KPIs}

\section{Enablers at the Spectrum Level}

\input{3_spectrum}

\subsection{Enablers for Terahertz Communications}

\input{3_THz}

\subsection{Enablers for Optical Wireless Communications}

\input{3_optical}

\section{Enablers at the Infrastructure Level} \label{sec:infrastructure}

Another key factor in the capacity formula in \eqref{eq:capacity-formula} is the multiplexing gain. There is a variety of ways that one can increase the number of simultaneously active UEs in 6G systems, including using larger antenna arrays and distributed arrays. This section will cover these technologies, as well as other infrastructure-related 6G enabling technology.

\subsection{Ultra-Massive MIMO and Holographic Radio}

\input{4_beamforming}

\subsection{Intelligent Reflecting Surfaces} \label{sec:4_IRSs}

\input{4_IRSs}

\subsection{Scalable Cell-Free Networking with User-Centric Association}

\input{4_cell_free}

\subsection{Integrated Access and Backhaul}

\input{4_integrated_access_backhaul}

\subsection{Integrated Space and Terrestrial Networks}

\input{4_integrated_space_terrestrial}

\subsection{Integrated Wideband Broadcast Networks} \label{section:broadcast_multicast}

\input{4_broadcast}

\section{Enablers at the Protocol/Algorithmic Level}
\label{sec:protocol}

The third factor in \eqref{eq:capacity-formula} is the spectral efficiency, which is closely related to the modulation and coding, interference management, and resource management in general. These are some of the enabling technologies covered in this section.

\subsection{Coding, Modulation, Waveform, and Duplex}

\input{5_coding_modulation_waveforms_duplex}

\subsection{Machine Learning-Aided Algorithms}

\input{5_machine_learning}

\subsection{Coded Caching}

\input{5_caching}

\section{6G Broadband connectivity for rural/underdeveloped areas} 

\input{6_full_coverage}

\section{Concluding remarks}

\input{7_summary}

\footnotesize
\bibliographystyle{IEEEtran}
\bibliography{IEEEabbr,bibliography}

\input{author_bio}

\EOD
\end{document}

%% file: 1_introduction.tex
\IEEEPARstart{W}{hile} fifth-generation (5G) wireless networks are being deployed in many parts of the world and 3rd generation partnership project (3GPP) is about to freeze the new long-term evolution (LTE) Release 16, providing enhancements to the first 5G new radio (NR) specifications published in 2018, the research community is starting to investigate the next generation of wireless networks, which will be designed during this decade and will shape the development of the society during the next decade. The following are key questions: \textit{(a)} What will beyond-5G (B5G) and the sixth-generation (6G) wireless networks look like? \textit{(b)} Which new technology components will be at their heart? (c) Will the 1~Tbps frontier be reached in practice? (d) Will uniform coverage and quality of service be fully achieved?

At the moment, it is very challenging and risky to provide definite answers to these questions, even though some key concepts and statements can be already made. A conservative and cautious answer is that \textit{6G networks will be based on a combination of 5G with other known technologies that are not mature enough for being included in 5G}. Based on this argument, any technology that will not be in the 3GPP standards by the end of 2020 will be a possible ingredient of future generations of wireless cellular systems. As an example, full massive multiple-input multiple-output (MIMO) with digital signal-space beamforming is one technology that, although known for some years now, has not been fully embraced by equipment manufacturers in favor of the more traditional codebook-based beamforming. A more daring answer is that \textit{6G networks will be based on new technologies that were not at all considered when designing and developing 5G combined with vast enhancements of technologies that were already present in the previous generation of wireless cellular networks}. As an example, the use of advanced massive MIMO schemes together with cell-free and user-centric network deployments will combine an advanced version of a 5G technology (i.e., massive MIMO), with the fresh concept of cell-free network architectures. Certainly, we can state that 6G wireless systems will:
\begin{itemize}
\item[$\bullet$] Be based on \textbf{extreme densification of the network infrastructure}, such as access points (APs) and intelligent reflecting surfaces (IRSs), which will cooperate to form a cell-free network with seamless quality of service over the coverage area.

\item[$\bullet$] Make intense use of \textbf{distributed processing and cache memories}, e.g., in the form of cloud-RAN technology.

\item[$\bullet$] Continue the trend of complementing the wide-area coverage achieved at sub-6 GHz frequencies by using \textbf{substantially higher carrier frequencies beyond the mmWave through the Terahertz (THz) band and up to visible light (VL)} to provide high-capacity point-to-point links.

\item[$\bullet$] Leverage \textbf{network slicing and multi-access edge computing} to enable the birth of new services with specialized performance requirements and to provide the needed resources to support vertical markets.

\item[$\bullet$] Witness an increasing \textbf{integration of terrestrial and satellite wireless networks}, with a big role played by unmanned aerial vehicles (UAVs) and low-Earth orbit (LEO) micro satellites, to fill coverage holes and offload the network in heavy-load situations.

\item[$\bullet$] Leverage \textbf{machine learning methodologies} to improve the efficiency of traditional model-based algorithms for signal processing and resource allocation.
\end{itemize}

The goal of this paper is to provide a vast and thorough review of the main technologies that will permit enhancing the broadband connectivity capabilities of future wireless networks. The paper indeed extends the huge collaborative effort that has led to the white paper \cite{6GFlagship_WP} with a more specific focus on broadband connectivity, which might be the most important use case in 6G (though far from the only one) and with abundant technical details and numerical study examples. The paper describes the physical-layer (PHY) and medium-access control (MAC) layer methodologies for achieving 6G broadband connectivity with very high data rates, up to the Tbps range.

\subsection{Related Works}

The existing literature contains several visionary articles that speculate around the potential 6G requirements, architecture, and key technologies~\cite{David20186G,9040264_Marco,8766143_Zhengquan,8792135_Emilio,rajatheva2020white}. One of the earliest articles with a 6G vision and related requirements is~\cite{David20186G}, which analyzed the need of 6G from a user's perspective and concluded that ultralong battery lifetime (maximum energy savings) will be the key focus of 6G rather than high bit rates. The authors envisioned that indoor communications will be completely changed by moving away from wireless radio communications to optical free-space communications. In~\cite{yang20196g}, the authors presented a vision for 6G mobile networks and an overview of the latest research activities on promising techniques that might be available for utilization in 6G. This work claimed that the major feature of 6G networks will be their flexibility and versatility, and that the design of 6G will require a multidisciplinary approach. New technology aspects that might evolve wireless networks towards 6G are also discussed in~\cite{9040264_Marco}. The authors analyzed the scenarios and requirements associated to 6G networks from a full-stack perspective. The technologies aiming to satisfy such requirements are discussed in terms of spectrum usage, PHY/MAC and higher layers, network architectures, and intelligence for 6G. In~\cite{8792135_Emilio}, the authors highlighted new services and core enabling technologies for 6G networks, identifying sub-THz and optical communication as the key physical layer technologies to achieve the 6G requirements. The authors also provided a 6G roadmap that starts with forming a 6G vision and ends in a 6G proof-of-concept evaluation using testbeds.

A vision for 6G that could serve as a research guide for the post-5G era was presented in~\cite{dang2020should}. Therein, it was suggested that high security, secrecy, and privacy will be the key features of 6G because human-centric communications will still be the most important application. The authors provided a framework to support this vision and discussed the potential application scenarios of 6G; furthermore, they examined the pros and cons of candidate technologies that recently appeared. In~\cite{viswanathan2020communications}, the communication needs and technologies in the time frame of 6G were presented. According to the authors, the future of connectivity will lie in unifying our experience across true representation of the physical and biological worlds at every spatial point and time instant. New themes such as man-machine interfaces, ubiquitous universal computing, multi-sensory data fusion, and precision sensing and actuation will drive the 6G system requirements and technologies. The authors predicted that artificial intelligence (AI) has the potential to become the foundation for the 6G air interface and discussed five more major technology transformations to define 6G. An analysis of 6G networks, extending to its relationships with B5G, core services, required key performance indicators (KPIs), enabling technologies, architecture, potential challenges, promising solutions, opportunities, and future developmental direction, was presented in~\cite{gui20206g}. The authors identified five core services as well as eight KPIs for 6G.

Networld2020 is a European Technology Platform for community discussion on the future research technologies in telecommunications. A white paper on the strategic research and innovation agenda for 2021--27 was published in 2018 and updated in 2020 by a large group of researchers in order to provide guidance for the development of the future European Union R\&D program, such as the 9-th EU Framework Program \cite{SRIA20}. In particular, the chapter ``Radio Technology and Signal Processing'' of the SRIA listed ten enabling technologies relevant to the 6G air interface design. Some prospective key enabling techniques for 6G and beyond have been surveyed in~\cite{akyildiz20206g}. The authors envisioned that the 6G wireless systems will be mainly driven by a focus of unrestricted availability of high quality wireless access. The authors highlighted technologies that are vital to the success of 6G and discussed promising early-stage technologies for beyond 6G communications such as Internet of NanoThings, the Internet of BioNanoThings, and Quantum Communications. Lastly, \cite{Bjornson2019d} envisioned that large antenna arrays will be omnipresent in B5G. This work described how the technology can evolve to include larger, denser, and more distributed arrays, which can be used for communication, positioning, and sensing.

\subsection{Paper contribution and organization}
This paper differs from previously cited related works on 6G papers for three main reasons: (i) it is not a general paper on 6G network but it has a special focus on the broadband connectivity, i.e. it describes the key technologies at the PHY and MAC layers, and the related research challenges, that will permit achieving broadband wireless connections at 1 Tbps in multiuser scenarios; (ii) the paper length and exhaustiveness makes it different from previously cited papers that have appeared in magazines and not in technical journals; and (iii) differently from many review papers that turn down the use of equations and illustrate concepts in a colloquial style, this paper contains several  technical dives into various key topics, providing in-depth details, equations, and illustrative numerical results to corroborate the technical exposition.

\begin{figure*}[t]
\centering
\includegraphics[scale=0.6]{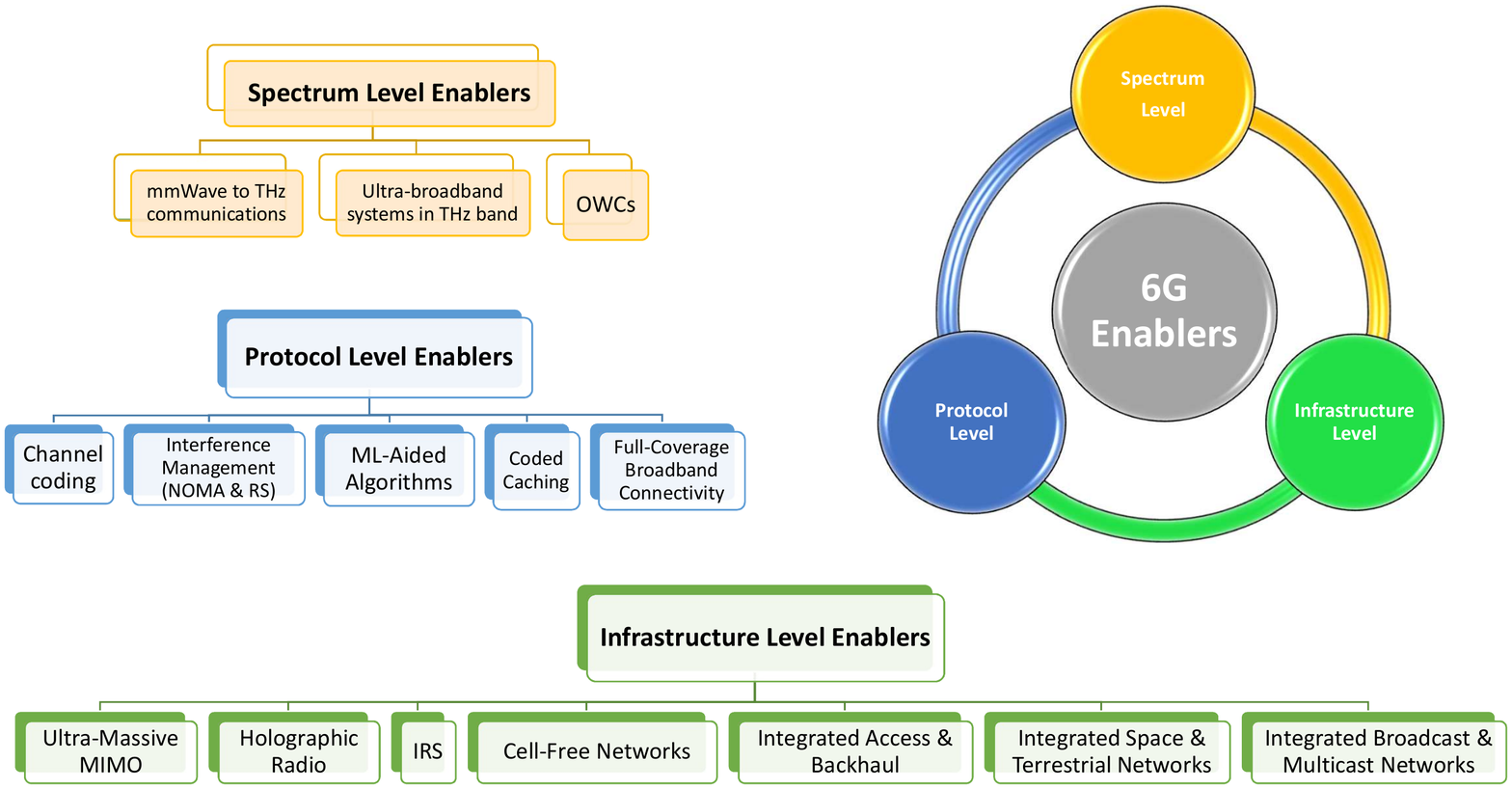}
\vspace*{-2.1cm}
\caption{The expected enablers for achieving 6G broadband access.}
\label{fig:6G_Enablers}
\end{figure*}

The rest of this paper is organized as follows. Section II provides a discussion about the technologies adopted and included in current 5G NR specifications. This helps shedding some light on the discrepancy between technologies and algorithms available in the scientific literature and the ones that are really retained by 3GPP for inclusion in their released specifications; moreover, it also helps to realize that some of the technologies that almost a decade ago were claimed to be part of future 5G systems now are claimed as being part of future 6G networks. An illustrative example of this phenomenon is the use of carrier frequencies above-30 GHz: for some reason, technology at such high frequencies was not yet mature for mass market production at a convenient cost, and so  the massive use of higher frequencies has been delayed and is not currently part of 5G NR specifications. Section II also provides a discussion of what the authors believe to be the most relevant use cases for future 6G networks. The detailed use cases describe applications and scenarios that cannot be implemented with current 5G technology, and that need a boost in broadband connectivity of 2-3 orders of magnitude. Finally, the section is concluded with a description of KPI values envisioned for the next generation of mobile networks. Sections III--V are the containers of the vast bulk of technologies that will enable Tbps wireless rates in 6G networks. We define these technologies as ``enablers'' for achieving 6G broadband connectivity, as illustrated in Fig.~\ref{fig:6G_Enablers}, and categorize them into enablers at the spectrum level (Section III), at the infrastructure level (Section IV), and at the protocol/algorithmic level (Section V). Section III contains thus a treatment of the enablers at the spectrum level. These include the use of THz, optical and VL carrier frequencies. The most promising bandwidths are discussed, along with their pros and cons. In particular, the section provides a detailed discussion of the challenges posed by the propagation channel at THz frequencies. Section IV is devoted to the enablers at the infrastructure level, and thus treats all the technological innovations that mobile operators will have to deploy in order to support Tbps connectivity. These include ultra-massive MIMO and holographic radio, IRSs, scalable cell-free networking with user-centric association, the use of integrated access and backhaul, the integration of terrestrial and non-terrestrial networks, and wideband broadcasting. The Section provides in-depth technical discussion, with equations and numerical results, about the advantages granted by the use of IRSs, by the adoption of cell-free network deployment in place of traditional massive MIMO networks, and by the use of integrated access and backhaul technologies. Then, Section V surveys the enablers at the protocol/algorithmic level. These include new coding, modulation, and duplexing schemes, new transceiver algorithms based on machine learning and  coded caching. The Section provides illustrative numerical results about the performance of polar-QAM constellations for THz communications, and of the spatial tuning methodology to design antenna arrays, still for THz communications. Numerical results are presented also to show the potential of the autoencoder methodology in the design of data detection algorithms based on machine learning. Section VI gives an overview of the technologies to provide full coverage in rural and underdeveloped areas. Indeed, one of the challenges of the wireless communications community for the next decade is to extend the benefits of broadband connections to the largest possible share of the world population. As discussed in the section, full coverage will be possible through a proper exploitation and combination of some of the  methodologies introduced in Sections III-V. Finally, Section VII contains concluding remarks and wraps up the paper. For the reader's ease, the list of acronyms used in this paper is reported in Table~\ref{table:acr}.

\begin{table*}[t!]
\begin{center}
\caption{List of Acronyms}
\begin{tabular}{|p{1.62cm}|p{6.3cm}|}
\hline
\multicolumn{1}{|c|}{\textbf{Acronym}}  & \multicolumn{1}{|c|}{\textbf{Definition}} \\ \hline \hline
3GPP                                    & 3rd generation partnership project \\ \hline
5G                                      & Fifth-generation \\ \hline
6G                                      & Sixth-generation \\ \hline
AI                                      & Artificial intelligence \\ \hline
AoA                                     & Angle of arrival \\ \hline
AoD                                     & Angle of departure \\ \hline
AP                                      & Access point \\ \hline
APSK                                    & Amplitude
and phase-shift keying \\ \hline
AR                                      & Augmented reality \\ \hline
ARQ                                     & Automatic repeat request \\ \hline
AWGN                                    & Additive white Gaussian noise \\ \hline
B5G                                     & Beyond 5G \\ \hline
BER                                     & Bit error rate \\ \hline
BiCMOS                                  & Bipolar complementary metal-oxide-semiconductor \\ \hline
BP                                      & Belief propagation \\ \hline
BPSK                                    & Binary phase-shift keying \\ \hline
BS                                      & Base station \\ \hline
CA-CSL                                  & CRC-aided successive cancellation list \\ \hline
CC                                      & Coded caching \\ \hline
CDF                                     & Cumulative distribution function \\ \hline
CF-UC                                   & Cell-free with user-centric \\ \hline
CMOS                                    & Complementary metal-oxide-semiconductor \\ \hline
CoMP                                    & Coordinated multi-point \\ \hline
CPRI                                    & Common public radio interface \\ \hline 
CPU                                     & Central processing unit \\ \hline
CRC                                     & Cyclic redundancy check \\ \hline
CSI                                     & Channel state information \\ \hline
DFT                                     & Discrete Fourier transform \\ \hline
DFT-s-OFDM                              & DFT-spread-orthogonal frequency-division multiplexing \\ \hline
DMRS                                    & Demodulation reference signals \\ \hline
EoA                                     & Elevation angle of arrival \\ \hline
EoD                                     & Elevation angle of departure \\ \hline
ETSI                                    & European Telecommunications Standards Institute \\ \hline
FCC                                     & Federal Communications Commission \\ \hline
FCF                                     & Full cell-free \\ \hline
FDD                                     & Frequency division duplex \\ \hline
FEC                                     & Forward error correction \\ \hline
FHPPP                                   & Finite homogeneous Poisson
point processes \\ \hline
FoV                                     & Field-of-view \\ \hline
FPA-DL                                  & Fractional power
allocation-dowlink \\ \hline
FPA-UL                                  & Fractional power
allocation-uplink \\ \hline
FSO                                     & Free-space optics \\ \hline
GEO                                     & Geostationary-Earth orbit \\ \hline
HAPS                                    & High-altitude platform station\\ \hline
HPC                                     & High performance computing \\ \hline
HST                                     & High-speed train \\ \hline
IAB                                     & Integrated access and backhaul \\ \hline
IBFD                                    &  In-band
full-duplex \\ \hline
I/O                                     & Input/output \\ \hline
IEEE                                    & Institute of Electrical and Electronics Engineers \\ \hline
IID                                     & Independent and identically distributed \\ \hline
IMT                                     & International Mobile Telecommunications \\ \hline
IoT                                     & Internet of things \\ \hline
IR                                      & Infrared \\ \hline
IRS                                     & Intelligent reflecting surface \\ \hline
ISTN                                    & Integrated space and terrestrial network \\ \hline
ITS                                     & Intelligent transportation systems \\ \hline
ITU                                     & International Telecommunication Union \\ \hline
\end{tabular}
\begin{tabular}{|p{1.62cm}|p{6.3cm}|}
\hline
\multicolumn{1}{|c|}{\textbf{Acronym}}  & \multicolumn{1}{|c|}{\textbf{Definition}} \\ \hline \hline
ITU-R                                   & ITU Radiocommunication Sector \\ \hline
KPI                                     & Key performance indicator \\ \hline
LDPC                                    & Low-density parity-check \\ \hline
LED                                     & Light emitting diode  \\ \hline
LEO                                     & Low-Earth orbit  \\ \hline
LiFi                                    & Light fidelity \\ \hline
LIoT                                    & Light-based Internet of things \\ \hline
LoS                                     & Line-of-sight \\ \hline
LTE                                     & Long-term evolution \\ \hline
M2M                                     & Machine-to-machine \\ \hline
MAC                                     & Medium access control \\ \hline
MBS                                     & Macro base station \\ \hline
MR                                      & Mixed reality \\ \hline
MEO                                     & Medium-Earth orbit \\ \hline
MIMO                                    & Multiple-input multiple-output \\ \hline
ML                                      & Machine learning \\ \hline
mMIMO                                   & Massive multiple-input multiple-output \\ \hline
MMSE                                    & Minimum mean square error \\ \hline
mmWave                                  & Millimeter wave \\ \hline
MS                                      & Multi-antenna scheme \\ \hline
NET                                     & Network layer \\ \hline
NGI                                     & Next generation of internet \\ \hline
NLoS                                    & Non-line-of-sight \\ \hline
NOMA                                    & Non-orthogonal multiple access\\ \hline
OFDM                                    & Orthogonal frequency-division multiplexing \\ \hline
NR                                      & New radio \\ \hline
OPEX                                    & Operating expense \\ \hline
OWC                                     & Optical wireless communications \\ \hline
PAPR                                    & Peak-to-average-power-ratio \\ \hline
PDF                                     & Probability density function  \\ \hline
PDP                                     & Power delay profile \\ \hline
PER                                     & Packet error rate \\ \hline
PHY                                     & Physical layer \\ \hline
PN                                      & phase noise \\ \hline
PPA                                     & Proportional power allocation \\ \hline
PPA-DL                                  & Proportional power allocation-dowlink \\ \hline
QAM                                     & Quadrature amplitude modulation \\ \hline
QPSK                                    & Quadrature phase-shift keying \\ \hline
RED                                     &  Reduced-subpacketization scheme \\ \hline
RF                                      & Radio frequency \\ \hline
RS                                      & Rate splitting \\ \hline
RT                                      & Ray tracing \\ \hline
SBS                                     & Small base station \\ \hline
SC                                      & Single-carrier \\ \hline
SNR                                     & Signal-to-noise ratio \\ \hline
SWIPT                                   & Simultaneous wireless information and power transfer \\ \hline
TDD                                     & Time division duplex \\ \hline
UAV                                     & Unmanned aerial vehicle  \\ \hline
UE                                      & User equipment \\ \hline
UPA-DL                                  & Uniform power allocation-downlink \\ \hline
UPA-UL                                  & Uniform power allocation-uplink \\ \hline
UV                                      & Ultraviolet \\ \hline
UW                                      & Ultra-wideband \\ \hline
V2X                                     & Vehicle-to-everything \\ \hline
VL                                      & Visible light \\ \hline
VLC                                     & Visible light communications \\ \hline
VR                                      & Virtual reality \\ \hline
WLAN                                    & Wireless local area network \\ \hline
\end{tabular} \vspace{1mm}
\end{center}
\label{table:acr}
\end{table*}

%% file: 2_5G_today.tex
The 5G NR sets the foundations of what is 5G today. In terms of modulation, 5G  uses the same waveforms as in 4G, except that orthogonal frequency-division multiplexing (OFDM) is mandatory now both in downlink and uplink. Moreover, discrete Fourier transform-spread-OFDM (DFT-s-OFDM) can be used in the uplink. This is a single carrier (SC) like transmission scheme mainly designed for coverage-limited cases due to its better power efficiency. The same waveforms are envisioned to be used up to the 71~GHz frequency band. At the moment, the maximum supported number of transmitted sub-carriers is 3168 which corresponds to a 400~MHz channel bandwidth using 120~kHz subcarrier spacing \cite{3gpp38.104} (or up to 100~MHz in sub-6\,GHz), which can be further aggregated to up to 800~MHz bandwidth. The current study for extending the NR specification up to 71~GHz will consider channel bandwidths up to 2~GHz. 
The 5G NR standard utilizes new channel coding schemes compared to earlier generations. The standard uses low-density parity-check (LDPC) coding for the data channels, while polar coding is used for the control channels when having more than 11 payload bits. The bits are mapped to modulation symbols using either QPSK, 16-QAM, 64-QAM and 256-QAM both in downlink and uplink, while $\pi/2$-BPSK is supported for the DFT-s-OFDM uplink, which can enable extreme coverage due to its low peak-to-average power ratio \cite{3gpp38.211}. For $\pi/2$-BPSK, the user equipment (UE) is allowed to use very power-efficient transmission by employing frequency-domain spectral shaping.

Multi-antenna techniques are already one of the key parts of the current 5G NR specification. Especially in the lower frequencies, MIMO technology is used to provide spatial multiplexing gains, either by multiplexing multiple users (multi-user MIMO) or increasing the throughput of a single user (single-user MIMO). At the moment, in the downlink, a maximum of two codewords mapped to a maximum of eight layers for a single UE and a maximum of four (orthogonal) layers per UE for multi-user MIMO are supported~\cite{3gpp38.300}. Basically, a maximum of 12 orthogonal DMRS ports are supported, but if the transmitter can perform some form of orthogonalization between the layers, such as using spatial separation, an even higher number of layers is possible~\cite{3gpp38.211}. In the uplink, a maximum of four layers are supported. Single-user MIMO is not supported for DFT-s-OFDM. In the uplink, both codebook and non-codebook based transmissions are supported. In the codebook based precoding, the gNB informs the UE of the transmit precoding matrix to use in the uplink, while in the non-codebook based precoding, the UE can decide the precoding based on feedback from its sounding reference signal. However, only wideband precoding is supported in both cases, which means that the same precoding must be utilized over all allocated subcarriers. While academia is often regarding 64 antenna ports has the minimum number that constitutes massive MIMO \cite{Bjornson2019d}, 3GPP defines massive MIMO as having more than eight ports, so massive MIMO is already widely supported. However, there have been plenty of commercial products with both 32 and 64 ports for use in sub-6 GHz bands \cite{Bjornson2019d}.

In the higher frequency bands, the number of antenna elements increase significantly to compensate for the element size shrinks and increasing signal penetration losses.
5G NR in mmWave frequencies has been designed relying on the assumption of many antenna elements which perform analog beamforming. Basically the beam sweeping phase in sub-6 GHz frequencies supports up to 8 beams in one dimension, while up to 64 different beams can be used in two dimensions (vertical/horizontal) in higher bands \cite{3gpp38.213}. Release 16 has also included support for transmissions from multiple transmission and reception points and several other enhancements to facilitate use of large number of antennas, and the improvements will further continue in Release 17 \cite{RP-193133}.

There are new features in Release 16 such as 5G in unlicensed bands, operations using multiple transmission and reception points, V2X applications, or IAB. The latter is a solution to provide backhaul or relay connectivity using the same resources as for UE access, without requiring additional sites, equipment, or resources. The donor gNB has a fiber connection to the core network. The radio connection from the donor gNB to the IAB node is the same as used to connect UEs \cite{3gpp38.874}. 

The work on Release 17 will start in the second half of 2020. It is expected to include enhancements of the broadband connectivity related to MIMO, dynamic spectrum sharing, coverage extension, dual connectivity, UE power saving, IAB, and data collection. As new features, there will be support for up to 71\,GHz frequencies, multicast and broadcast services \cite{RP-193248}, multi-SIM devices, non-terrestrial networks, and sidelink relaying.

%% file: 2_use_cases.tex
\begin{figure*}[t]
\centering
\includegraphics[scale=0.85]{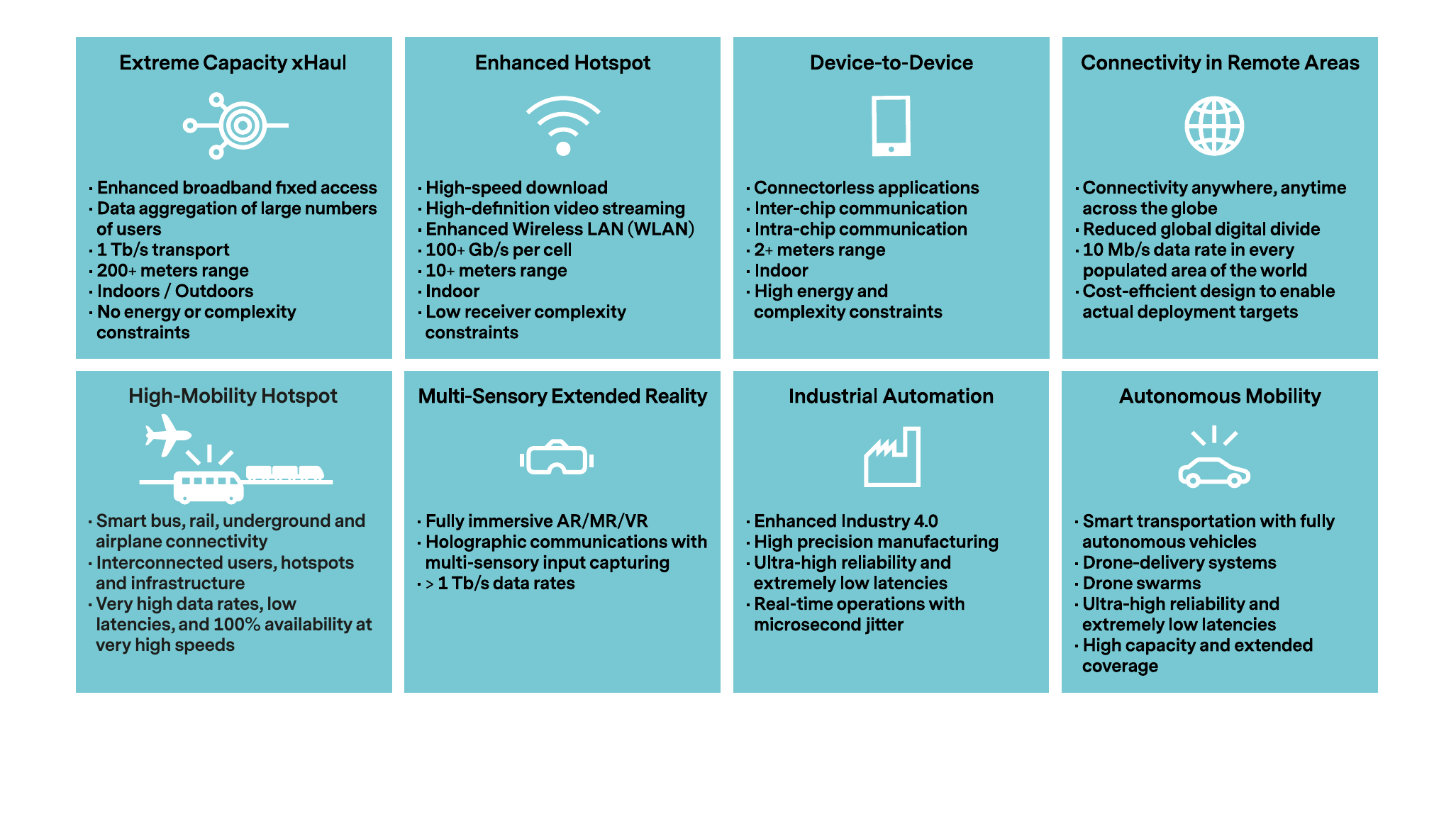}
\vspace{-10mm}
\caption{Different 6G use cases and applications.}
\label{fig:Use1}
\end{figure*}

The emergence and need for 6G technology will be governed by  unprecedented performance requirements arising from exciting new applications foreseen in the 2030 era, which existing cellular generations will not be able to support. This paper focuses on applications that require broadband connectivity with high data rates and availability, in combination with other specialized characteristics. The following is a list of the potential new use cases and applications in 6G which will help to understand the key requirements of future 6G systems. The list should not be seen as a replacement but complement to existing 5G use cases. A summary of the use cases discussed is presented in Fig.~\ref{fig:Use1}.

\begin{itemize}
\item \textbf{Extreme capacity xHaul:} This use case refers to a fixed symmetric point-to-point link targeting high data rates without energy or complexity constraints since no user devices are involved. This can only be enabled using a combination of high bandwidth and high spectral efficiency. The envisioned ultra-dense network topology in urban areas with extreme capacity and latency requirements makes fiber-based backhauling highly desirable but complicated due to the limited fiber network penetration (variable from country to country) and related extension costs. Hence, wireless infrastructure is needed as a flexible means to complement optical fiber deployment, both indoors and outdoors, to avoid bottlenecks in the backhaul (or xHaul). Ultra-high speed is required since the backhaul aggregates the data rates of many user devices. The xHaul can also provide efficient access to computing resources at the edge or in the cloud. Fig.~\ref{fig:Use2} depicts an example setup of an extreme capacity xHaul network in an indoor environment.

\begin{figure*}[t]
\centering
\includegraphics[scale=1]{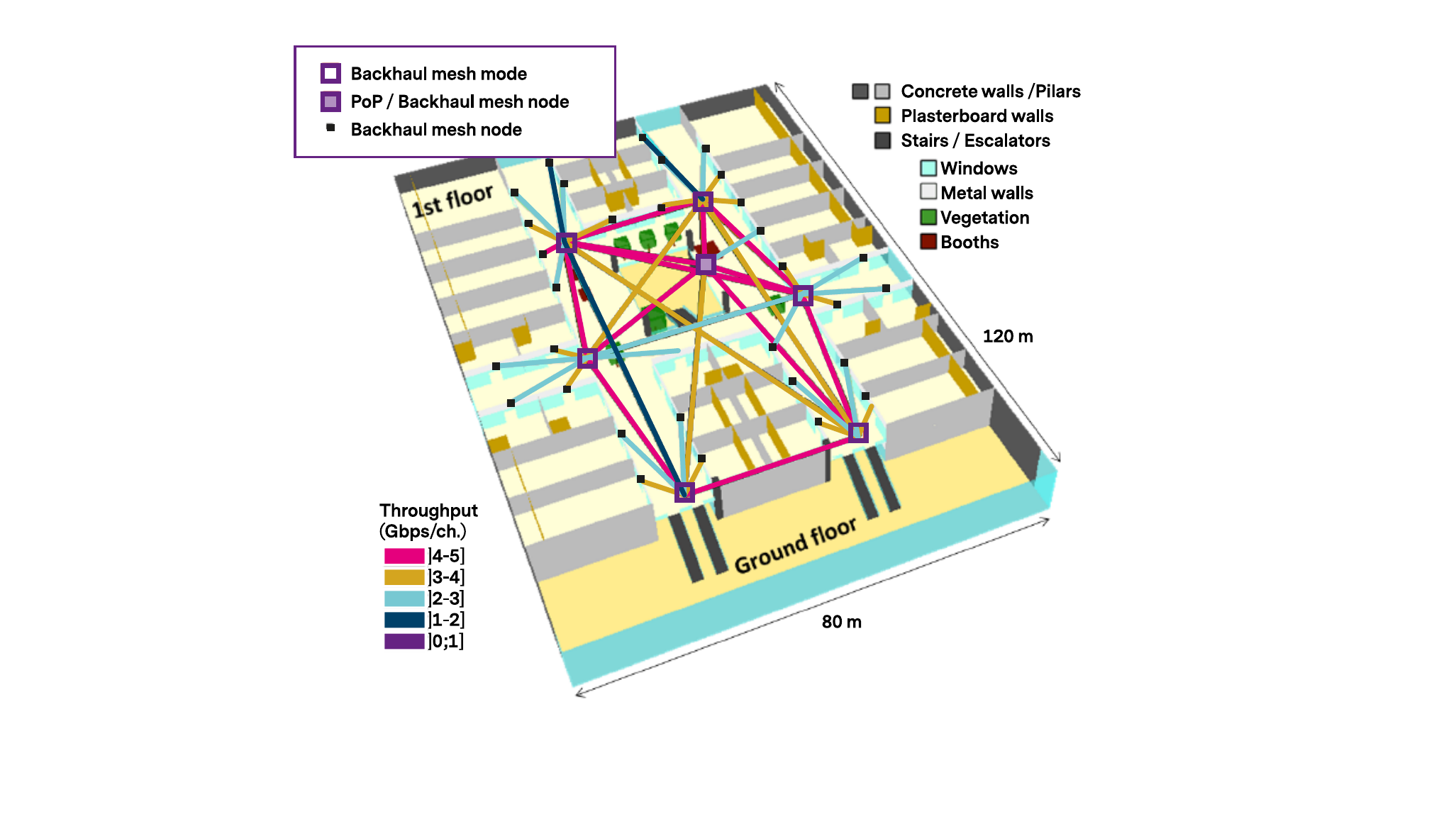}

\vspace{-8mm}

\caption{An example setup of an extreme capacity xHaul network in an indoor environment. Source: \cite{9083890_jean}.}
\label{fig:Use2}
\end{figure*}

\item \textbf{Enhanced hotspot:} An enhanced hotspot entails a high-rate downlink from the AP to several user devices, with short coverage and low receiver complexity constraints. The envisaged applications are hotspots delivering high-speed data to demanding applications such as high-definition video streaming and enhanced WLAN technology.

\item \textbf{Short-range device-to-device communications:} A symmetric high-rate point-to-point link with very stringent energy and complexity constraints is considered here. This use case focuses on data exchange between user devices that are within a short distance, with limited involvement of the network infrastructure. This use case also includes inter/intra-chip communications and wireless connectors, among others.

\item \textbf{High-mobility hotspot:} This use case is an umbrella for multiple applications, such as smart bus, rail, underground, or even airplane connectivity. High-speed mobile communication while on mass public transportation systems has drastically evolved in the last few years and it is expected by passengers. As part of these applications, there are different scenarios to be considered, including user to hotspot connectivity (e.g., inside the train/bus/plane), mobile hotspot to fixed infrastructure (e.g., from trains/bus/plane to fixed networked stations), moving base station (providing access on the move to users outside the vehicle), or hotspot to hotspot, among others. The biggest challenge is posed by the mobile hotspot to fixed infrastructure scenario because this link needs to achieve very high data rates, low latencies, as well as close to 100\% availability while traveling at speeds potentially exceeding hundreds of km/h. To support all users' aggregated traffic, a bandwidth of several GHz (or even higher) is needed to accommodate up to 100 Gbps data rates. Such high-data rate and huge bandwidth requirements are a strong motivation to explore the mmWave and THz bands~\cite{GuanITSM}.


\item \textbf{Multi-sensory extended reality:} AR/MR/VR applications, capturing multi-sensory inputs and providing real-time user interaction are considered under this use case. Extremely high per-user data rates in the Gbps range and exceptionally low latencies are required to deliver a fully immersive experience \cite{9040264_Marco}. Remote connectivity and interaction powered by holographic communications, along with all human sensory input information, will further push the data rate and latency targets. Multiple-view cameras used for holographic communications will require data rates in the order of terabits per second \cite{8792135_Emilio}.

\item \textbf{Industrial automation and robotics:} Industry 4.0 envisions a digital transformation of manufacturing industries and processes through cyber-physical systems, IoT networks, cloud computing, and artificial intelligence. In order to achieve high-precision manufacturing, automatic control systems, and communication technologies are utilized in the industrial processes. Ultra-high reliability in the order of $1$-$10^{-9}$ and extremely low latency around 0.1-1 ms round-trip time are expected in communications, along with real-time data transfer with guaranteed microsecond delay jitter in industrial control networks \cite{8792135_Emilio}. While 5G initiated the implementation of Industry 4.0, 6G is expected to reveal its full potential supporting these stringent requirements by using novel, disruptive technologies brought by 6G.

\item \textbf{Autonomous mobility:} The smart transportation technologies initiated in 5G are envisioned to be further improved towards fully autonomous systems, providing safer and more efficient transportation, efficient traffic management, and improved user experiences. Connected autonomous vehicles demand reliability above 99.99999\% and latency below 1 ms, even in very high mobility scenarios up to 1000 km/h \cite{8792135_Emilio, 9040264_Marco}. Moreover, higher data rates are required due to the increased number of sensors in vehicles that are needed to assist autonomous driving. Other autonomous mobility solutions such as drone-delivery systems and drone swarms are also evolving in different application areas such as construction, emergency response, military, etc. and require improved capacity and coverage \cite{9040264_Marco}.

\item \textbf{Connectivity in remote areas:} Half of the world's population still lacks basic access to broadband connectivity. The combination of current technologies and business models have failed to reach large parts of the world. To reduce this digital divide, a key target of 6G is to guarantee 10 Mbps in every populated area of the world, using a combination of ground-based and spaceborne network components. Importantly, this should not only be theoretically supported by the technology but 6G must be designed in a sufficiently cost-efficient manner to enable actual deployments that deliver broadband to the entire population of the world.

\item \textbf{Other use cases:} Some other applications that 6G is expected to enable or vastly enhance include: internet access on planes, wireless brain-computer interface based applications,
broadband wireless connectivity inside data centers, Internet of Nano-Things and Internet of Bodies through smart wearable devices and intrabody communications achieved by implantable nanodevices and nanosensors \cite{8766143_Zhengquan}.
\end{itemize}

%% file: 2_KPIs.tex
The KPIs have to a large extent stayed the same for several network generations \cite{ITU-IMTA,ITU-IMT2020}, while the minimum requirements have become orders-of-magnitude sharper. One exception is the energy efficiency, which was first introduced as a KPI in 5G, but without specifying concrete targets. We believe 6G will mainly contain the same KPIs as previous generations but with much higher ambitions. However, it is of great importance that the existing KPIs be critically reviewed and new KPIs be seriously considered covering both technology and productivity related aspects and sustainability and societal driven aspects \cite{6GFlagship_WP}. While the KPIs were mostly independent in 5G (but less stringent at high mobility and over large coverage areas), a cross-relationship is envisaged in 6G through a definition of groups. All the indicators in a group should be fulfilled at the same time, but different groups can have diverse requirements. The reason for this is that we will move from a situation where broadband connectivity is delivered in a single way to a situation where the requirements of different broadband applications will become so specialized that their union cannot be simultaneously achieved. Hence, 6G will need to be configurable in real-time to cater to these different groups.

The following are the envisaged KPIs.


\begin{table}[t]
\begin{center}
\caption{A Comparison of 5G and 6G KPIs \cite{8766143_Zhengquan, 8792135_Emilio, 9040264_Marco,Bjornson2018b}}
\label{table:kpi}
\begin{tabular}{|p{3cm}|p{2cm}|p{2cm}|}
\hline
\multicolumn{1}{|c|}{\textbf{KPIs}} & \multicolumn{1}{|c|}{\textbf{5G}} & \multicolumn{1}{|c|}{\textbf{6G}} \\ \hline \hline
Peak data rate                      & $20$ Gb/s                         & $1$ Tb/s \\ \hline
Experienced data rate               & $0.1$ Gb/s                        & $1$ Gb/s\\ \hline
Peak spectral efficiency            & $30$ b/s/Hz                       & $60$ b/s/Hz \\ \hline
Experienced spectral efficiency     & $0.3$ b/s/Hz                      & $3$ b/s/Hz \\ \hline
Maximum bandwidth                   & $1$ GHz                           & $100$ GHz \\ \hline
Area traffic capacity               & $10$ Mb/s/m$^{2}$                 & $1$ Gb/s/m$^{2}$ \\ \hline
Connection density                  & $10^{6}$ devices/km$^{2}$         & $10^{7}$ devices/km$^{2}$ \\ \hline
Energy efficiency                   & Not specified                     & $1$ Tb/J \\ \hline
Latency                             & $1$ ms                            & $100$ $\mu$s \\ \hline
Reliability                         & $1-10^{-5}$                       & $1-10^{-9}$ \\ \hline
Jitter                              & Not specified                     & $1$ $\mu$s \\ \hline
Mobility                            & $500$ km/h                        & $1000$ km/h \\ \hline
\end{tabular}  \vspace{1mm}

\end{center}
\end{table}

\begin{itemize}
\item \textit{Extreme data rates}: Peak data rates up to 1~Tb/s are envisaged for both indoor and outdoor connectivity. The user-experienced data rate, which is guaranteed to 95\% of the user locations, is envisioned to reach 1 Gb/s.

\item \textit{Enhanced spectral efficiency and coverage}: The peak spectral efficiency can be increased using improved MIMO technology and modulation schemes, likely up to 60 b/s/Hz. However, the largest envisaged improvements are in terms of the uniformity of the spectral efficiency over the coverage area. The user-experienced spectral efficiency is envisaged to reach 3 b/s/Hz. Moreover, new physical layer techniques are needed to allow for broadband connectivity in high mobility scenarios and more broadly in scenarios for which former wireless networks generations do not fully meet the needs. 

\item \textit{Very wide bandwidths}: To support extremely high peak rates, the maximum supported bandwidth must greatly increase. Bandwidths up to 10 GHz can be supported in mmWave bands, while more than 100 GHz can be reached in sub-THz, THz, and VL bands.

\item \textit{Enhanced energy efficiency}: Focusing on sustainable development, 6G technologies are expected to pay special attention in achieving better energy efficiency, both in terms of the absolute power consumption per device and the transmission efficiency. In the latter case, the efficiency should reach up to  1 terabit per Joule. Hence, developing energy-efficient communication strategies is a core component of 6G.

\item \textit{Ultra-low latency}: The use of bandwidths that are wider than 10 GHz will allow for latency down to 0.1 ms. The latency variations (jitter) should reach down to 1 $\mu$s, to provide an extreme level of determinism.

\item \textit{Extremely high reliability}: Some new use cases require extremely high reliability up to $1$-$10^{-9}$ to enable mission and safety-critical applications. 

\end{itemize} 

It is unlikely that all of these requirements will be simultaneously supported, but different use cases will have different sets of KPIs, and only some will have the maximum requirements mentioned above. A comparison of the 5G and 6G KPIs is shown in Table~\ref{table:kpi}, where also the area traffic capacity and connection density are considered. The KPIs presented there are not limited only to the broadband scenario, but provides an overall idea of expected KPIs for 6G in a broader view. More detailed KPIs for different  verticals and use cases can be found in \cite{6GFlagship_MTC_WP} and \cite{6GFlagship_Trials_WP}.

It is highly likely that 6G will to a large extent carry information related also to non-traditional applications of wireless communications, such as distributed caching, computing, and AI decisions. Thus, we need to investigate whether there is a need to introduce new KPIs for such applications, or if the traditional KPIs are sufficient.

\subsubsection{Scoring the Terabit/s goal}

As indicated by the title of this paper, our main focus is to identify different technologies that can be utilized to reach 1 Tb/s. Such a high KPI requirement can appear in point-to-point use cases, between a single transmitter and a single receiver, for extreme capacity xHaul and device-to-device communications. Alternatively, 1 Tb/s can be the accumulate capacity requirement in point-to-multipoint use cases, for example, hotspots where a base station is spatially multiplexing a larger number of devices.
To give an indication of how one can reach 1 Tb/s, the following high-level capacity formula can be utilized:
\begin{align} \notag
     &\textrm{Multiplexing gain} \times \textrm{Bandwidth} \times \textrm{Spectral efficiency}\\ &= 1 \, \mathrm{ Tb/s}. \label{eq:capacity-formula}
\end{align}
Hence, there are three multiplicative factors that we can play with: 1) The multiplexing gain, which represents the number of devices that are spatially multiplexed (i.e., transmitted at the same time over the same frequency band); 2) The bandwidth (Hz) of the frequency band utilized for data transmission; 3) The spectral efficiency (b/s/Hz) of the transmission to/from the individual devices, which can be obtained using one or multiple data streams/layers per device.

Suppose we can reach a peak spectral efficiency of 60 b/s/Hz and a maximum bandwidth of 100 GHz, as listed in Table~\ref{table:kpi}. If these KPIs are achievable simultaneously by the 6G system, then we reach 6 Tb/s. As indicated earlier, it is unlikely that the KPIs can reach their peak values simultaneously. However, with 100 GHz of spectrum, it is sufficient with 10 b/s/Hz to reach 1 Tb/s. This spectral efficiency requirement is rather easy to achieve; two parallel streams using 16-QAM is sufficient.

Is the 1 Tb/s unreachable when operating in a conventional millimeter-wave band with ``only'' 1 GHz of spectrum and with a spectral efficiency of 10 b/s/Hz? No, this is when the multiplexing gain is the design variable that can be pushed to its limit instead. By spatial multiplexing of 100 devices, the accumulate capacity is 1 Tb/s, even if each device only achieves 10 Gb/s. It is even possible to reach 1 Tb/s in sub-6 GHz bands by an even more aggressive spatial multiplexing.

These widely different ways of scoring the terabit/s goal call for a variety of different technical enablers, at the spectrum level, infrastructure level, as well as protocol/algorithmic level.

%% file: 3_spectrum.tex
\begin{figure*}[t]
\centering
\includegraphics[scale=0.8]{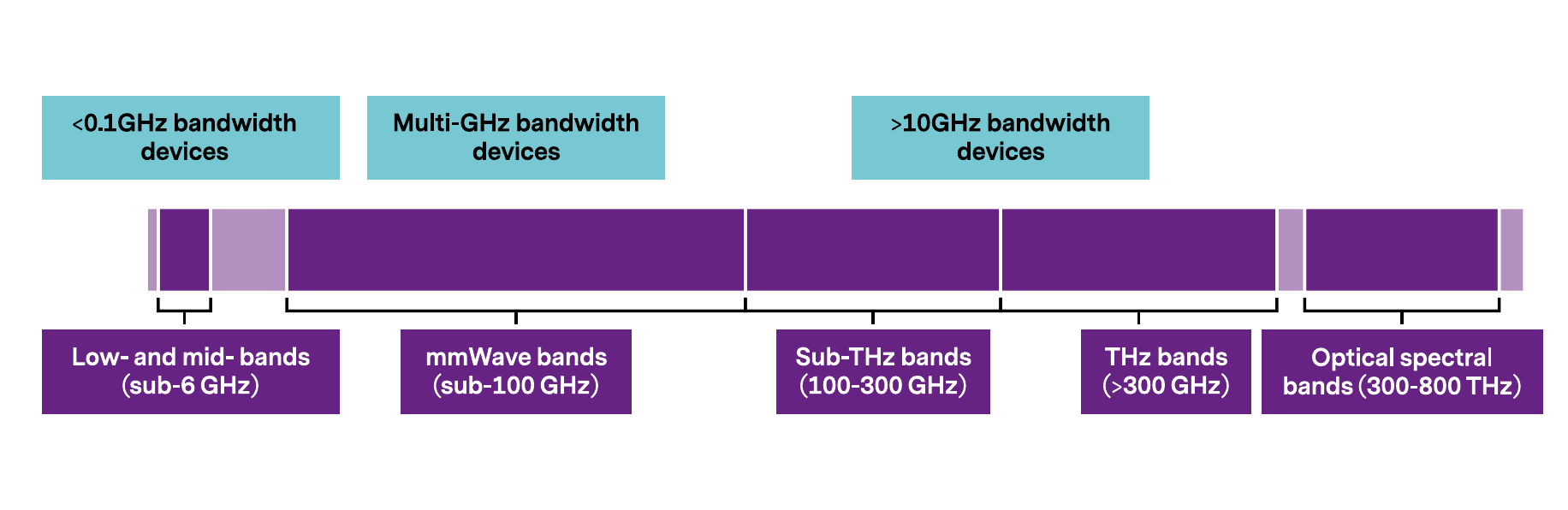}
\caption{Potential spectrum regions for 6G.}
\label{fig:SpectrumBands}
\end{figure*}

One of the three key factors in \eqref{eq:capacity-formula} is the bandwidth. Reaching 1 Tb/s over a point-to-point link will require not only enhanced utilization of current wireless spectrum, but also the adoption of additional spectrum bands for communications. Currently, 5G defines operations separately for sub-6~GHz and 24.25 to 52.6~GHz. Release 17 is expected to extend the upper limit to 71~GHz \cite{RP-193259} and band options up to 114.25~GHz were included in the Release 16 pre-study on NR beyond 52.6~GHz. 
In the 6G era, we expect to see an expansion of the spectrum into many new bands including additional mmWave bands, the THz band (0.1-10~THz) and the optical wireless spectrum (including infrared and visible optical). The potential spectrum regions are illustrated in Fig.~\ref{fig:SpectrumBands}. 
The communication bandwidth is expected to increase at higher frequencies. For example, up to 18~GHz aggregated bandwidth is available for fixed communications in Europe in the frequency band 71-100~GHz, while in the USA, both mobile and fixed communications are allowed. Beyond mmWave, there are also tens of GHz wide bands between 95~GHz and 3~THz recently opened by the FCC for experimental use to support the development of innovative communication systems~\cite{FCC:SpectrumHorizons}.

When moving to higher frequencies in 6G, the intention is not to achieve a gradual increase in operational frequency, as was done in 5G. Instead, we envision a convergence of existing technologies in these different bands into a joint wireless interface that enables seamless handover between bands.
The operation in existing bands will be enhanced in 6G with respect to the KPIs described earlier, but not all targets are expected to be reached in all frequency bands. For example, low frequency bands are often preferable in terms of spectral efficiency, reliability, mobility support, and connectivity density. In contrast, high frequency bands are often preferable in terms of peak data rates and latency. It is not a question of one or the other band, but a dynamic utilization of all bands. When a UE can access several bands, the network can allocate it to the ones that are most suited for its current service requests.

There is plenty of experience in operating wireless communication systems in the sub-6 GHz spectrum. In 5G, moving from sub-6 GHz to mmWave has introduced several technical challenges ranging from initial access to beamforming implementation since fully digital solutions take time to develop. The development of 5G has led to large innovations in these respects. Now, for 6G, all these become even more challenging when going to higher frequencies and, thus, new solutions are needed. In this section, we describe the enablers for THz communications and the enablers for optical wireless communications in detail.



%% file: 3_THz.tex
THz communication~\cite{akyildiz2014teranets,kurner2014towards} is envisioned as a 6G technology able to simultaneously support higher data rates (in excess of 1 Tbps) due to the larger bandwidth and denser networks (hundreds to thousands of spectrum sharing users) due to the shorter ranges.

\subsubsection{THz technologies} \label{terahertz_technologies} For many years, the lack of compact, energy-efficient device technologies (which are able to generate, modulate, detect, and demodulate THz signals) has limited the feasibility of utilizing this frequency range for communications. 
However, many recent advancements with several different device technologies are closing the so-called THz gap~\cite{sengupta2018terahertz}. In an \textit{electronic approach}, the limits of standard silicon CMOS technology~\cite{nikpaik2018219}, silicon-germanium BiCMOS technology~\cite{aghasi20170} and III--V semiconductor transistor~\cite{deal2017660} and Schottky diode~\cite{mehdi2017thz} technologies are being pushed to reach the 1~THz mark. In a \textit{photonics approach}, uni-traveling carrier photodiodes~\cite{song2012uni}, photo-conductive antennas~\cite{huang2017globally}, optical down-conversion systems~\cite{nagatsuma2016advances} and quantum cascade lasers~\cite{lu2016room} are being investigated for THz systems. 

More recently, the use of nanomaterials such as graphene is enabling the development of plasmonic devices~\cite{ferrari2015science}. These devices are intrinsically small, operate efficiently at THz frequencies, and can support large modulation bandwidths. Examples of such devices are, graphene-based plasmonic THz sources~\cite{jornet2014graphene,nafari2018plasmonic}, modulators~\cite{sensale2012broadband,singh2016graphene}, antennas~\cite{llatser2012graphene,jornet2013graphene} and antenna arrays~\cite{hum2013reconfigurable,singh2020operation_tx}. Moreover, graphene is just the first of a new generation of two-dimensional materials, which can be stacked to create new types of devices that leverage new physics. 

Clearly, the technology readiness level of the different approaches will determine the timeline for their adoption in practical communication systems. The majority of  THz technology demonstrators and communication testbeds are largely based on electronic systems~\cite{jastrow2010wireless, moeller20112, kallfass2012,deal2017666, merkle2017testbed,sen2020teranova} consisting of frequency multiplying chains able to generate a (sub-)THz carrier signal from a mmWave oscillator, followed by a frequency mixer that combines the THz carrier with an information-bearing intermediate frequency. Purely photonic~\cite{belem2019300} and hybrid electronic/photonic systems~\cite{koenig2013wireless} have also been demonstrated. Ultimately, the adoption of one or other technology will depend on the application too. For example, THz transceivers for user-equipment will require large levels of integration, achievable at this stage only by CMOS technology. In applications related to backhaul connectivity, high-power III-V semiconductor systems are  likely to be adopted instead.

\subsubsection{THz Wave Propagation and Channel Modeling}
In parallel to device technology developments, major efforts have been devoted to characterize the propagation of THz waves and to correspondingly develop accurate channel models in different scenarios.

In the case of LoS propagation in free space, the two main phenomena affecting the propagation of THz signals are spreading and molecular absorption.
The \emph{spreading loss} accounts for the attenuation due to expansion of the wave as it propagates through the medium, and it is determined by the spreading factor and the antenna effective area, which measures the fraction of the power a receiving antenna can intercept. As the carrier frequency increases, the wavelength becomes smaller (sub-millimetric at THz frequencies), which leads to smaller antennas (when comparing equal-gain antennas) and this results in a lower received power. 
The \emph{molecular absorption loss} accounts for the attenuation that a propagating electromagnetic wave suffers because a fraction of its energy is converted into vibrational kinetic energy in gaseous molecules. THz waves can induce internal resonances in molecules, but are not ionizing. In our frequency range of interest, water vapor is the main absorber.

\begin{figure*}
\centering
\includegraphics[width=0.9\textwidth]{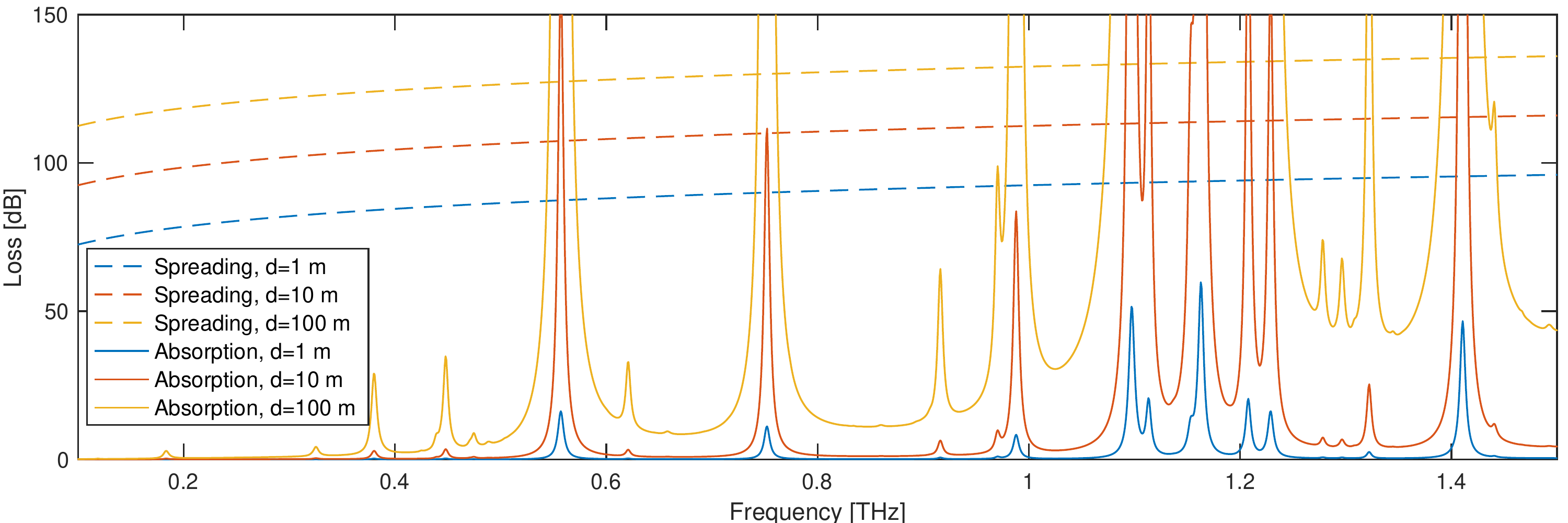}
\caption{Spreading loss (dashed lines) and molecular absorption loss (solid lines) for frequencies ranging from 0.1 to 1.1~THz and three different distances (1, 10 and 100~m).}
\label{fig:thzchannel}
\end{figure*}

In Fig.~\ref{fig:thzchannel}, the spreading and absorption losses are illustrated as functions of the frequency for different distances for standard atmospheric conditions. These results are obtained with the model presented in~\cite{jornet2011channel}, which combines tools from radiative transfer theory, electromagnetics and communication theory, and leverages the contents of the high-resolution transmission molecular absorption database. Molecular absorption defines multiple transmission windows with bandwidths ranging from tens of GHz up to hundreds of GHz depending on the transmission distance and the molecular composition of the medium. For the time being, the majority of THz efforts have been focused on the transmission windows at 120-140~GHz~\cite{xing2018propagation,abbasi2020measurement}, 240~GHz~\cite{koenig2013wireless,kallfass201564}, 300~GHz~\cite{jastrow2010wireless,priebe2011channel,ma2017frequency} and 650~GHz~\cite{moeller20112,deal2017666}. Within these transmission windows, the molecular absorption losses are much lower than the spreading loss. The latter requires the use of high-gain directional antenna systems (or arrays of low-gain antennas with the same physical aperture), to compensate for the limited transmission power of THz sources.

Other meteorological factors, such as rain and snow, cause extra attenuation on the THz wave propagation. In \cite{ishii2016rain}, the the rain attenuation was calculated at 313 GHz and 355 GHz by the four raindrop size distributions Marshall-Palmer, Best, Polyakova-Shifrin, and Weibull raindrop-size distributions, and a specific attenuation model using the prediction method recommended by ITU-R P.838-3 \cite{ITU-R2005}. The results show that the propagation experiment results are in line with the specific attenuation prediction model recommended by ITU-R. In \cite{norouziari2018low}, snow events were categorized as dry snow and wet snow. The comparison between the measured attenuation through wet and dry snowfall showed that a larger attenuation happened through wet snowfall. The general conclusion was that as the snowfall rate increases, the attenuation also increases for both wet and dry snow. Higher attenuation is expected for snowflakes with higher water content. Yet, till now, few experimental studies on snow effect on the wireless channel have been reported systematically compared to the rain effect, not to mention in the THz band range. Thus, there is no recommendation in ITU-R to predict the snow attenuation. In \cite{MaSnowAtt}, an extra loss of 2.8 dB was measured for a distance of 8 m in a LoS link at 300 GHz under the condition of the most significant snowstorm. This implies that the link length of THz outdoor communications can differ considerably under various meteorological conditions.

Beyond free-space propagation, the presence of different types of elements (e.g., objects, furniture, walls, plants, animals, and human beings) affect the propagation of THz signals in realistic scenarios. Depending on the material, shape and dimensions, THz signals might be transmitted, absorbed, reflected or diffracted. For example, THz signals propagate well across common plastics. In the case of paper, cloth and wood, THz signals are partially reflected, partially absorbed and partially transmitted. Metals, glass, and tiles with different coatings are mainly reflectors. Such reflection can be of two types---specular or diffused---depending on the roughness of the surfaces relative to the signal wavelength~\cite{piesiewicz2007scattering,jansen2008impact,jansen2011diffuse,kokkoniemi2016wideband}. 

For the time being, a few multi-path channel models for the THz band have been developed~\cite{priebe2013stochastic,han2015multi,kim2016statistical,hossain2017stochastic}. In~\cite{priebe2013stochastic}, Rayleigh, Gaussian and negative exponential PDFs were fitted to extensive data points obtained by means of ray-tracing simulations of an office scenario. 
In~\cite{han2015multi}, a ray-tracing approach was followed to characterize the multi-path THz channel again in an office-like scenario. For these two works, the main constraint in their approach is the need to re-run extensive simulations for each possible scenario. In~\cite{kim2016statistical}, a stochastic multi-path channel model for an infinite field of scatterers was analytically derived and semi-closed form expressions for the frequency autocorrelation function were computed. 
In~\cite{hossain2017stochastic}, an analytical model for the number of single bounce multi-path components and the power delay profile in a rectangular deployment scenario is derived, considering the density of obstacles, variable geometry of the rectangle, the signal blocking by obstacles and the propagation properties of THz signals.


To experimentally characterize the THz channel, channel sounding techniques are needed. However, channel sounding at the THz band is more challenging than that at traditional cellular bands, mainly because the directionality and attenuation of THz signals are much greater than those at frequencies below 6~GHz \cite{ElayanTerahertz}. In order to overcome high path loss and capture energy from all the directions, channel sounders for high frequencies (such as millimeter wave or THz) often resort to manually or mechanically rotate high-gain steerable horn antennas \cite{metisReport,Impulse2010Sawada}. In such measurements, only a relatively small number of channel samples with limited degrees of freedom (e.g., two-dimensional channel) can be obtained, because the measurements are very costly and time-consuming. Thus, the relatively small measurement data sets must be extended using simulation-based analysis to extract spatial and temporal channel parameters, like the authors of \cite{Hur2016Proposal,Overview_RappaportJianhuaZhang} did for urban cellular channels at mmWave bands. Thus, employing the ray-tracing simulators which are calibrated by measurements becomes an alternative to extending the sparse empirical data sets and analyzing the three-dimensional channel characteristics at mmWave bands \cite{Hur2016Proposal,Wei_12,CXWangSurvey2018}.

\begin{figure}[ht]
\center
\includegraphics[width=0.9\columnwidth,draft=false]{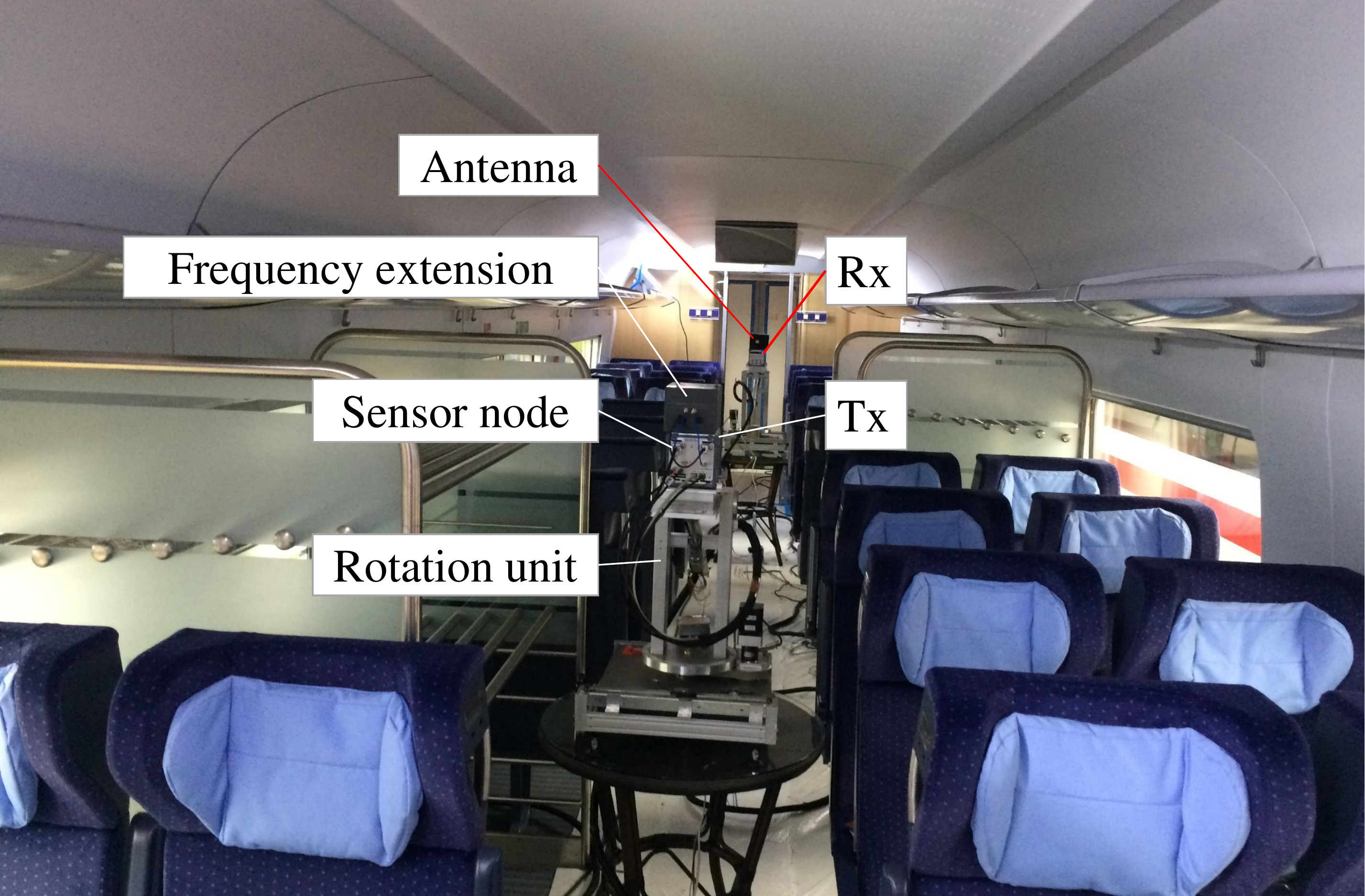}\\
\begin{center}
\caption{\normalsize Measurement campaign in a real HST wagon, originally from Fig. 1 of \cite{GuanIntraTVT}. \label{fig:MeasurementCampagin_1}}\vspace{-0.1in}
\end{center}
\end{figure}

\begin{figure}[ht]
\center
\includegraphics[width=0.9\columnwidth,draft=false]{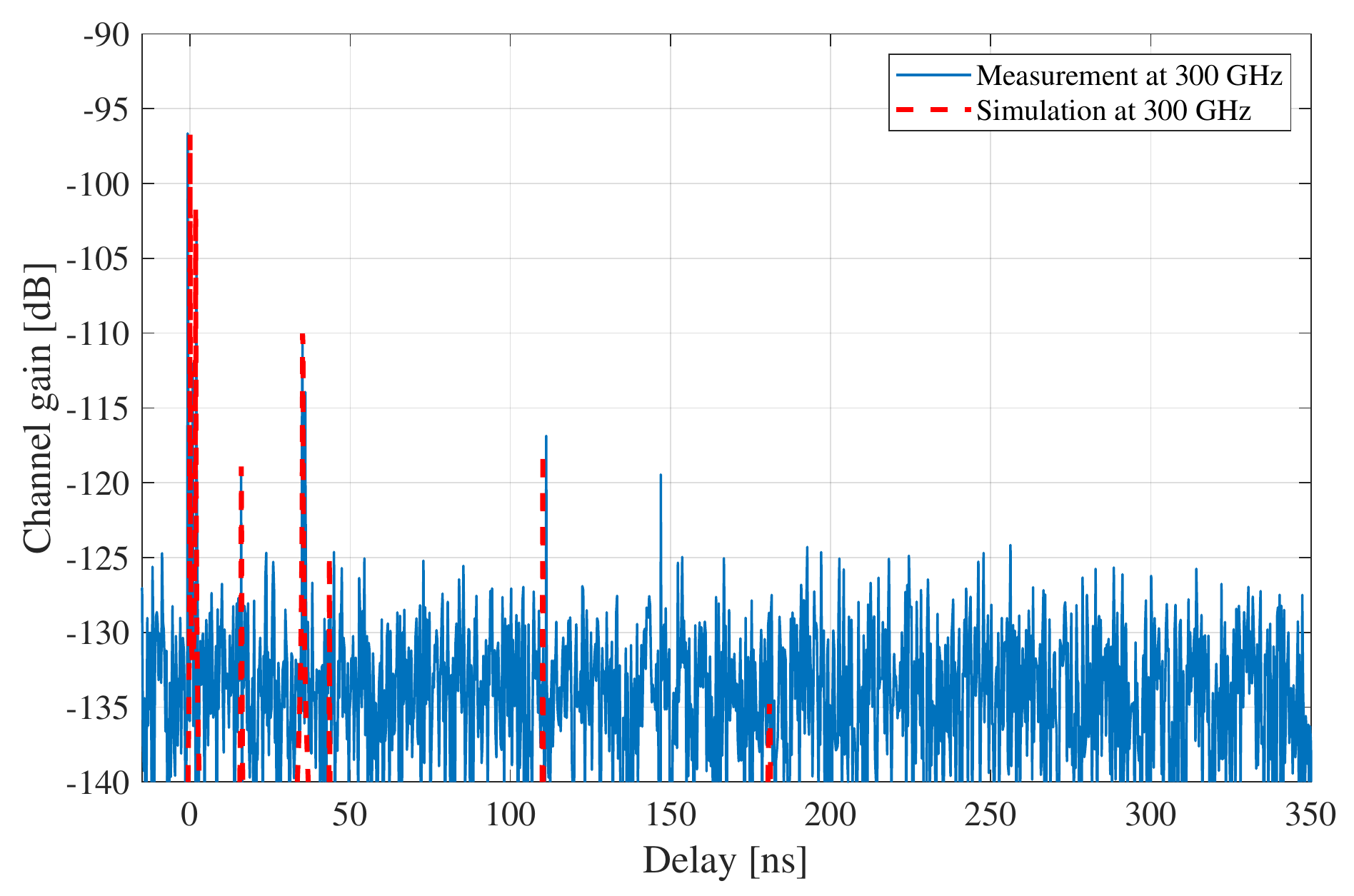}\\
\begin{center}
\caption{\normalsize Comparison of PDP between measurement and RT simulation at 300 GHz, originally from Fig. 12 of \cite{GuanIntraTVT}. \label{fig:300GHz_mea_sim_afterCal_cir}}\vspace{-0.1in}
\end{center}
\end{figure}

As an example in this direction, an \textit{M}-sequence (a kind of pseudo random sequence) correlation based THz UWB channel sounder is customized \cite{Reysounder}. In order to extend the limited channel measurements to more general cases, an in-house-developed HPC cloud-based 3D RT simulator -- CloudRT -- is jointly being developed by Beijing Jiaotong University and Technische Universit\"{a}t Braunschweig, integrating a V2V RT simulator \cite{Guan2013Deterministic,J.Nuckelt2013} and a UWB THz RT simulator \cite{PriebeTHz}. In our recent work \cite{5GmmWaveHSTPartI,5GmmWaveHSTPartII}, CloudRT has so far been validated at 30 GHz and 90 GHz in HST outdoor and tunnel environments, respectively. Based on image theory, CloudRT can output more than 10 properties of every ray, including the type of the ray, reflection order, time of arrival, complex amplitude, AoA, AoD, EoA, EoD, and so on. More information on CloudRT can be found in tutorial \cite{DanpingCloudRT} and \url{http://www.raytracer.cloud}. 

For THz channel characterization, the THz UWB channel sounder and the CloudRT have been utilized in multiple scenarios such as smart rail mobility. Fig. \ref{fig:MeasurementCampagin_1} shows the measurement campaign in a real HST wagon between 300-308 GHz with an 8 GHz bandwidth. The comparison of PDP between the measured and simulated PDP of the intra-wagon scenario is shown in Fig. \ref{fig:300GHz_mea_sim_afterCal_cir}, where the good agreements of the significant paths are achieved both in terms of power and delay. This measurement-validated RT simulator can be utilized to generate more realistic channel data with various setups for comprehensive characterization. More information can be found in \cite{GuanIntraTVT}.

While more extensive multi-path propagation modeling studies are needed, some of the key properties of the THz channel are already clear. While multiple paths are expected, the combination of high-gain directional antennas at the transmitter and the receiver to overcome the otherwise very large propagation losses and the fact that many materials behave as THz (partial) absorbers limits the number of signal copies at the receiver. It is likely that the strongest copies of the signal are generated in the vicinity of the receiver, presumably by metallic objects. It is worth noting that, as a result of the very small wavelength of THz signals, very small metallic obstacles (such as a paper clip) can significantly reflect or diffract THz signals (a phenomenon that can be leveraged for eavesdropping~\cite{ma2018security}). Ultimately, the THz channel is extremely scenario-dependent, which makes the development of real-time ultra-broadband channel estimation and equalization a key task at the physical layer.

\subsubsection{Signal Processing, Communications and Networking Challenges} 
The capabilities of the THz devices and the peculiarities of the THz channel introduce opportunities and challenges for THz communication networks.

On the one hand, the much higher spreading losses  combined with the low power of THz transmitters makes increasing the communication distance \emph{the biggest challenge}~\cite{akyildiz2018combating}. Despite their quasi-optical propagation traits, THz communications possess several microwave characteristics. As a result, the enablers at the infrastructure and algorithm levels introduced in Sec.~\ref{sec:infrastructure} and Sec.~\ref{sec:protocol} are highly relevant for THz communications. 
For example, the term ultra-massive MIMO was first introduced precisely in the context of THz communications~\cite{akyildiz2016realizing}. Since then, many potential architectures have been proposed, ranging from adaptive arrays-of-subarrays antenna architectures in which each subarray undergoes independent analog beamforming~\cite{yan2019dynamic,hadi}, to fully digital beamforming architectures with thousands of parallel channels~\cite{singh2020operation_tx} enabled by the aforementioned graphene-based plasmonic devices. Similarly, IRSs at THz frequencies have also been proposed to overcome obstacles through directed NLoS propagation as well as for path diversity~\cite{nie2019intelligent, ma2020intelligent, singh2020operation}. 

On the other hand, at higher frequencies, molecular absorption has a higher impact. As already explained, the absorption defines multiple transmission windows, tens to hundreds of GHz wide each. As a result, simple single-carrier modulations can already enable very high-speed transmissions, exceeding tens of Gbps. Beyond the traditional schemes, dynamic-bandwidth algorithms that can cope with the distance-dependent absorption-defined transmission channel bandwidth have been proposed for short~\cite{han2016distance} and long~\cite{hossain2019hierarchical} communication distances. Ultimately, resource allocation strategies to jointly orchestrate frequency, bandwidth and antenna resources need to be developed.

An additional challenge in making use of the most of the THz band is related to the digitalization of large-bandwidth signals. While the THz-band channel supports bandwidth in excess of 100~GHz, the sampling frequency of state-of-the-art digital-to-analog and analog-to-digital converters is in the order of 100 Gigasamples-per-second. Therefore, high-parallelized systems and efficient signal processing are needed to make the most out of the THz band. Since channel coding is the most computationally demanding component of the baseband chain, efficient coding schemes need to be developed for Tbps operations. Nevertheless, the complete chain should be efficient and parallelizable. Therefore, algorithm and architecture co-optimization of channel estimation, channel coding, and data detection is required. The baseband complexity can further be reduced by using low-resolution digital-to-analog conversion systems, and all-analog solutions should also be considered. 

Beyond the physical layer, new link and network layer strategies for ultra-directional THz links are needed. Indeed, the necessity for very highly directional antennas (or antenna arrays) simultaneously at the transmitter and at the receiver to close a link introduces many challenges and requires a revision of common channel access strategies, cell and user discovery, and even relaying and collaborative networks. For example, receiver-initiated channel access policies based on polling from the receiver, as opposed to transmitter-led channel contention, have been recently proposed~\cite{xia2019link}. Similarly, innovative strategies that leverage the full antenna radiation pattern to expedite the neighbor discovery process have been experimentally demonstrated~\cite{xia2019expedited}.
All these aspects become more challenging for some of the specific use cases defined in Section~\ref{sec:use_cases}, for example, in the case of wireless backhaul, for which very long distances lead to very high gain directional antennas and, thus, ultra-narrow beamwidths, or smart rail mobility, where ultra-fast data-transfers can aid the intermittent connectivity in train-to-infrastructure scenarios.

Last but not least, it is relevant to note that there is already an active THz communication standardization group, IEEE 802.15 IGTHz, which lead to the first standard IEEE 802.15.3d-2017.

%% file: 3_optical.tex
OWC is an efficient and mature technology that has been developed alongside the cellular technology, which has only used radio spectrum. OWC can potentially satisfy the demanding requirements at the backhaul and access network levels in beyond 5G networks. As the 6G development gains momentum,  comprehensive research activities are being carried out  on the development of OWC-based solutions capable of delivering ubiquitous, ultra-high-speed, low-power consumption, highly secure, and low-cost wireless access in diverse application scenarios \cite{Agiwal}. In particular, this includes the use of hybrid networks that combine OWC with radio frequency or wired/fiber-based technologies. Solutions for IoT connectivity in smart environments is being investigated for developing flexible and efficient backhaul/fronthaul OWC links with low latency and support for access traffic growth \cite{Uysal1}.

The OWC technology covers the three optical bands of infrared (IR: 187-400 THz, 750-1600 nm wavelength), visible light (VL: 400-770 THz, 390-750 nm) and ultraviolet (UV: 1000-1500 THz, 200-280 nm). FSO and VLC are commonly used terms to describe various forms of OWC technology \cite{Uysal2}.  FSO mainly refers to the use of long-range, high speed point-to-point outdoor/space laser links in the IR band \cite{Fawas}, while VLC relies on the use of LEDs operating in the VL band, mostly in indoor and vehicular environments \cite{Pathak}.

In comparison to RF, OWC systems offer significant technical and operational advantages including, but not limited to: i) huge bandwidth, which leads to high data rates; e.g., a recent FSO system achieved a world record data rate of 13.16 Tbps over a distance of 10 km \cite{Uysal2}, and multiple Gbps in indoor VLC setups \cite{Haas1}; However, it is difficult to make fair comparisons in terms of how many bits/Joule that are delivered by VLC systems. This is mainly due to the fact that  VLC is also employed for illumination purpose. In this respect, if the power spent for illumination is considered as a completely wasted power, the power efficiency value of VLC would be around of, or lower than, the 1-10 Gb/Joule value reached by RF in 5G. On the other hand, it is possible to reach high bit rate values per Joule by using small energies in VLC's imaging and coherent applications. ii) operation in unregulated spectrum, thus no license fees and associated costs; iii) immunity to the RF electromagnetic interference; iv) a high degree of spatial confinement, offering virtually unlimited frequency reuse capability, inherent security at the physical layer, and no interference with other devices; v) a green technology with high energy efficiency due to low power consumption and reduced interference. With such features, OWC is well positioned to be a prevailing complement to RF wireless solutions from micro- to macro-scale applications, including intra/inter-chip connections, indoor WA and localization, ITS, underwater, outdoor and space point-to-point links, etc. Beyond the state-of-the-art, however, the dominance of RF-based WA technologies will be challenged and, in order to release the pressure on the RF spectrum, utilization of the optical transmission bands will be explored to address the need of future wireless networks beyond 5G. However, some disadvantages of OWC, compared to RF, needs to be addressed  here for completeness. These are:
\begin{itemize}
\item Tx-Rx alignment/collimation requirements (for long-range free-space OWC; this can be a disadvantage compared to sub-6 GHz RF systems but this is also a requirement for mmWave RF systems).
\item Limitations on receive sensitivity due to the existing detector technology (it is not an issue in RF systems).
\end{itemize}

Optimized solutions need to be devised for the integration of OWC in heterogeneous wireless networks as well as for enhanced cognitive hybrid links in order to improve reliability in coexisting optical and RF links. Moreover, more efforts are needed in characterizing the propagation channel considering the additional practical constraints imposed by the optical front-ends,  as well as developing optimal physical layer design and custom-designed MAC and NET layers.  In   addition, standardization efforts need to be undertaken in the area of OWC to standardization bodies such as IEEE, ITU, ETSI, 3GPP and international forums such as WWRF, Photonics21 and 5G-PPP before it is utilized in 6G.

\begin{figure}[t]
\includegraphics[scale=0.58]{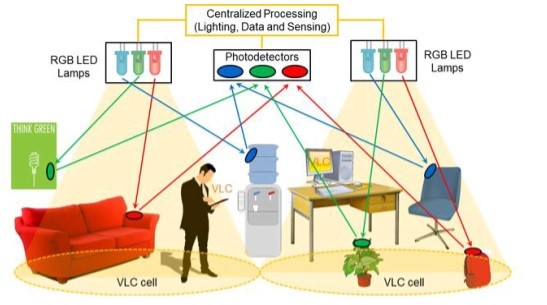}
\caption{An indoor VLC system configuration.} \label{fig:IndoorVLC_systemconfig}
\end{figure}

VLC, a special form of OWC, is also called LiFi \cite{Haas1} in the same way that WLAN systems are today widely known as WiFi. LiFi is a promising technology to provide  local broadband connectivity \cite{Pathak}. As shown in Fig.~\ref{fig:IndoorVLC_systemconfig}, VLC provides high-speed, bi-directional, networked delivery of data through the lighting infrastructure.  When a device moves out of the light cone of one light source, the services can be handed over to the next light source, or eventually, the device can be connected and handed over to an RF-based system, if optical access is not any longer provided.  The former case corresponds to a horizontal handover while the latter is known as vertical handover, and these handovers are needed to provide seamless connectivity to the end-user device. In VLC, all the baseband processing at the transmitter and the receiver is performed in the electrical domain and intensity modulation/direct detection is the most practical scheme. LEDs with a large FoV or laser diodes with small FoV are used to encode and transmit data over the LoS/NLoS optical channel. Photodetectors at the receiver convert data-carrying light intensity back to electrical signals for baseband processing.  However, the intensity-modulating data signal must satisfy a positive-valued amplitude constraint. Hence, it is not possible to straightforwardly apply techniques used in RF communications. A VLC-enabled device inside a pocket or briefcase cannot be connected optically, which is one example of why a hybrid optical-radio wireless network is needed. A reconfigurable optical-radio network is a high-performance and highly flexible communications system that can be adapted for changing situations and different scenarios   \cite{Saud1}, \cite{Saud2}.
 
Performance-wise, data throughput below 100 Mbps can be achieved with relatively simple optical transceivers and off-the-shelf components. Data rates of up to hundreds of Gbps have been demonstrated in laboratory conditions, and it is expected that even Tbps-communications will be achieved in the future. Key optical components for VLC, such as LEDs and photodetectors, have been developed for decades, and they are typically standard low-cost components. VLC is not intended to replace but complement existing technologies. When it comes to vertical applications, VLC can be used for both conventional data services for consumers and support emerging applications and services such as smart cities, smart buildings, e-health, factories of the future, ITS, smart grids, mining, and the IoT. The concept of LIoT exploits light not only to create optical links but also to harvest its energy \cite{Katz1}, \cite{Katz2}.  Thus, an LIoT node can be energy autonomous.

Open research directions in OWC  as well as in VLC toward 6G include:
\begin{itemize}
\item  Accurate VLC channel modeling and characterization for various deployment scenarios with a particular emphasis on user-dense environments. Incorporating user mobility and device orientation into the VLC channel models and combining VLC and RF systems \cite{Murat,Uysal3,Uysal4}.
\item New non-coherent physical-layer transmission schemes such as spatial modulation, and its variations can be used, as well as non-orthogonal communication schemes such as MIMO  \cite{Erdal1,Erdal2,Erdal3,Erdal4,poor1,poor2}.     
\item Exploiting red-green-blue LEDs, development of new materials and optoelectronic devices (e.g., fast non-phosphorous LEDs, micro-LEDs), very fast switching mechanisms between optical and radio systems, etc. \cite{Rajagopal}.
\item Use of OWC to provide secure and safe connectivity in in-body communications applications, including communications to and from the body, communications between sensors inside the body, etc. Recent results have shown that near infrared light can be used for this purpose  \cite{Ahmed1,Ahmed2}.    
\item Design of new and novel  optical IoT, new devices and interfaces to tackle the efficient generation, routing, detection and processing of optical signals \cite{Chowdhury}.
\item For ultra-dense IoT scenarios, there are a number of open challenges that appeal for radical rethinking of network topologies and the design of media access control and network layers in OWC \cite{Chen}.
\item In VLC, to account for multi-user scenarios and user mobility, robust low-complexity multiple access techniques need to be designed together with efficient cellular architectures with user-scheduling and intra-room handover capability, achieving high capacity, low latency, and fairness \cite{Uysal1,Cossu}.   
\item At the MAC layer, due to the small packet sizes used in M2M applications and constraints on sensor devices, robust link quality estimators will be developed and routing algorithms will be devised taking into account the optimal tradeoff between the link capacity, connectivity, latency and energy consumption \cite{Pathak,Gomez,Cailean}.
\item In medium-range OWC, the effects of weather and environmental conditions, ambient noise, and link misalignments need to investigated and to enable connectivity between distant vehicles, physical-layer designs need to be built upon multi-hop transmission to reduce the delay, which is very important in the transmission of road safety related information \cite{Pathak,Cailean,Uysal1}.    
\item For long-range links, extensive research should be carried out for the minimization of the terminal size to enable the technology to be integrated in small satellites, e.g., CubeSats, with data rates up to 10 Gbps and to investigate how to deal with cloud obstruction, site diversity techniques and smart route selection algorithms will be investigated for satellite links and airborne networks, respectively. Also, hybrid RF/FSO and optimized multi-hop transmission techniques will also be investigated to improve link reliability between satellites or HAPS \cite{Kaushal,Sharma,Hughes}.
\end{itemize}

%% file: 4_beamforming.tex
Massive MIMO is a cellular technology where the APs are equipped with a large number of antennas, which are utilized for spatial multiplexing of many data streams per cell to one or, preferably, multiple users. Massive MIMO has become synonymous with 5G, but the hardware implementation and algorithms that are used in practice differ to a large extent from what was originally proposed in~\cite{Marzetta2010a} and then described in textbooks on the topic~\cite{Marzetta2016a,massivemimobook}. For example, compact $64$-antenna rectangular panels with limited angular resolution in the azimuth and elevation domains are adopted instead of physically large horizontal uniform linear arrays with hundreds of antennas \cite{Bjornson2019d}, which would produce very narrow azimuth beams. Moreover, a beam-space approach is taken by dividing the spatial domain into grids of $64$ beams pointing to predetermined angular directions using two-dimensional DFT codebooks. Only one of these predefined beams is selected for each user, thus the approach is only appropriate for LoS communications with calibrated planar arrays and widely spaced users. In general, NLoS channels contain arbitrary linear combinations of these beams, the arrays might have different geometries, and the array responses of imperfectly calibrated arrays cannot be described by DFT codebooks. A practical reason for these design simplifications is that analog and hybrid beamforming were needed to come quickly to market in 5G. However, fully digital arrays will be available for a wide range of frequencies (including mmWave bands) when 6G arrives and, therefore, we should utilize them to implement something capable of approaching the theoretical performance of massive MIMO~\cite{Marzetta2016a,massivemimobook}. Since the massive MIMO terminology has become diluted by many suboptimal design choices in 5G, we will use the term \emph{ultra-massive MIMO}~\cite{akyildiz2016realizing} in this paper to describe a potential 6G version of the technology.

\medskip

\subsubsection{Beamforming Beyond the Beam-Space Paradigm}

Spectrum resources are scarce, particularly in the sub-6 GHz bands that will always define the baseline coverage of a network. Hence, achieving full utilization of the spatial dimensions (i.e., how to divide the available spatial resources between concurrent transmissions) is particularly important. The current beam-space approach describes the signal propagation in three dimensions and, although current planar arrays are capable of generating a set of beams with varying azimuth and elevation angles, it remains far from utilizing all the available spatial dimensions. On the other hand, a $64$-antenna array is capable of creating beams in a $64$-dimensional vector space (where most beams lack a clear angular directivity but can anyway match a physical channel) and utilize the multipath environment to focus the signal at certain points in space.

The low-dimensional approximation provided by the two-dimensional DFT-based beam-space approach can be made without loss of optimality only in a propagation environment with no near-field scattering, predefined array geometries, perfectly calibrated arrays, and no mutual coupling. These restrictions are impractical when considering ultra-massive MIMO arrays that are likely to interact with devices and scattering objects in the near-field, have arbitrary geometries, and feature imperfect hardware calibration and coupling effects. In particular, the latter stems from the non-zero mutual reactance between (hypothetical) isotropic radiators whose spacing is a multiple of half a wavelength~\cite{LNBX18}. That is why the massive MIMO concept was originally designed to use uplink pilots and uplink-downlink channel reciprocity to estimate the entire channel instead of its three-dimensional far-field approximation~\cite{Marzetta2016a,massivemimobook}. Mutual coupling at the transmit side affects not only the overall channel but also the amount of radiated power, which in turn impacts the reciprocity of the so-called information theoretic channel~\cite{LNBX20}. The effects of mutual coupling can be taken into account using models based on circuit theory~\cite{WallaceJ2004,WaldschmidtSW2004,IvrlacNTCAS2010}.

The beamforming challenge for 6G is to make use of physically large panels, since the dimensionality of the beamforming is equal to the number of antennas and the beamwidth is inversely proportional to the array aperture. With an ultra-high spatial resolution, each transmitted signal can be focused on a small region around the receiver, leading to a beamforming gain proportional to the number of antennas as well as enhanced spatial multiplexing capabilities. The latter is particularly important to make efficient use of the low frequency bands where the channel coherence time is large and thereby accommodate the channel estimation overhead for many users. With a sufficient number of antennas, 1~MHz of spectrum in the 1~GHz band can give the same data rate as 100~MHz of spectrum in the 100~GHz band. The reason for this is that the lower frequency band supports spatial multiplexing of 100~times more users since the coherence time is 100~times larger. Recently, novel solutions to reduce the channel estimation overhead and support more users in higher frequency bands have been proposed \cite{BX18,BSL19}.

Ideally, ultra-massive MIMO should be implemented using fully digital arrays with hundreds or thousands of phase-synchronized antennas. This is practically possible in both sub-6 GHz and mmWave bands~\cite{Bjornson2019d}, although the implementation complexity grows with the carrier frequency. On the one hand, new implementation concepts are needed that are not stuck in the suboptimal beam-space paradigm, as in the case of hybrid beamforming, but can make use of all the spatial dimensions. On the other hand, new device technologies, possibly leveraging new materials, can be utilized to implement on-chip compact ultra-massive antenna arrays that can potentially enable fully digital architectures~\cite{singh2020operation_tx}. Doubly massive MIMO links, wherein a large number of antennas are present at both sides of the communication link, will also be very common at mmWave frequencies~\cite{Buzzi_DMIMO1,Buzzi_DMIMO2,8269171,Buzzi_DMIMO3_chapter}.
Note that orbital angular momentum methods cannot be used to increase the channel dimensionality of MIMO links since these dimensions are already implicitly utilized by the MIMO technology \cite{OAMedfors}.


Continuous-aperture antennas can be considered to improve the beamforming accuracy. While the beamwidth of the main-lobe is determined by the array size, a continuous aperture gives cleaner transmissions with smaller side-lobes~\cite{Asilomar2019a}. One possible method to implement such technology is to use a dense array of conventional discretely spaced antennas, but the cost and energy consumption would be prohibitive if every antenna element has an individual RF chain. Another method is to integrate a large number of antenna elements into a compact space in the form of a meta-surface, but this would be limited to the passive setup described in Section~\ref{sec:4_IRSs}. A possible solution to implement active continuous-aperture antennas is given by the holographic radio technology described next.

\medskip

\subsubsection{Holographic Radio}

Holographic radio is a new method to create a spatially continuous electromagnetic aperture to enable holographic imaging-level, ultra-high density spatial multiplexing with pixelated ultra-high resolution~\cite{Zong20196G}. In general, holography records the electromagnetic field in space based on the interference principle of electromagnetic waves. The target electromagnetic field is reconstructed by the information recorded by the interference of reference and signal waves. The core of holography is that the reference wave must be strictly coherent as a reference and the holographic recording sensor must be able to record the continuous wave-front phase of the signal wave so as to record the holographic electromagnetic field with high accuracy. Because radio and light waves are both electromagnetic waves, holographic radio is very similar to optical holography~\cite{Bjornson2019d}. For holographic radios, the usual holographic recording sensor is the antenna.

\smallskip

\noindent \textbf{Realization of holographic radio}

To achieve a continuous-aperture antenna array, one ingenious method is to use an ultra-broadband tightly coupled antenna array  based on a current sheet. In this approach, uni-traveling-carrier photodetectors are bonded to the antenna elements using flip-chip technology and form a coupling between antenna elements~\cite{Konkol2017High}. In addition, patch elements are directly integrated into the electro-optic modulator. The current output by the photodetectors directly drives the antenna elements, so the entire active antenna array has a very large bandwidth (about 40~GHz). Moreover, this innovative continuous-aperture active antenna array does not require an ultra-dense RF feed network at all, which means not only that it is achievable but also that is has clear implementation advantages.

\begin{figure*}[t!]
\includegraphics[scale=0.7]{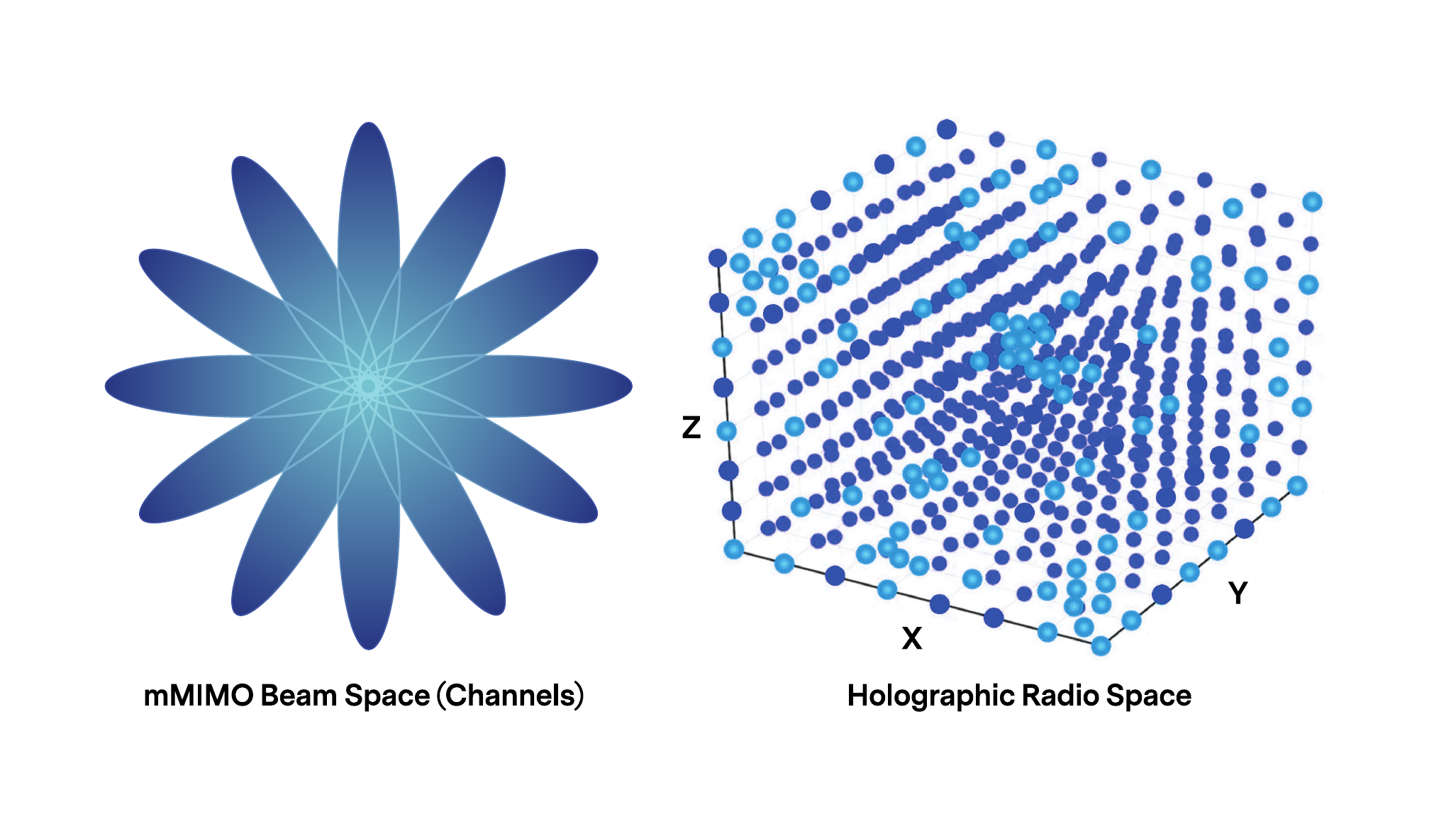}
\centering \label{fig:radioSpace}
\vspace{-8mm}
\caption{Comparison between the beam-space channel in 5G and the holographic radio space. The beam-space approach can only describe channels that comply with a given angular parametrization, while the holographic radio technology can focus the signal at any point in space.}
\end{figure*}

Unlike the beam-space approach to massive MIMO, which dominates in 5G, holographic radios are capable of making use of all available spatial dimensions. The signal wavefront is generated by exploiting a diffraction model based on the Huygens-Fresnel principle. Each point on the array is emitting a spherical wave and the interference pattern between these waves is creating the emitted waveform. The set of possible waves that can be generated is called the holographic radio space and is illustrated in Fig.
~\ref{fig:radioSpace}. Correspondingly, accurate computation of the communication performance requires detailed electromagnetic numerical computations for the radio space; that is, algorithms and tools related to computational electromagnetics and computational holography. The spatial channel correlation is described based on the Fresnel-Kirchhoff integral. Moreover, holographic radios use holographic interference imaging to obtain the RF spectral hologram of the RF transmitting sources (e.g., UEs). In other words, the CSI is implicitly acquired by sending pilot signals from the UE and apply holographic image recording techniques to find the interference pattern that maximizes the received signal power.
A three-dimensional constellation of distributed UEs in the RF phase space can be obtained through spatial spectral holography, providing precise feedback for spatial RF wave field synthesis and modulation in the downlink. Spatial RF wave field synthesis and modulation can obtain a three-dimensional pixel-level structured electromagnetic field (similar to an amorphous or periodic electromagnetic field lattice), which is the high-density multiplexing space of holographic radio and different from sparse beam-space considered in 5G.

It is worth noting that all waveforms that can be generated using a holographic array can in theory  be also generated by replacing the holographic array with a sphere (that would have encapsulated it) whose surface area consists of half-wavelength-spaced discrete antennas. We can achieve the same communication performance using such a spherical array, along with the classical theory for channel estimation and signal processing in the massive MIMO literature~\cite{massivemimobook}. However, this is a hypothetical comparison and not a practical alternative. The purpose of holographic radios is to enable array capabilities that are not practical to implement using other hardware technologies.
On the other hand, since holographic radio utilizes the advantages of optical processing, spectrum computing, large-scale photon integration, electro-optical mixing, and analog-digital photon hybrid integration technologies, the physical layer technology must be adapted to make use of these new methods.

\smallskip

\noindent \textbf{Signal processing for holographic radio}

There are different ways to implement holographic radios for the purpose of joint imaging, positioning, and wireless communications~\cite{Xu2017Holographic}. However, extreme broadband spectrum and holographic RF generation and sensing will produce massive amounts of data, which are challenging to process for critical tasks with low latency and high reliability. Thus, machine learning might be required to operate the system effectively and bridge the gap between theoretical models and practical hardware effects. To meet the 6G challenges of energy efficiency, latency, and flexibility, a hierarchical heterogeneous optoelectronic computing and signal processing architecture will be an inevitable choice~\cite{Baiqing2017Photonics}. Fortunately, holographic radios achieve ultra-high coherence and high parallelism of signals through coherent optical up-conversion of the microwave photonic antenna array, which also facilitate the signal processing directly in the optical domain. However, it is challenging to adapt the signal processing algorithms of the physical layer to fit the optical domain.

How to realize holographic radio systems is a wide-open area. Due to the lack of existing models, in future work, holographic radio will need fully featured theory and modeling converging the communication and electromagnetic theories. Moreover, performance estimation of communication requires dedicated electromagnetic numerical computation, such as the algorithms and tools related to computational electromagnetics and computational holography. The massive MIMO theory can be extended to make optimal use of these propagation models. As mentioned above, a hierarchical and heterogeneous optoelectronic computing architecture is a key to holographic radio. Research challenges related to the design of the hardware and physical layer include the mapping from RF holography to optical holography, integration between photonics-based continuous-aperture active antennas, and high-performance optical computing.\footnote{The authors would like to thank Danping~He ({\em Beijing Jiaotong University}) for the fruitful discussions on the topics of this section.}

%% file: 4_IRSs.tex
When the carrier frequency is increased, the wireless propagation conditions become more challenging due to the larger penetration losses and lower level of scattering, leading to fewer useful propagation paths between the transmitter and the receiver. Moreover, designing coherently operating antenna arrays becomes more difficult since the size of each antenna element shrinks with the wavelength. In such situations, an IRS can be deployed and utilized to: \textit{i)}~improve the propagation conditions by introducing additional scattering, and \textit{ii)}~control the scattering characteristics to create passive beamforming towards the desired receivers to achieve high beamforming gain and suppress the co-channel interference~\cite{Liaskos2018a,Wu2019a}. Ideally, an IRS would create a smart, programmable, and controllable wireless propagation environment, which brings new degrees of freedom to the optimization of wireless networks (in addition to the traditional transceiver design).

An IRS is a type of relay with innovative hardware features~\cite{bjornson2020myths}. Instead of being a compact unit, it consists of a thin two-dimensional surface with a large area. Currently, the most promising way of implementing an IRS is by means of a meta-surface consisting of meta-materials with unusual electromagnetic properties that can be controlled without the need for traditional RF chains. A large IRS can be potentially produced at very low cost, complexity, and energy consumption since no RF components are required unlike conventional active MIMO arrays and holographic radio~\cite{Wu2019a,zhang2016fundamental,wu2016overview}. From an implementation standpoint, IRSs can be conveniently coated on facades of outdoor buildings or indoor walls/ceilings, making them deployable with low complexity and potentially invisible to the human eye. An IRS can be flexibly fabricated and to be mounted on arbitrarily shaped surfaces and, therefore, to be straightforwardly integrated into different application scenarios. The integration of an IRS into a wireless network can be made transparent to the users, thus providing high deployment flexibility \cite{Wu2019a}.

The IRS concept has its origin in reflectarray antennas, which are a class of directive antennas that are flat but can be configured to act as parabolic reflectors or convex RF mirror~\cite{Tretyakov2016,Headland2017}. A characteristic feature of an IRS is that it is neither co-located with the transmitter nor the receiver, thus making it a relay rather than a transceiver.\footnote{An IRS that is co-located with the transmitter or receiver falls more into the category of holographic radio.} The IRS is envisaged to be reconfigurable in real-time so it can be adapted to small-scale fading variations (e.g., due to mobility of the users). An IRS contains a large number of sub-wavelength-sized elements with controllable properties (e.g., impedance) that can be tuned to determine how an incoming signal is scattered (e.g., in terms of phase delay, amplitude, and polarization). As a consequence, an incoming waveform can be reflected in the shape of a beam whose direction is determined by the phase-delay pattern over the elements and can, thus, be controlled~\cite{Ozdogan2019a}. In order words, the individually scattered signals will interfere constructively in the desired directions and destructively in other directions.

To exemplify the basic operation, let us consider a system with a single-antenna transmitter, a single-antenna receiver, and an IRS with $N$ elements. The direct path is represented by the channel coefficient $\sqrt{\beta_\mathrm{d}} e^{j \psi_\mathrm{d}}$, where $\beta_\mathrm{d}>0$ is the channel gain and $\psi_\mathrm{d} \in [0,2\pi)$ is the phase-delay. There are also $N$ scattered paths, each travelling via one of the IRS elements. The $n$th such path is represented by the channel coefficient $\sqrt{\beta_\mathrm{IRS}}e^{j(\psi^\mathrm{TX}_n+ \psi^\mathrm{RX}_n - \phi^\mathrm{IRS}_n) }$, where $\beta_\mathrm{IRS}>0$ is the end-to-end channel gain (i.e., the product of the channel gain from the transmitter to the IRS and the channel gain from the IRS to the receiver), $\psi^\mathrm{TX}_n \in [0,2\pi)$ is the phase-delay between the transmitter and the IRS, and $\psi^\mathrm{RX}_n \in [0,2\pi)$ is the phase-delay between the IRS and the receiver. These parameters are determined by the propagation environment and thus uncontrollable. However, each element in the IRS can control the time delay that is incurred to the incident signal before it is re-radiated, leading to a controllable phase-delay $\phi^\mathrm{IRS}_n \in [0,2\pi]$ for each element $n$.
The received signal is
\begin{equation} \label{eq:received-signal-IRS}
	y = \left( \underbrace{\sqrt{\beta_\mathrm{d}} e^{j \psi_\mathrm{d}}}_{\textrm{Direct path}} +  \sum_{n=1}^{N} \underbrace{\sqrt{\beta_\mathrm{IRS}} e^{j(\psi^\mathrm{TX}_n+ \psi^\mathrm{RX}_n - \phi^\mathrm{IRS}_n) }}_{\textrm{Scattered path from $n$th IRS element}}
	  \right) x + w,
\end{equation}
where $x$ is a desired signal with power $P$ and $w\sim \mathcal{CN}(0,\sigma^2)$ is the receiver noise. The signal-to-noise ratio (SNR) in \eqref{eq:received-signal-IRS} is
\begin{align} \nonumber
    \mathrm{SNR} &= \frac{P}{\sigma^2} \left| \sqrt{\beta_\mathrm{d}} e^{j \psi_\mathrm{d}} +  \sum_{n=1}^{N} \sqrt{\beta_\mathrm{IRS}} e^{j(\psi^\mathrm{TX}_n+ \psi^\mathrm{RX}_n - \phi^\mathrm{IRS}_n) }
	  \right|^2 \\ 
	  & \leq \frac{P}{\sigma^2} \left| \sqrt{\beta_\mathrm{d}} +  N \sqrt{\beta_\mathrm{IRS}} 
	  \right|^2,
\end{align}
where the upper bound is achieved when all the terms inside the absolute value have the same phase, which can be obtained by tuning the phase-delays of the IRS elements. In particular, we need to set $\phi^\mathrm{IRS}_n = \psi^\mathrm{TX}_n+ \psi^\mathrm{RX}_n - \psi_\mathrm{d}$ so that each of the scattered paths has a phase-delay that matches with the direct path. In a more advanced setup with multiple antennas at the transmitter and/or receiver, the reflection coefficients (e.g., phase-delays) can be jointly optimized with transmitters and receivers to maximize performance metrics such as the spectral efficiency or the energy efficiency of the end-to-end link~\cite{Huang2018a}.

\begin{figure}[t!]
	\centering 
	\begin{overpic}[width=\columnwidth,tics=10]{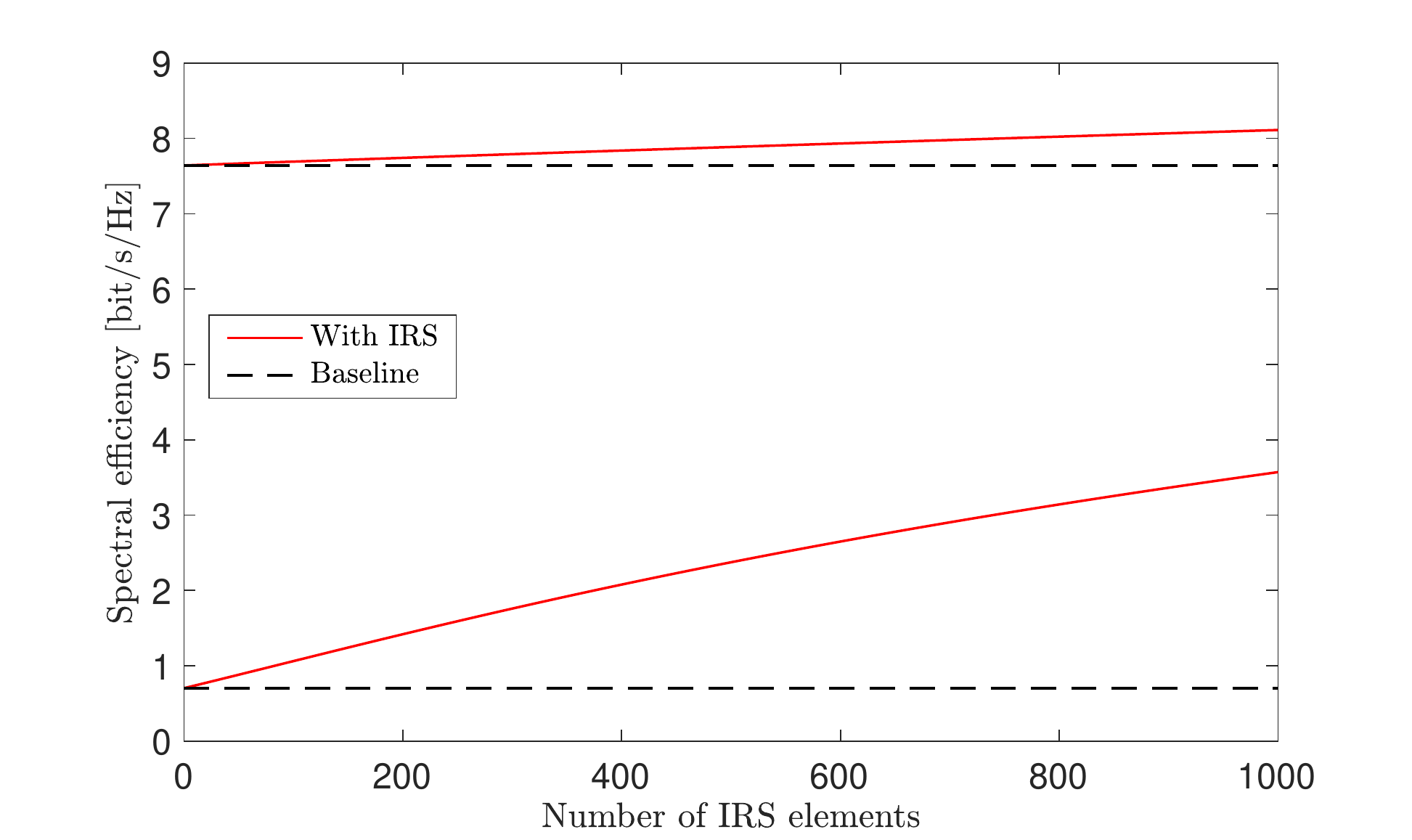}
	\put(32,12){\footnotesize \textbf{Case 1}: $\beta_\mathrm{d}=-100$~dB}
	\put(32,44){\footnotesize \textbf{Case 2}: $\beta_\mathrm{d}=-75$~dB}
\end{overpic} 
	\caption{By adding an IRS to a baseline communication system, the spectral efficiency can be greatly improved. The gain is particularly large when the direct path is weak, as in Case 1.}
	\label{figure:simulationIRS} 
\end{figure}

The benefit of adding an IRS to a single-antenna communication link is illustrated in Fig.~\ref{figure:simulationIRS}, where the spectral efficiency $\log_2(1+\mathrm{SNR})$ is used as performance metric. The end-to-end channel gain via each IRS element is $\beta_\mathrm{IRS}=-150$~dB, which corresponds to having LoS links with $-75$~dB from the transmitter to the IRS and from the IRS to the receiver. We then consider two cases: \textit{1)}~a weak direct path with $\beta_\mathrm{d}=-100$~dB, and \textit{2)}~a strong direct path with $\beta_\mathrm{d}=-75$~dB. The transmit power is selected to correspond to 10 mW per 20 MHz. Fig.~\ref{figure:simulationIRS} shows the spectral efficiency as a function of the number of IRS elements $N$; the baseline without any IRS is also shown for reference. We notice that the IRS greatly improves the performance when the direct path is weak (Case~1), while the performance gains are more modest when the direct path is strong (Case~2). In any case, we need hundreds of elements for the IRS to be effective; the end-to-end channel gain of the path via a single IRS element is typically very small but the phase-aligned combination of many such paths can give considerable gains. The large number of elements might seem like a showstopper but it is not since the elements are spread out over two dimensions and are sub-wavelength-sized, thus more than 1000 elements fits into a $1 \,\textrm{m} \times 1
 \,\textrm{m}$ surface.

When the IRS elements are configured to maximize the SNR, all the propagation paths will interfere constructively at the location of the receiver. If the receiver is in the vicinity of the IRS, the reflected signal will be focused on the particular spatial location of the receiver. As this location is moved further away, the focusing operation gradually becomes equivalent to forming a beam in the angular direction leading to the point, as illustrated in Fig.~\ref{fig:metasurface}. Since an IRS can change not only the direction of the reflected wave but also the shape of the waveform, it should be viewed as an electronically reconfigurable curved/concave mirror. There are special cases when an IRS can approximate a specular reflector (i.e., a flat mirror) but this is generally suboptimal~\cite{Bjornson2020IRS,bjornson2020myths}.

The IRS is a full-duplex transparent relay that does not amplify the incident signal but reflects it without causing any noticeable propagation delays (except that the delay spread of the channel might increase)~\cite{JR:wu2018IRS,Matthiesen2020IRS}. Since the surface needs to be physically large to beat a classical half-duplex regenerative relay and is subject to beam-squinting~\cite{Bjornson2020b,Bjornson2020IRS}, the most promising use case for IRS is to increase the propagation conditions in short-range communications, particularly in sub-THz and THz frequency bands where conventional relaying technology is unavailable. For example, an IRS can provide an additional strong propagation path that remains available even when the LoS path is blocked. An IRS should ideally be deployed to have an LoS path to either the transmitter and/or receiver. There are also possible use cases in NLOS scenarios, for example, to increase the rank of the channel to achieve the full multiplexing gain.

\begin{figure}[t!]
	\centering 
	\begin{overpic}[width=\columnwidth,tics=10]{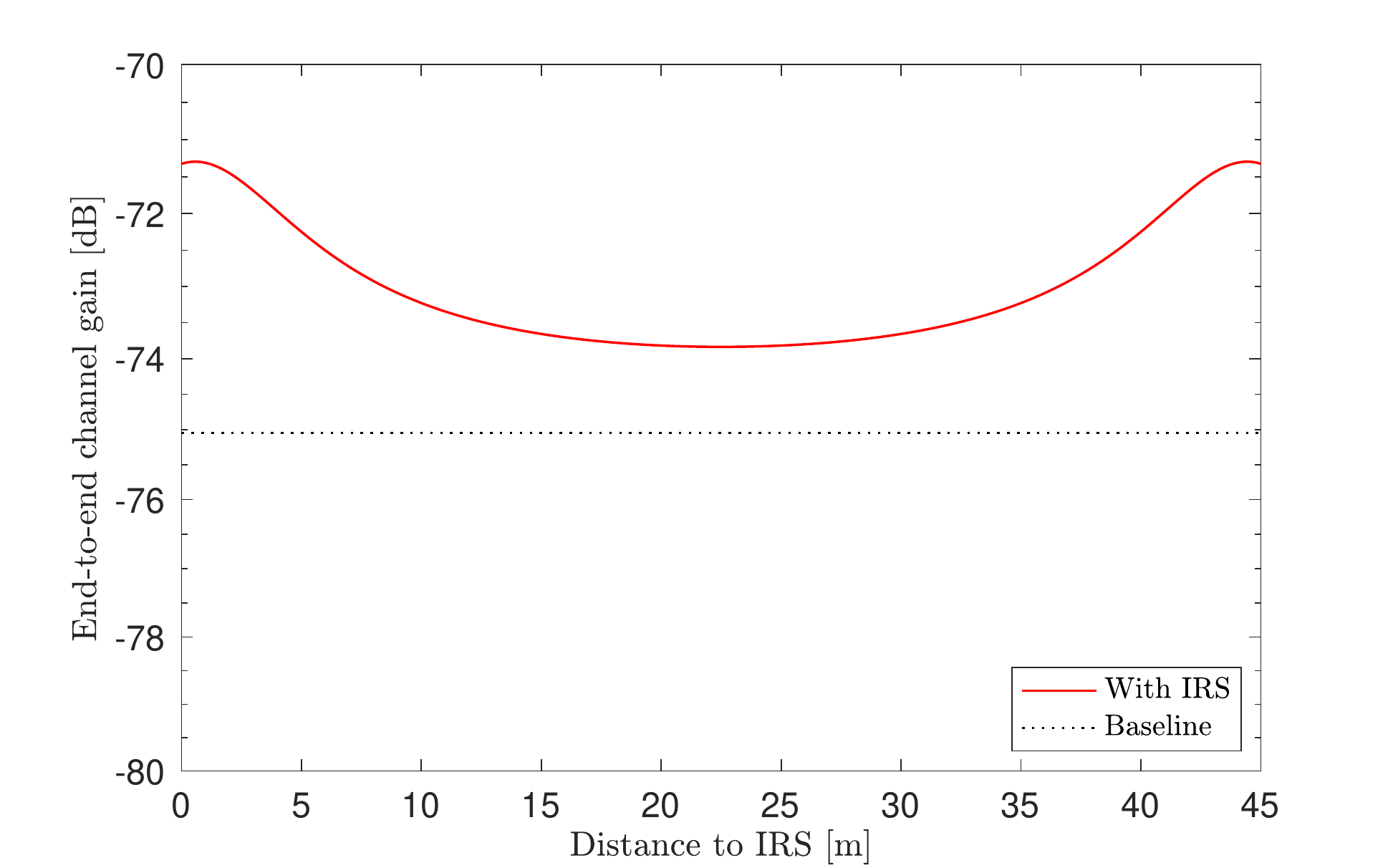}
	\put(15,51){\footnotesize $\leftarrow$ Close to the transmitter}
	\put(57,51){\footnotesize Close to the receiver $\rightarrow$}
\end{overpic} 
	\caption{End-to-end channel gain between a transmitter and receiver that are $45$~m apart depends on where the IRS is deployed. It is largest when the IRS is either close to the transmitter or the receiver.}
	\label{figure:simulationIRS2} 
\end{figure}

To further elaborate on how to deploy an IRS, we continue the example from Fig.~\ref{figure:simulationIRS} and focus on Case~2, where the direct LOS path between the transmitter and receiver has a channel gain of $-75$~dB. The distance between the transmitter and receiver is $45$~m. An IRS with $N=1024$ elements is deployed somewhere along a line that is $5$~m from the line between the transmitter and the receiver. We compute the channel gains of the reflected LoS paths using the formulas from~\cite{Bjornson2020IRS}. Fig.~\ref{figure:simulationIRS2} shows the end-to-end channel gain $| \sqrt{\beta_\mathrm{d}} +  N \sqrt{\beta_\mathrm{IRS}} |^2$ achieved for different locations of the IRS. The baseline channel gain $\beta_\mathrm{d}=-75$~dB is also shown as a reference. We notice that the IRS improves the channel gain the most when it is either close to the transmitter or close to the receiver, but there is always a substantial gap over the baseline case with no IRS. Hence, two preferable types of IRS deployments are envisioned:

\begin{figure}[ht]
\centering
\includegraphics[width=0.45\textwidth]{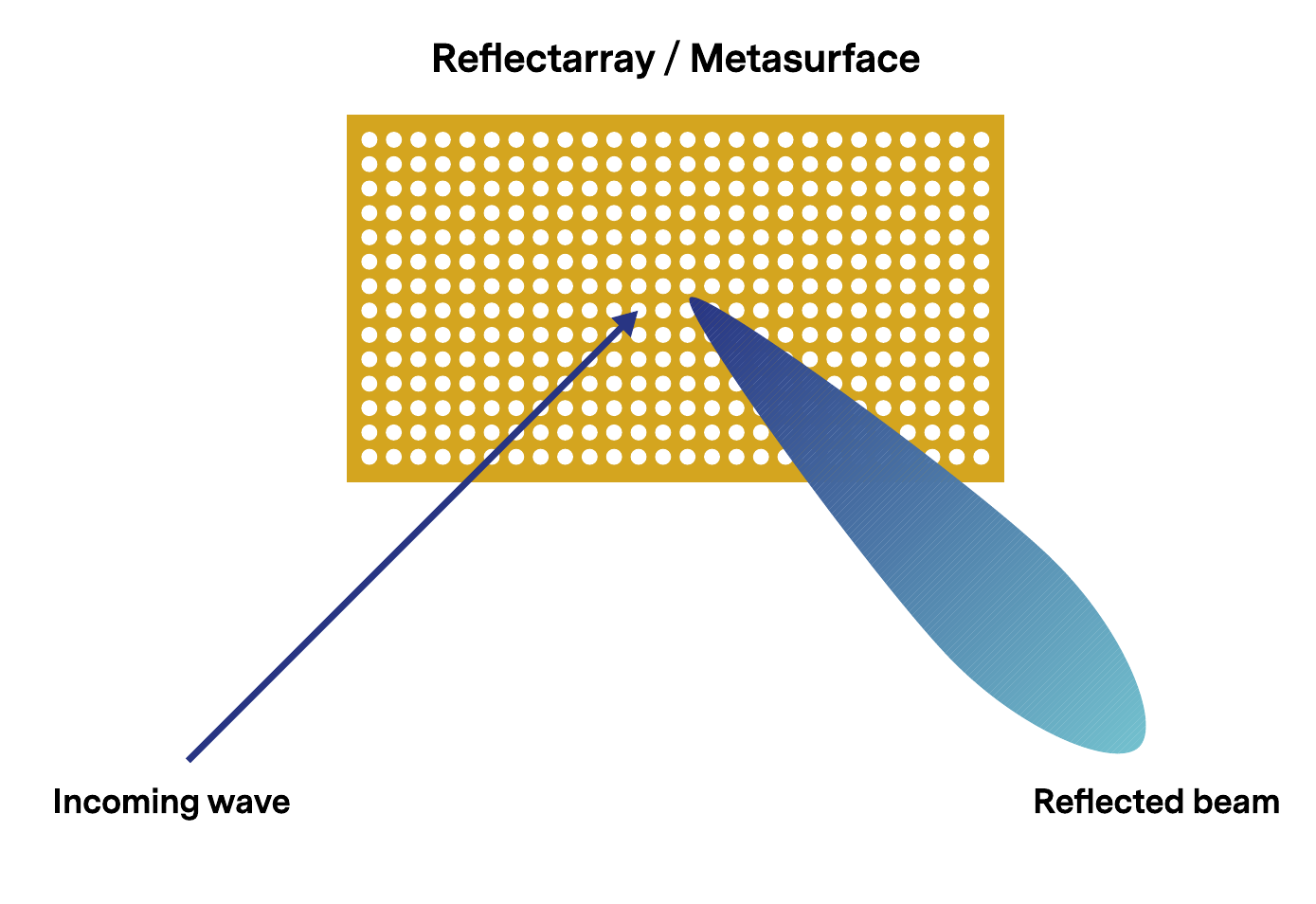}

\vspace{-4mm}

\caption{An IRS takes an incoming wave and reflects it as a beam in a particular direction or towards a spatial point.} \label{fig:metasurface}
\end{figure}

\begin{itemize}
\item[\textit{i)}] A large IRS can be deployed close to a small AP to control how the transmitted/received signals are interacting with the propagation environment and focused in different directions depending on where the users are located for the moment. Such an IRS can be thought of as an additional antenna array that helps the AP to cover a geographical region.
\item[\textit{ii)}] A large IRS can be deployed in a room where prospective users are located so it has a good channel to both the AP and the UE, even if the direct path between them happens to be weak.
\end{itemize}

In addition to increasing the signal strength of a single user, an IRS can improve the channel rank (for both single- and multi-user MIMO), suppress the interference~\cite{Yang2020TCOM}, and enhance the multicasting performance~\cite{QQW2020TR}. Essentially anything that can be done with traditional beamforming can also be implemented using an IRS, and they can be deployed as an add-on to many existing systems. The beamwidth of the scattered signal from the IRS will also match that of an equal-sized antenna array.
Other prospective use cases are cognitive radio~\cite{guan2020joint,yuan2019intelligent}, wireless power transfer~\cite{wu2019weighted,mishra2019channel,wu2019joint}, physical layer security~\cite{guan2020intelligent,yang2020intelligent,lu2020robust,yang2020deep}, NOMA~\cite{zheng2020intelligent,li2019joint}, vehicular networks with predictive mobility~\cite{makarfi2020a,makarfi2019a,Matthiesen2020IRS}, coordinated multipoint transmission (CoMP)~\cite{hua2020intelligent}, UAV communications~\cite{zhang2019reflections,hua2020uav,ma2020enhancing,li2020reconfigurable,zeng2019accessing} and backscattering, where IoT devices near the surface can communicate with an AP at zero energy cost~\cite{Liaskos2018a}. It still remains to identify which use cases lead to the largest improvements over the existing technology~\cite{bjornson2020myths}.

The main open research challenges are related to the signal processing, hardware implementation, channel modeling, experimental validation, and real-time control. Although software-controllable meta-surfaces exist~\cite{Liaskos2018a}, the accuracy of the reconfigurability is currently limited due to, for example, the small number of possible phase values per element, fixed amplitudes for each phase delay~\cite{abeywickrama2019intelligent}, and by only having control over groups of elements. Although semi-accurate physical channel models exist for LoS scenarios, these omit mutual coupling and other hardware effects that are inevitable in practice. Therefore, experimentally validated channel models are strongly needed. The performance loss caused by using practical low-resolution hardware (such as 1-bit delay resolution) also needs to be carefully studied, especially for the case when a large number of reflecting elements is required to achieve a considerable beamforming gain~\cite{wu2019beamformingICASSP,JR:wu2019discreteIRS}. Furthermore, wireless networks, in general, operate in broadband channels with frequency selectivity. As such, the phase-delay pattern of the IRS needs to strike a balance between the channels of different frequency sub-bands, which further complicates the joint active and passive beamforming optimization. Besides, most of the existing works on IRS assume that the maximum amount of power is scattered by the IRS, while only the IRS phase shifts are fine-tuned to improve the system performance. However, it remains unknown in which circumstances one can achieve higher performance by jointly optimizing the amplitudes and phase shifts, and how to balance the performance gain and practical hardware cost as well as complexity \cite{zhao2020two,zhao2020exploiting}. 
Lastly, the control interface and the channel estimation are difficult when using a passive surface that cannot send or receive pilot signals. Hence, channel measurements can only be made from pilots sent by other devices and received at other locations.

The radio resources required for the channel estimation and control will grow linearly with the number of elements and may become huge in cases of interest unless a clever design that utilizes experimentally validated channel models can be devised. To reduce the channel estimation overhead and implementation complexity, two-timescale beamforming optimization can be a potential approach~\cite{zhao2019intelligent}. An anchor-assisted two-phase channel estimation scheme is proposed in \cite{guan2020channelestimation}, where two anchor nodes are deployed near the IRS for helping the BS to acquire the cascaded BS-IRS-user channels. 
The power consumption of the control interface will likely dominate the total energy consumption of an IRS since there are no power amplifiers or RF chains~\cite{bjornson2020myths}. One channel estimation approach is to repeatedly transmit pilot signals and change the IRS configuration according to a codebook to identify the preferred operation~\cite{Zheng2020}. Another potential solution is to have a small number of active elements in the IRS and then use deep learning or compressive sensing to extrapolate the estimates made in those element to an estimate of the entire channel \cite{Taha2019a}. Finally, proper deployment of the IRS with active BSs in such hybrid networks to maximize the system performance is another important problem to solve, especially in practical multi-cell scenarios. The design considerations include the rank of the channel, the beamforming gain, LoS versus non-LoS propagation, and so on. 

Recently, IRSs and their various equivalents have also drawn significant attention from the industry~\cite{QQW2020TR}. In November 2018, NTT DoCoMo and Metawave jointly demonstrated that by properly deploying a meta-structure based reflecting surface, $560$~Mbps communication rate can be achieved in the 28~GHz band as compared to 60~Mbps without it. There are startups such as Greenerwave, Echodyne, Kymeta that attempt to commercialize IRS-type technologies for consumer-grade use cases.

%% file: 4_cell_free.tex

\begin{figure*}[t]
\begin{center}
\begin{equation}
{\cal{R}}_{k}^{\rm DL} = \displaystyle B \frac{\tau_{ d}}{\tau_c} \log_2 \left( \displaystyle 1 + 
\frac{N_{\rm AP} \left( \ds \sum_{m \in {\cal M}_k} {\ds \sqrt{\eta_{k,m}^{\rm DL} \gamma_{k,m}}} \right)^2}{\ds \sum_{j \in \mathcal{K}} \sum_{m \in {\cal M}_j} \eta_{j,m}^{\rm DL} \beta_{k,m} + \ds \sum_{j \in \mathcal{P}_k \backslash \{k\}} N_{\rm AP} \left( \ds \sum_{m \in {\cal M}_j} \sqrt{\eta_{j,m}^{\rm DL} \gamma_{k,m}} \right)^2 + \sigma^2_{z,k}} \right), \label{eq:rateDL}
\end{equation}
\begin{equation}
{\cal{R}}_{k}^{\rm UL} = \ds B \frac{\tau_{ u}}{\tau_c} \log_2 \left( \ds 1 + \frac{\eta_{k}^{\rm UL} N_{\rm AP} \left( \ds \sum_{m \in {\cal M}_k} {\ds \gamma_{k,m}} \right)^2}{\ds \sum_{j \in \mathcal{K}} \eta_{j}^{\rm UL} \sum_{m \in {\cal M}_k} \beta_{j,m} \gamma_{k,m} + \ds \sum_{j \in \mathcal{P}_k \backslash \{k\}} N_{\rm AP} \eta_{j}^{\rm UL}\left( \ds \sum_{m \in {\cal M}_k} \gamma_{k,m} \sqrt{\frac{\eta_{j}}{\eta_{k}}} \frac{\beta_{j,m}}{\beta_{k,m}} \right)^2 + \!\!\!\! \sum_{m \in {\cal M}_k} {\!\! \sigma^2_{w,m} \gamma_{k,m}}} \right) \label{eq:rateUL}
\end{equation}
\hrulefill
\end{center}
\end{figure*}

The large performance disparity between cell-center and cell-edge UEs is one of the main drawbacks of the traditional cellular network topology. The concept of cell-free massive MIMO has recently emerged to overcome this long-standing issue by providing consistent performance and seamless handover regardless of the UEs' positions~\cite{uniformlygreatperformance, Ngo17,Nayebi2017a,BuzziUCletter}. By conveniently combining elements from massive MIMO, small cells, and user-centric CoMP with joint transmission/reception~\cite{Boldi2011a,Bjornson2013d,Int19,Liu18,Buz20}, cell-free massive MIMO gives rise to a cell-less architecture characterized by almost uniform achievable rates across the coverage area. For this reason, it is widely regarded as a potential physical-layer paradigm shift for 6G wireless systems.

\begin{figure*}[t]
\centering
\includegraphics[width=0.9\textwidth]{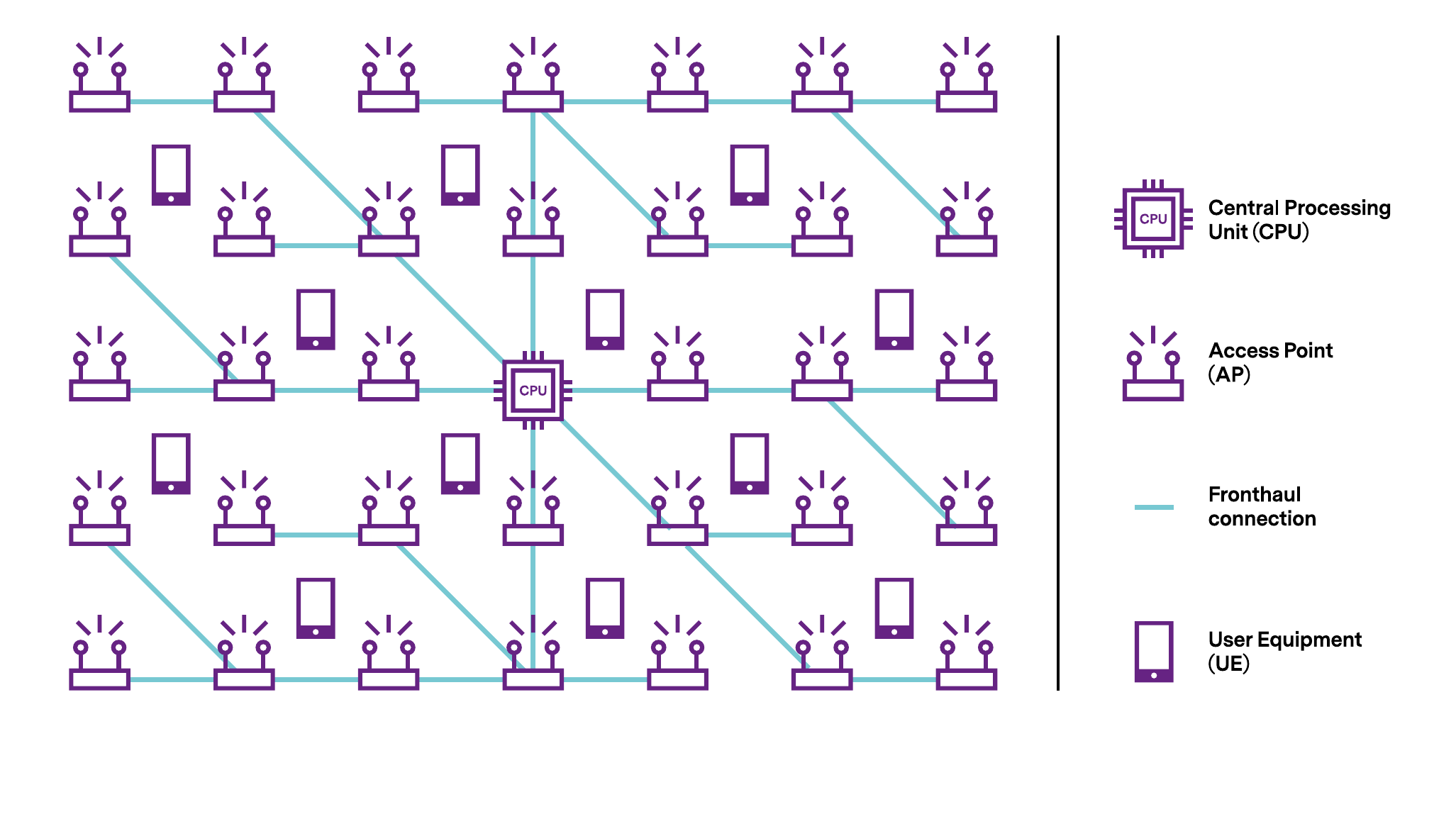} 
\vspace{-6mm}
\caption{A cell-free massive MIMO system consists of distributed APs that jointly serve the UEs. The cooperation is facilitated by a fronthaul network and a CPU.}
\label{fig:cell-free}
\end{figure*}

In cell-free massive MIMO, conventional APs equipped with massive co-located antenna arrays are replaced by a large number of low-cost APs equipped with few antennas, which cooperate to jointly serve the UEs. To enable such cooperation, the APs are connected to one or more CPUs via fronthaul links, as illustrated in Fig.~\ref{fig:cell-free}. Distributing the transmit/receive antennas over a large number of cooperating APs has a two-fold advantage over cellular massive MIMO. First of all, it enables each UE to be located near one or a few APs with high probability and, as a consequence, to be jointly served by a reasonable number of favorable antennas with reduced path loss. In addition, as the serving antennas for each UE belong to spatially separated APs (which are usually seen with different angles), it brings an improved diversity against blockage and large-scale fading.

Unlike CoMP with joint transmission/reception, which is traditionally implemented in a network-centric fashion with well-defined edges between clusters of cooperating APs, cell-free massive MIMO adopts a user-centric approach, where the above clusters are formed so that each UE is served by its nearest APs~\cite{Bjornson2013d,Bjornson2019,Buz20}. Furthermore, as for CoMP systems, the cell-free approach greatly benefits from TDD operations, which allow to use uplink pilot signals for both upink and downlink channel estimation. While CoMP was designed as an add-on to an existing cellular network, the deployment architecture and protocols are co-designed in cell-free massive MIMO to deliver uniform service quality \cite{uniformlygreatperformance,Ngo17,Nayebi2017a,BuzziUCletter}.

\begin{table}[t]
\begin{center} \label{table:cell-free_parameters}
\caption{Symbols Definition for a Cell-Free Massive MIMO Deployment}
\begin{tabular}{|p{1cm} | p{6.5cm}|}
\hline
\multicolumn{1}{|c|}{\textbf{Symbol}} & \multicolumn{1}{c|}{\textbf{Description}} \\ \hline
$K$                          & Number of UEs in the system \\ \hline
$M$                          & Number of APs \\ \hline
$N_{\rm AP}$                 & Number of antennas at each AP \\ \hline
$\beta_{k,m}$                & scalar coefficient representing large-scale fading between the $k$th UE and the $m$th AP \\ \hline
$\mathbf{g}_{k,m}$           & $N_{\rm AP}$-dimensional uplink channel between the $k$th UE and the $m$th AP \\ \hline
$\sigma^2_{z,k},\sigma^2_{w,m}$ & Receiver noise power at the $k$th UE and $m$th AP \\ \hline
$\tau_c$                     & Length, in discrete-time samples, of the channel coherence block \\ \hline
$\tau_d$ ($\tau_u$)          & Length, in discrete-time samples, of the downlink (uplink) data transmission phase \\ \hline
$\tau_p$                     & Length of and number of mutually orthogonal pilot sequences \\ \hline
$\mathcal{P}_k$              & Set of UEs using the same pilot sequence as the $k$th UE \\ \hline
$\eta_k$                     & Uplink power for the $k$th UE during channel training phase \\ \hline
$\eta_{k,m}^{\rm DL}$        & Power used on the downlink by the $m$th AP to transmit data to the $k$th UE \\ \hline
$\eta_{k}^{\rm UL}$          & Uplink power for the $k$th UE during uplink data transmission phase \\ \hline
${\cal M}_k$                 & Set of APs serving the $k$th UE \\ \hline
${\cal K}_m$                 & Set of users served by the $m$th AP \\ \hline
\end{tabular} \vspace{2mm}
\end{center}
\end{table}

Let us consider a cell-free massive MIMO deployment with single-antenna UEs, where MMSE channel estimation is performed based on the transmission of uplink pilots, maximum ratio transmission/combining is used at the APs, and the small-scale fading coefficients are i.i.d. complex Gaussian distributed (i.e., the usual Rayleigh fading assumption). Considering the notation reported in Table~\ref{table:cell-free_parameters}, a lower bound on the uplink and downlink capacities for the $k$th UE can be expressed as in \eqref{eq:rateDL} and \eqref{eq:rateUL} at the top of the page, which are based on the use-and-then-forget bounding technique~\cite{massivemimobook}. In these expressions, $B$ represents the signal bandwidth, whereas $\sigma^2_{z,k}$ and $\sigma^2_{w,m}$ are the receiver noise powers at the $k$th UE and at the $m$th AP, respectively. Moreover, the mean-square of the channel estimates is
\begin{align}
\gamma_{k,m} = \frac{\eta_k \beta_{k,m}^2}{\ds \sum_{i \in \mathcal{P}_k} \eta_i \beta_{i,m}  +\sigma^2_{z,k}}.
\end{align}

The expressions in \eqref{eq:rateDL} and \eqref{eq:rateUL} are general enough to describe the performance of three relevant systems.
\begin{itemize}
\item[\textit{i)}] Letting $\mathcal{K}_m=\{1, 2, \ldots, K\}, \forall m$ yields an FCF association, where all APs transmit to all the UEs in the network. This is the originally conceived version of cell-free massive MIMO, which is practically applicable only to a system deployed in a limited area~\cite{Bjornson2019}.
\item[\textit{ii)}] A more practically relevant scenario is obtained by using a CF-UC association, where each UE is served only by a limited number of APs~\cite{BuzziUCletter,Bjornson2019,Buz20}. The optimal UE-AP association is a complex combinatorial problem that depends on the type of the adopted precoding and hardware constraints. However, one viable and simple solution is to let each UE be served only by a predetermined number of APs, which can be selected based on the large-scale fading coefficients. For example, the $k$th user can be served by the $N_{\rm UC}$ APs with the best average channel conditions. Let $O_k \, : \, \{1,\ldots, M \} \rightarrow \{1,\ldots, M \}$ denote the sorting operator for the vector $\left[\beta_{k,1},\ldots, \beta_{k,M}\right]$, such that $\beta_{k,O_k(1)} \geq \beta_{k,O_k(2)} \geq \ldots \geq \beta_{k,O_k(M)}$. The set $\mathcal{M}_k$ of the $N_{\rm UC}$ APs serving the $k$th UE is then given by
\begin{equation}
\mathcal{M}_k=\{ O_k(1), O_k(2), \ldots , O_k(N_{\rm UC}) \}.
\end{equation}
Consequently, the set of UEs served by the $m$th AP is defined as $\mathcal{K}_m=\{ k: \,  m \in \mathcal{M}_k \}$.
\item[\textit{iii)}] Assuming a small number of APs with a large number of antennas each and supposing that each UE can be associated to only one AP (so that the sets $\mathcal{M}_k, \forall k$ have cardinality 1) yields a traditional multi-cell mMIMO deployment.
\end{itemize}

\begin{figure*}[t!]
  \centering
  \includegraphics[scale=0.6]{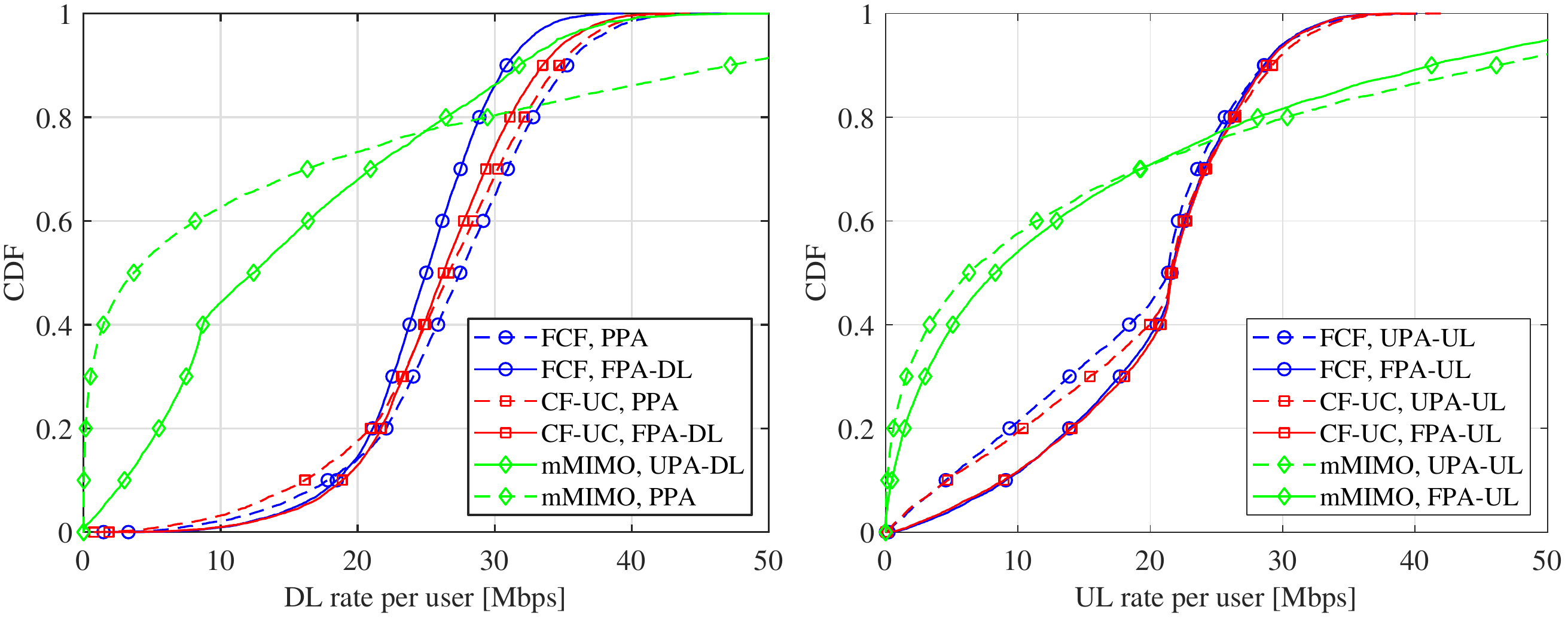}  
  \caption{Cumulative distribution functions (CDFs) of the downlink and uplink rate per UE of the FCF, CF-UC and mMIMO with different power allocation strategies. Parameters: $M=100$, $N_{\rm AP}=4$, $K=30$, $N_{UC}=10$, $\tau_p=16$, $\alpha_{\rm DL}=-0.5$, $P_0=-10$ dBm, and $\alpha_{\rm UL}=0.5$.}
  \label{Fig_CF_UC_CDF}
\end{figure*}

\begin{figure*}[t!]
	\centering
	\includegraphics[scale=0.5]{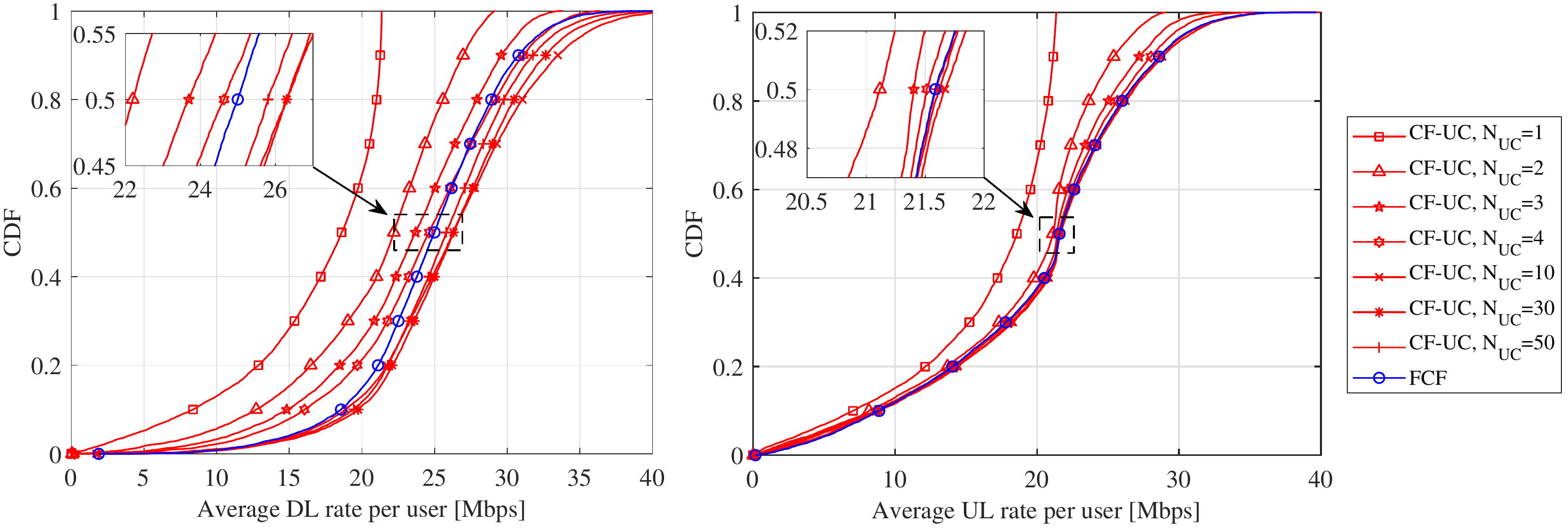}  
	\caption{Cumulative distribution functions (CDFs) of the downlink and uplink rate per UE of the CF-UC with different values of $N_{UC}$ compared with the FCF with FPA-DL and FPA-UL. Parameters: $M=100$, $N_{\rm AP}=4$, $K=30$,  $\tau_p=16$,  $\alpha_{\rm DL}=-0.5$, $P_0=-10$ dBm,  $\alpha_{\rm UL}=0.5$.}
		\label{Fig_CF_UC_CDF_N}
\end{figure*}

Some numerical results illustrating the performance of the FCF, CF-UC and mMIMO deployments 
are reported in Figs.~\ref{Fig_CF_UC_CDF}-\ref{Fig_CF_UC_CDF_N} and in Tables \ref{Table_PPA_Uni}-\ref{Table_FPA_DL_UL}. 
The simulation setup is the following: communication bandwidth $B = 20$ MHz; carrier frequency $f_0=1.9$ GHz; antenna height at the APs $10$ m and at the UEs $1.65$ m; thermal noise with power spectral density $-174$ dBm/Hz; front-end receiver at the APs and at the UEs with noise figure of $9$ dB. 
The number of APs is $M=100$ and each has $N_{\rm AP}=4$ antennas and the number of UEs simultaneously served in the system on the same time-frequency coherence block is $K=30$. 
The APs and UEs are deployed at random positions on a square area of $1000 \times 1000$ square meters. The large-scale fading coefficients $\beta_{k,m}$ are evaluated as in~\cite{DAndrea_ICC2019,DAndrea_OJCOMS20}. In order to avoid boundary effects, the square area is wrapped around~\cite{Ngo17,BuzziUCletter}. The mutually orthogonal pilot sequences have length $\tau_p=16$; the downlink and uplink data transmission phases of each coherence block contain  $\tau_{u}=\tau_{ d}=\frac{\tau_c-\tau_p}{2}$ samples, where $\tau_c=200$ is the length of the coherence block in samples. The uplink transmit power for channel estimation is $\eta_k=\tau_p p_k$, with $p_k=100$ mW, $\forall k=1,\ldots,K$. With regard to power control, the results have been obtained for the cases of PPA and FPA-DL for the downlink. For the uplink, instead, UPA-UL and FPA-UL has been considered. More precisely, for the downlink, denoting as $P_{\rm max, AP}^{\rm DL}$ the maximum power available at each AP, the power coefficients for the PPA are set as  $\eta_{k,m}^{\rm DL}=\gamma_{k,m}P_{\rm max, AP}^{\rm DL}/(\sum_{k \in \mathcal{K}(m)} \gamma_{k,m})$ and for the FPA-DL rule as $\eta_{k,m}^{\rm DL}=\gamma_{k,m}^{-\left(\alpha_{\rm DL}+1\right)}P_{\rm max, AP}^{\rm DL}/(\sum_{k \in \mathcal{K}(m)} \gamma_{k,m}^{-\alpha_{\rm DL}})$, with $\alpha_{\rm DL}=-0.5$. On the uplink, instead, denoting by $P_{\rm max}^{\rm UL}$ the maximum power available at each UE, for the UPA-UL we let $\eta_{k}^{\rm UL}=P_{\rm max}^{\rm UL}$,  $\forall \; k=1,\ldots,K$, while for the FPA-UL we have  $\eta_{k}^{\rm UL}=\min\left(P_{\rm max}^{\rm UL}, P_0 \bar{\gamma}_k^{- \alpha_{\rm UL}}\right)$, $\forall \; k=1,\ldots,K$, with $\bar{\gamma}_k=\sqrt{\sum_{m \in \mathcal{M}_k} \gamma_{k,m}}$.
The performance of the FCF and CF-UC are compared with a mMIMO system with 4 APs and 100 antennas each covering the same area. The same transceiver signal processing is assumed for the APs and the APs, i.e., the performance of the mMIMO, FCF and CF-UC systems are compared assuming MR processing in both uplink and downlink. For the mMIMO system, in order to consider a fair comparison, we set  $P_{\rm max, AP}^{\rm DL}=M P_{\rm max, AP}^{\rm DL}/4$ and present results for the PPA-DL and for the UPA-DL, with the power coefficients set as $\eta_{k,m}^{\rm DL}=P_{\rm max, AP}^{\rm DL}/\left|\mathcal{K}_m \right|\gamma_{k,m}$. We assume , $P_{\rm max, AP}^{\rm DL}=200$ mW, $P_{\rm max, AP}^{\rm DL}=5$ W, and $P_{\rm max}^{\rm UL}=100$ mW.

\begin{table*}[t]
	\centering
	\caption{Performance of FCF, CF-UC and mMIMO with UPA-UL and PPA in Downlink. Parameters: $M=100$, $N_{\rm AP}=4$, $K=30$, $N_{UC}=10$, $\tau_p=16$}
	\label{Table_PPA_Uni}
	\label{table:Cell_free_UPA_PPA}
	\begin{tabular}{|l|l|l|l|l|l|l|l|l|}
		\hline
		& \textbf{DL 1\%-rate} & \textbf{DL 5\%-rate} & \textbf{DL 50\%-rate} & \textbf{DL 90\%-rate}  & \textbf{UL 1\%-rate} & \textbf{UL 5\%-rate} & \textbf{UL 50\%-rate} & \textbf{UL 90\%-rate} \\ \hline
		FCF    &        7.38 Mbps                   &        13.9 Mbps    &        27.4 Mbps                   &        35.2 Mbps&        0.47 Mbps                   &        2.1 Mbps&        21.3 Mbps                   &        28.5 Mbps         \\ \hline
		CF-UC     &        5.8  Mbps                  &          12.1 Mbps  &        26.6  Mbps                  &          34.7 Mbps&        0.42  Mbps                  &          2 Mbps&        21.5  Mbps                  &          29.2 Mbps       \\ \hline
		mMIMO    &           0 Mbps             &            0 Mbps  &           0.2 Mbps             &            38.6 Mbps &           0.0004 Mbps             &            0.02 Mbps &           6.3 Mbps             &            46.1 Mbps           \\ \hline
	\end{tabular} \vspace{2mm}
\end{table*}

\begin{table*}[t]
\centering
\caption{Performance of FCF, CF-UC with FPA-DL and mMIMO with UPA-DL and FPA-UL. Parameters: $M=100$, $N_{\rm AP}=4$, $K=30$, $N_{UC}=10$, $\tau_p=16$, $\alpha_{\rm DL}=-0.5$, $P_0=-10$ dBm, and $\alpha_{\rm UL}=0.5$}
\label{Table_FPA_DL_UL}
\label{table:Cell_free_FPA}
\begin{tabular}{|l|l|l|l|l|l|l|l|l|}
\hline
                        & \textbf{DL 1\%-rate} & \textbf{DL 5\%-rate} & \textbf{DL 50\%-rate} & \textbf{DL 90\%-rate}  & \textbf{UL 1\%-rate} & \textbf{UL 5\%-rate} & \textbf{UL 50\%-rate} & \textbf{UL 90\%-rate} \\ \hline
FCF    &        10.2 Mbps                   &        15.6 Mbps    &        25 Mbps                   &        30.8 Mbps&        1.9 Mbps                   &        5.6 Mbps&        21.6 Mbps                   &        26.2 Mbps         \\ \hline
CF-UC     &        10  Mbps                  &          15.9 Mbps  &        12  Mbps                  &          33.4 Mbps&        1.7  Mbps                  &          5.2 Mbps&       21.6  Mbps                  &          28.5 Mbps       \\ \hline
mMIMO    &           0.01 Mbps             &            0.09 Mbps  &           2.6 Mbps             &            23.3 Mbps &           0.002 Mbps             &            0.1 Mbps &           8.3 Mbps             &            41.2 Mbps           \\ \hline
\end{tabular} \vspace{2mm}

\end{table*}

Fig.~\ref{Fig_CF_UC_CDF} reports the CDF of the rate per UE for the three considered deployments. It is clearly seen that mMIMO, while providing very large rates to the lucky UEs that are very close to the macro AP site (this corresponds to the upper-right part of the figure), is largely outperformed by FCF and CF-UC deployments when considering the vast majority of all the UEs. Indeed, the figure reveals that about 90$\%$ of the UEs on the downlink and 80 $\%$ of the UEs on the uplink enjoy much better rates with FCF and CF-UC deployments than with mMIMO. This situation is again clearly represented by the numbers reported in Tables \ref{Table_PPA_Uni}-\ref{Table_FPA_DL_UL}, where we have reported the 1$\%$, 5$\%$, 50$\%$ and 90$\%$ likely per-user rates. It is indeed seen that mMIMO outperforms novel FCF and CF-UC deployments only when considering the 90$\%$-likely per user rate, while being practically not-able to serve a good share of the 
UEs.\footnote{The reader should however not be led to draw the conclusion that ``massive MIMO does not work''. Indeed, we are considering here an extreme 
situation with simple MR processing and with a large number of UEs being served on the coherence block. In a less loaded scenario, with a smaller number of UEs and more advanced processing schemes such as regularized zero-forcing beamforming~\cite{massivemimobook}, the performance of mMIMO is restored and the gap with FCF and CF-UC deployments gets reduced.}

Fig.~\ref{Fig_CF_UC_CDF} shows the performance of FCF versus a CF-UC deployment where each UE is connected to the $N_{UC}=10$ APs with the largest large-scale fading coefficients. It is seen that the two systems perform quite similarly, especially in the uplink. In order to give a deeper look at the comparison between FCF and CF-UC deployments, in  
Fig.~\ref{Fig_CF_UC_CDF_N} we report  the performance of the FCF deployment and of the CF-UC deployment for several values of $N_{UC}$.  Results show that taking $N_{UC}\geq 10$ provides performance levels comparable or superior to the FCF deployment. Since the FCF deployment corresponds to an CF-UC deployment with $N_{UC}=M$, the figure shows that there is an ``optimal'' value of $N_{UC}$ that maximizes the system performance. Besides this, the main conclusion that can be drawn from this figure is that CF-UC deployments offer performance levels comparable or better than FCF deployments, but with much lesser signaling and backhaul overhead, so special care should be devoted to the design of the UE-AP association rules since they have a crucial impact on the system performance. 

\begin{figure*}[t!]
\centering
\includegraphics[scale=0.7]{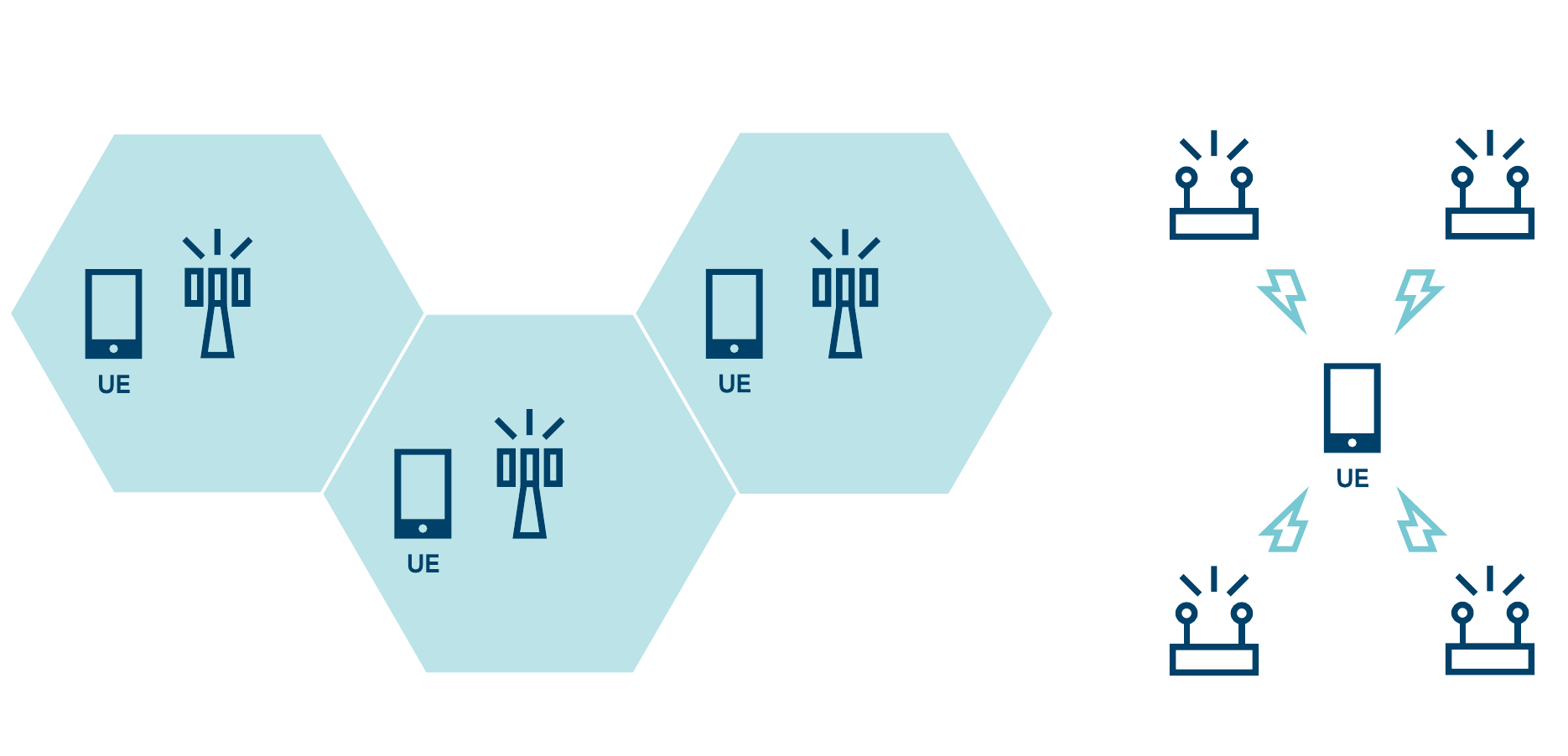}

\vspace{-6mm}

\caption{Initial access based on cellular system (left) and cell-free system (right).}
\label{fig:cfac}
\end{figure*}

The cell-free massive MIMO architecture is not meant to increase the peak rates in broadband applications, since these can only be achieved in extreme cases, but has been shown to vastly outperform traditional small-cell and cellular massive MIMO for the majority of UEs~\cite{Ngo17,Nayebi2017a,Bjornson2020a}. A cell-free massive MIMO deployment can also provide support for the implementation of low-latency, mission-critical applications. The availability of many APs, coupled with the rapidly decreasing cost of storage and computing capabilities, permits using cell-free massive MIMO deployments for the caching of content close to the UEs and for the realization of distributed computing architectures, which can be used to offload network-intensive computational tasks. Moreover, in low-demand situations, some APs can be partially switched off (control signals may still be transmitted) with a limited impact on the network performance, thus contributing to reducing the OPEX of network operators and their carbon footprint.

\subsubsection{Cell-Free Initial Access}

The basic connection procedures of cell search and random access determine the network performance in terms of latency, energy consumption, and the number of supported UEs~\cite{3gpp300}. These basic functionalities are currently tailored to the cellular architecture. Fig.~\ref{fig:cfac} illustrates initial access based on different mechanisms. As shown in the figure, in the cellular system, a UE attempts to access a cell in a cellular network. The UE usually chooses the strongest cell based on the measurement of received synchronization signals and will be subject to interference from neighboring cells. 
On the contrary, in a cell-free system, a UE attempts to access a cell-free network where all the neighboring APs support the UE's access to the network. To enable this, the traditional cell identification procedure must be re-defined, including the synchronization signals and how system information is broadcast.
Similarly, a new random access mechanism suitable for cell-free networks is needed such that some messages in the random access procedure can be transmitted and processed at multiple APs.

In summary, it is desirable that idle/inactive UEs can harness the benefits of the cell-free architecture as much as the active UEs can. To realize this, it is imperative to redesign the procedures of cell search and random access. To reduce the latency and improve the resource utilization efficiency, NOMA-enabled two-step random-access channel~\cite{3gpp812} or autonomous grant-free data transmission~\cite{yuan2018} should be investigated in cell-free networks.

\subsubsection{Implementation Challenges}

The huge amount of CSI that needs to be exchanged over the fronthaul to implement centralized joint precoding/combining is a long-standing scalability bottleneck for the practical implementation of CoMP and network MIMO~\cite{Bjornson2013d,Kal18}. In its original form~\cite{Ngo17,Nayebi2017a}, cell-free massive MIMO avoids CSI exchange by only performing signal encoding/decoding at the CPU, while combining/precoding (such as matched filtering, local zero-forcing, and local MMSE processing) is implemented at each AP using locally acquired CSI. By synchronizing the APs, the signals can be coherently combined without the need for CSI sharing. In this case, each AP consists of antennas and UE-grade RF modules that perform digital operations such as channel estimation, combining/precoding, interpolation/decimation, digital pre-distortion, and DFT~\cite{Int19}.

However, the performance of cell-free massive MIMO systems can be sensibly boosted by increasing the level of coordination among the APs, in particular, by sending the CSI to the CPU and performing joint combining/precoding~\cite{Nayebi2016a,Bjornson2020a} as in conventional CoMP and network MIMO. In this case, the APs can be made smaller since most of the digital operations are carried out at the CPU. While the data rates can be increased, the drawback is higher computational complexity and latency, and the fronthaul signaling might also increase. If there are 64 antennas taking 20-bit samples, the total fronthaul capacity will far exceed 1 Gbps at the sampling rates of interest. This imposes significant challenges for interconnection technologies, such as CPRI and on the I/O interface of processing circuits~\cite{Li18}. A potential solution to this problem could be to form separate antenna clusters and have a separate CPRI for each cluster. However, this increases the overall complexity of the system. Over-the-air bi-directional signaling between the APs and the UEs might be utilized as a flexible alternative to fronthaul signaling~\cite{Tol19,Atz20,Atz20a}. Furthermore, it is necessary to investigate scalable device coordination and synchronization methods to implement CSI and data exchange~\cite{Bjornson2019}.

To provide the aforementioned gains over cellular technology,  cell-free massive MIMO requires the use of a large number of APs and the attendant deployment of suitable fronthaul links. Although there are concepts (e.g., radio stripes where the antennas are integrated into cables~\cite{Int19}) to achieve practically convenient deployment, the technology is mainly of interest for crowded areas with a huge traffic demand or robustness requirements. A cell-free network will probably underlay a traditional cellular network, and is likely to use APs with large co-located arrays. The potential offered by integrated access and backhaul techniques could also be very helpful in alleviating the fronthaul problem and in reducing the cost of deployment.

Cell-free massive MIMO can be deployed in any frequency band, including below-6 GHz, mmWave, sub-THz, and THz bands. In the latter cases, the APs can serve each UE using a bandwidth of $100$~GHz or higher, which yields extremely high data rates over short distances and low mobility. The spatial diversity gains of the cell-free architecture become particularly evident in such scenarios because the signal from a single AP is easily blocked, but the risk that all neighboring APs are simultaneously blocked is vastly lower.

%% file: 4_integrated_access_backhaul.tex
Considering dense networks, we need to provide multiple APs of different types with backhaul connections. There are different backhauling methods today, among which wireless microwave and optical fiber are the dominant ones. Fiber provides reliable transport with demonstrated Tbps rates. For this reason, the use of (dark) fiber will continue growing in 5G and beyond. However, fiber deployment requires a noteworthy initial investment for trenching/installation, may incur long installation delays, and even may be not allowed in some metropolitan areas. 

Wireless backhaul using microwave is a competitive alternative to fiber, supporting up to 100 Gbps \cite{sagIAB1}. Particularly, compared to fiber, microwave backhaul is an economical and scalable option, with considerably lower cost and flexible/timely deployment (e.g., no digging, infrastructure displacement, and possible to deploy in principle everywhere) \cite{MSIAB2}, \cite{refi1ercIAB}. Today, microwave backhauling operates mainly in the 4–70/80 GHz range. Then, given the fact that 6G networks will see an even greater level of densification and spectrum heterogeneity, compared to 4G and even 5G networks, wireless backhauling is expected to play a major role, especially at mmWave and THz carrier frequencies \cite{refi1ercIAB}.

\begin{figure}
\centerline{\includegraphics[width=3.5in]{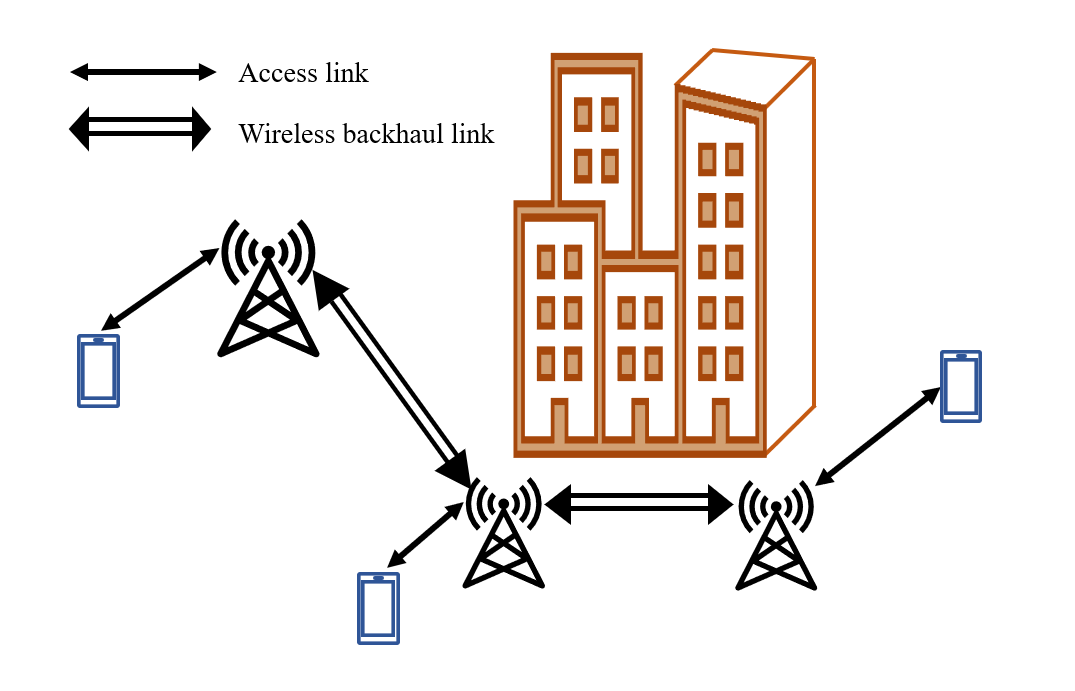}}
\caption{
An example of a multi-hop IAB network.\label{iab3gpp1}}
\end{figure}

Following the same reasoning, IAB networks, where the operator uses part of the spectrum resources for wireless backhauling, has recently received considerable attention \cite{madapatha2020integrated}, \cite{refioumer6}. IAB aims to provide flexible low-cost wireless backhaul using 3GPP NR technology in IMT bands, and provide not only backhaul but also the cellular services in the same node. This will be a complement to existing microwave point-to-point backhauling in suburban and urban areas (see Fig. \ref{iab3gpp1}).

Wireless backhaul was initially studied in 3GPP in the scope of LTE relaying \cite{refiIAB5}. However, there have been only a few commercial LTE relay deployments with separate bands for access and backhaul as well as only one-hop relaying. This is mainly because the existing LTE spectrum is very expensive to be used for backhauling, and also network densification did not reach the expected potential in the 4G timeline. As opposed, IAB is expected to be more commercially successful, compared to LTE relaying, mainly because: 
\begin{itemize}
    \item The large bandwidth available in mmWave (and, possibly, higher) frequencies creates more economically viable opportunities for backhauling.
    \item The limited coverage of high frequencies creates a growing demand for AP densification, which, in turn, increases the backhauling requirement. 
    \item Advanced spatial processing features, such as massive MIMO, enables the same bandwidth to be simultaneously used for both access and backhaul \cite{massivemimobook}.
\end{itemize}
\subsubsection{On the Performance of IAB Networks}
IAB supports both sub-6 GHz and mmWave spectrum as well as both inband and outband backhauling, where the wireless backhaul links operate, respectively, in the same and different frequency bands, as the access links. In this subsection, we present a comparison between the performance of IAB and fiber-connected networks using mmWave spectrum and inband backhauling.

Assume an outdoor two-tier HetNet with multiple MBSs (M: macro), SBSs (S: small) and UEs. Here, both the MBSs (IAB donor, in the 3GPP terminology) and the SBSs (IAB node, in the 3GPP terminology) use wireless connections for both backhaul and access. Moreover, only the IAB donors are fiber-connected while the IAB nodes, i.e., SBSs, receive backhaul data from the IAB donors wirelessly. 

We model the network as an FHPPP, e.g., \cite{refnewbIAB}, \cite{refIAB1}, in which the MBSs, the SBSs and the UEs are randomly distributed within a finite region according to mutually-independent FHPPPs having densities $\phi_\text{M}$, $\phi_\text{S}$ and $\phi_\text{U}$, respectively.
In this way, following the mmWave channel model in \cite{refIAB1}, the  power received by each node can be expressed as
\begin{equation}
    P_{\text{r}}=P_{\text{t}}h_{\text{t,r}}G_{\text{t,r}}{L}_{(1\text{m})}L_{\text{t,r}}\left|\left|\mathbf{x}_{\text{t}}-\mathbf{x}_{\text{r}}\right|\right|^{-1}.
\end{equation}
where $\mathbf{x}_{\text{t}},\mathbf{x}_{\text{r}}$ are the locations of the transmitter and receiver, respectively.
Moreover, $h_{t,r}$ denotes the small-scale fading, $P_{\text{t}}$ is the transmit power, and $G_{t,r}$ represents the combined antenna gain of the transmitter and the receiver of the considered link. Also, $L_{t,r}$ and ${L}_{(1m)}$ are the path loss due to propagation and the reference path loss at 1 meter distance, respectively. In the simulations, we model the small-scale fading by a normalized Rayleigh random variable. Using the 5GCM UMa close-in model \cite{refIAB3}, the path loss, in dB, is given by 
\begin{equation}
\text{PL}=32.4+10 \alpha \log_{10} \left(\frac{r}{1 \textrm{ m}} \right) +20\log_{10} \left(\frac{f_\text{c}}{1 \textrm{ GHz}} \right),
\end{equation}
with $r$ denoting the propagation distance between the nodes, $f_\text{c}$ being the carrier frequency, and $\alpha$ is the path loss exponent. Depending on the blockage, NLoS and LoS links are affected by different path loss exponents, and the propagation loss in the link between nodes $i$ and $j$ is obtained by
\begin{equation}L_{\text{i,j}} = \begin{cases} (r / 1 \textrm{ m})^{\alpha _{\text{{N}}}},&\text {if NLoS,} \\ (r / 1 \textrm{ m})^{\alpha_{{{\text{L}}}}},&\text {if NLoS.}{} \end{cases}
\end{equation}
For the blockage, we use the germ grain model \cite[Chapter 14]{refIAB2}, which is an accurate model for environments with large obstacles because it takes the obstacles induced blocking correlation into account. Here, the blockages are distributed according to an FHPPP distributed with density $\lambda_{\text{B}}$ in the same area as the other nodes. Also, all blockings are assumed to be walls of length $l_{\text{B}}$ and orientation $\theta$, which is an IID uniform random variable in $[0, 2\pi)$. 

Modeling the beam pattern as a sectored-pattern antenna array, the antenna gain between two nodes is obtained by \cite{refIAB1}
\begin{equation}G_{i,j }(\varphi)= \begin{cases} G_0&\frac{-\theta_{\text{HPBW}}}2\leq\varphi\leq\frac{\theta_{\text{HPBW}}}2 \\ g(\varphi)&\text{otherwise,} \end{cases}
\end{equation}
with $i, j$ denoting the indices of the transmit and receive nodes, $\varphi$ representing the angle between them and $\theta_\text{{HPBW}}$ being the half power beamwidth of the antenna. Also, $G_0$ denotes the directional antenna’s maximum gain and $g(\varphi)$ gives the side lobe gain. 

In our analysis, the inter-UE interference is neglected, motivated by the low power of the devices and with the assumption of sufficient isolation. Also, motivated by the high beamforming capability in the IAB-IAB backhaul links, we ignore the interference and assume them to be noise-limited. Then, the UEs are served by either an MBS or an SBS following open access strategy and based on the maximum average received power rule. Also, we follow the same approach as in \cite{madapatha2020integrated} to determine the bandwidth allocation in the access and backhaul of each SBS proportional to its load and the number of UEs in the access link such that the network coverage probability is maximized.

\begin{figure}
\centerline{\includegraphics[width=3.5in]{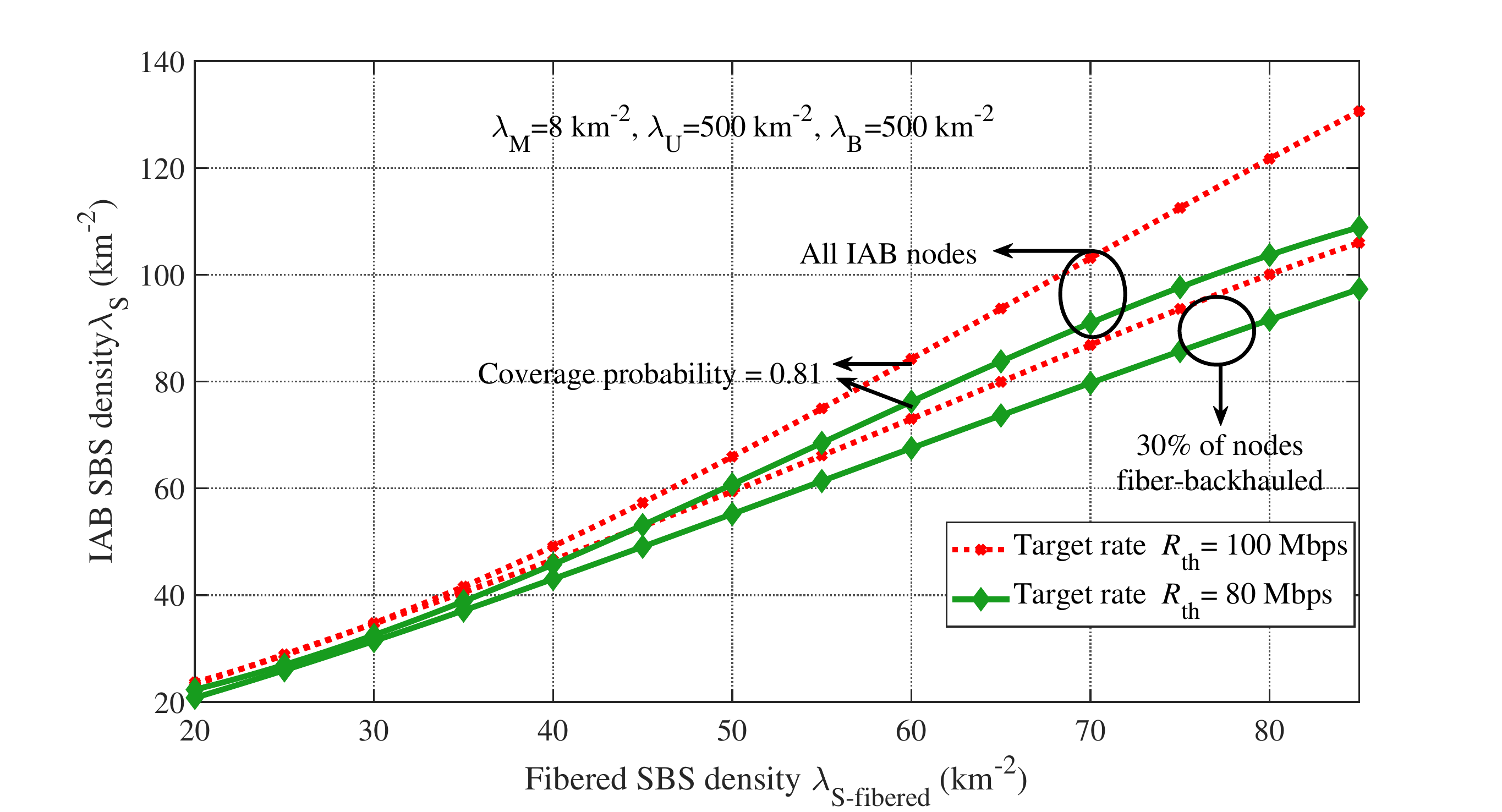}}
\caption{
 Density of IAB nodes providing the same coverage probability as in fiber-backhauled networks. The parameters are set to $(\alpha_\text{LoS}, \alpha_\text{NLoS}) = (2, 3)$, $l_\text{B} = 5$ m, $f_\text{c} = 28$ GHz, bandwidth$= 1$ GHz and ${P_{\text{MBS}}, P_{\text{SBS}}, P_{\text{UE}}} = (40, 24, 0)$ dBm. \label{iabeq}}
\end{figure}

\begin{figure}
\centerline{\includegraphics[width=3.5in]{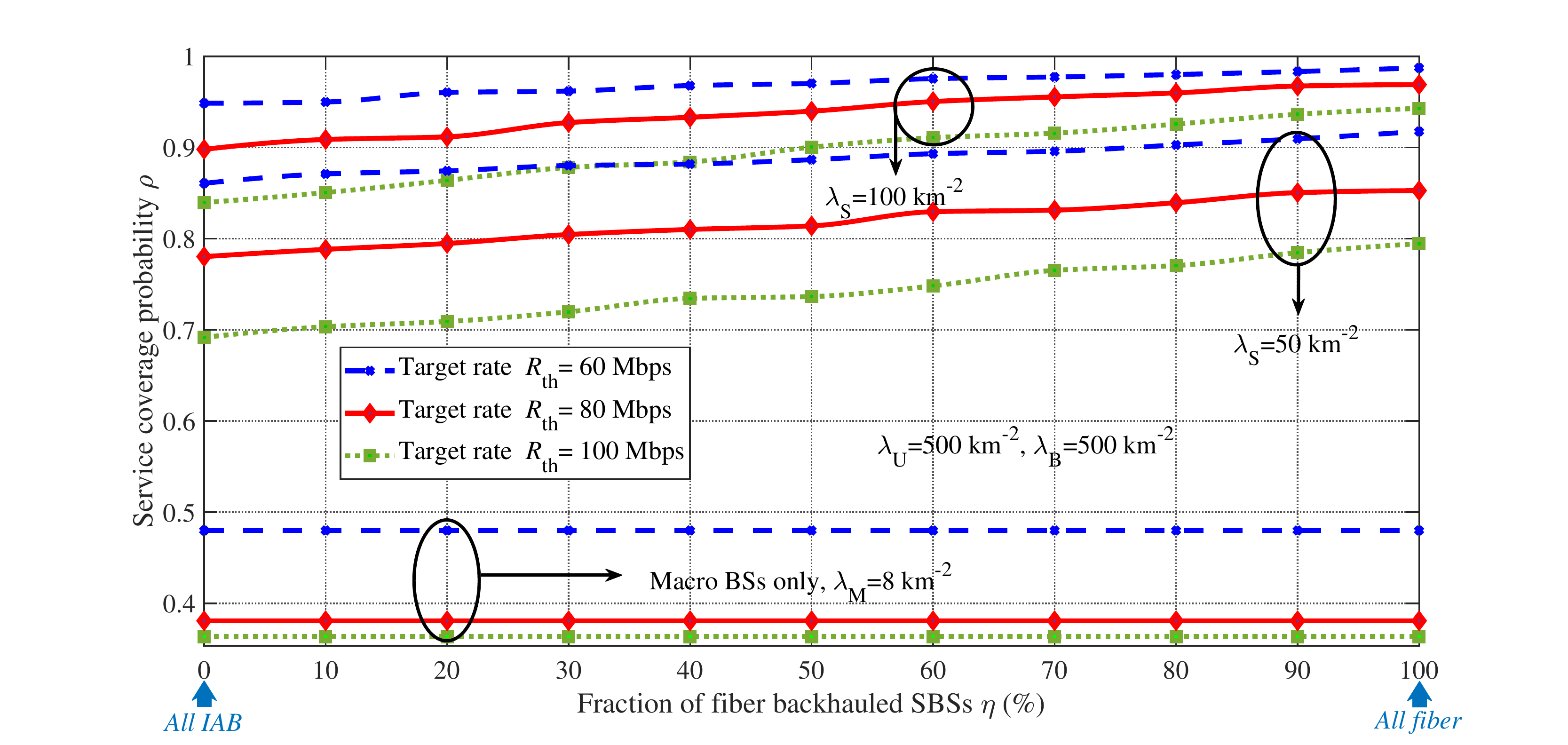}}
\caption{Service coverage probability as a function of percentage of fiber-backhauled SBSs, $(\alpha_\text{LoS}, \alpha_\text{NLoS}) = (2, 3)$, $l_\text{B} = 5$ m, $f_\text{c} = 28$ GHz, bandwidth$= 1$ GHz and ${P_{\text{MBS}}, P_{\text{SBS}}, P_{\text{UE}}} = (40, 24, 0)$ dBm. \label{hybrid}}
\end{figure}

In Figs.~\ref{iabeq} and \ref{hybrid}, we compare the performance of IAB and fiber-connected networks, in terms of coverage probability, i.e., the probability of the event that the UEs' minimum target rate requirements are satisfied. Note that, in practice, a number of SBSs may have access to fiber. For this reason, the figures also present the results for the cases with a fraction of SBSs being fiber-connected. In this case, we assume the fiber-connected SBSs to be randomly distributed, and adapt the association, the resource allocation rule and the achievable rates correspondingly.

Fig.~\ref{iabeq} shows the required number of IAB nodes that guarantee the same coverage probability as in the cases with (partially) fiber-connected SBSs. Moreover, Fig. \ref{hybrid} demonstrates the coverage probability versus the fraction of fiber-connected SBSs. The figure also compares the performance of the IAB network with the cases having only MBSs. In each simulation setup, we have followed the same approach as in \cite{madapatha2020integrated} to optimize different parameters, such as resource allocation and UE association, such that the coverage probability is maximized. The details of the simulation parameters are given in the figures’ captions. 

As can be seen in the figures, IAB network increases the coverage probability, compared to the cases with only MBSs, significantly (Fig.~\ref{hybrid}). For a broad range of parameter settings, the IAB network can provide the same coverage probability as in the fiber-connected network with relatively small increment in the number of IAB nodes. For instance, consider the parameter settings of Fig. \ref{iabeq} and the UEs' target rate threshold of 100 Mbps. Then, a fully fiber-connected network with the SBSs density $\lambda_\text{s} = 60 \text{ km}^{-2}$ corresponds to an IAB network with density $\lambda_{s}= 85 \text{ km}^{-2}$ resulting in coverage probability $0.81$. Also, interestingly, providing $30\%$ of the SBSs with fiber connection reduces the required density to $\lambda_\text{s}=70 \text{ km}^{-2}$, i.e., only $16\%$ increment in the number of SBSs. Then, as the network density increases, the relative performance gap of the IAB and fiber-connected networks decrease (Fig. \ref{iabeq}). Finally, it is interesting to note that our results, which are based on FHPPP, give a pessimistic performance of the IAB
networks. The reason is that, in practice, the network will be fairly well planned, resulting in even less gap between the performance of IAB and fiber-connected networks.

Such a small increment in the number of fiber-free SBSs, i.e., IAB nodes, leads to the following benefits:
\begin{itemize}
    \item \emph{Network flexibility increment}: As opposed to fiber-connected networks, where the APs can be installed only in the places with fiber connection, the IAB nodes can be installed in different places as long as they have fairly good connection to their parent nodes. This increases the network flexibility and the possibility for topology optimization remarkably.
    \item \emph{Network cost reduction}: An SBS is much cheaper than fiber\footnote{In general, the fiber cost varies vastly in different regions. However, for different areas, fiber installation accounts to a significant fraction, in the order of $80\%,$ of the total network cost}. For instance, as illustrated in \cite[Table 7]{refIAB8}, in an urban area the fiber cost is in the range of 20000 GBP/km, while an SBS in 5G is estimated to cost around 2500 GBP per unit \cite{refIAB9}. Moreover, different evaluations indicate that, for dense urban/suburban areas, even in the presence of dark fiber, the IAB network deployment is an opportunity to reduce the total cost of ownership.
    \item \emph{Time-to-market reduction}: Fiber laying may take a long time, because it requires different permissions and labor work. In such cases, IAB can establish new radio sites quickly. Consequently, starting with IAB and, if/when required, replacing it by fiber is expected to be a common setup.
\end{itemize}

Along with these benefits, different evaluations indicate that, unless for the cases with suburban areas and moderate/high tree foliage, with proper antenna height and network density the performance of the IAB network is fairly robust to blockage, rain and tree foliage, which introduces it as a reliable system in different environments. These are the reasons that different operators have shown interest to implement IAB in 5G networks \cite{refIAB12}, \cite{refIAB13}, and the trend is expected to increase in beyond 5G/6G networks.

%% file: 4_integrated_space_terrestrial.tex
\begin{figure*}[t]
\centering
\includegraphics[width=\textwidth]{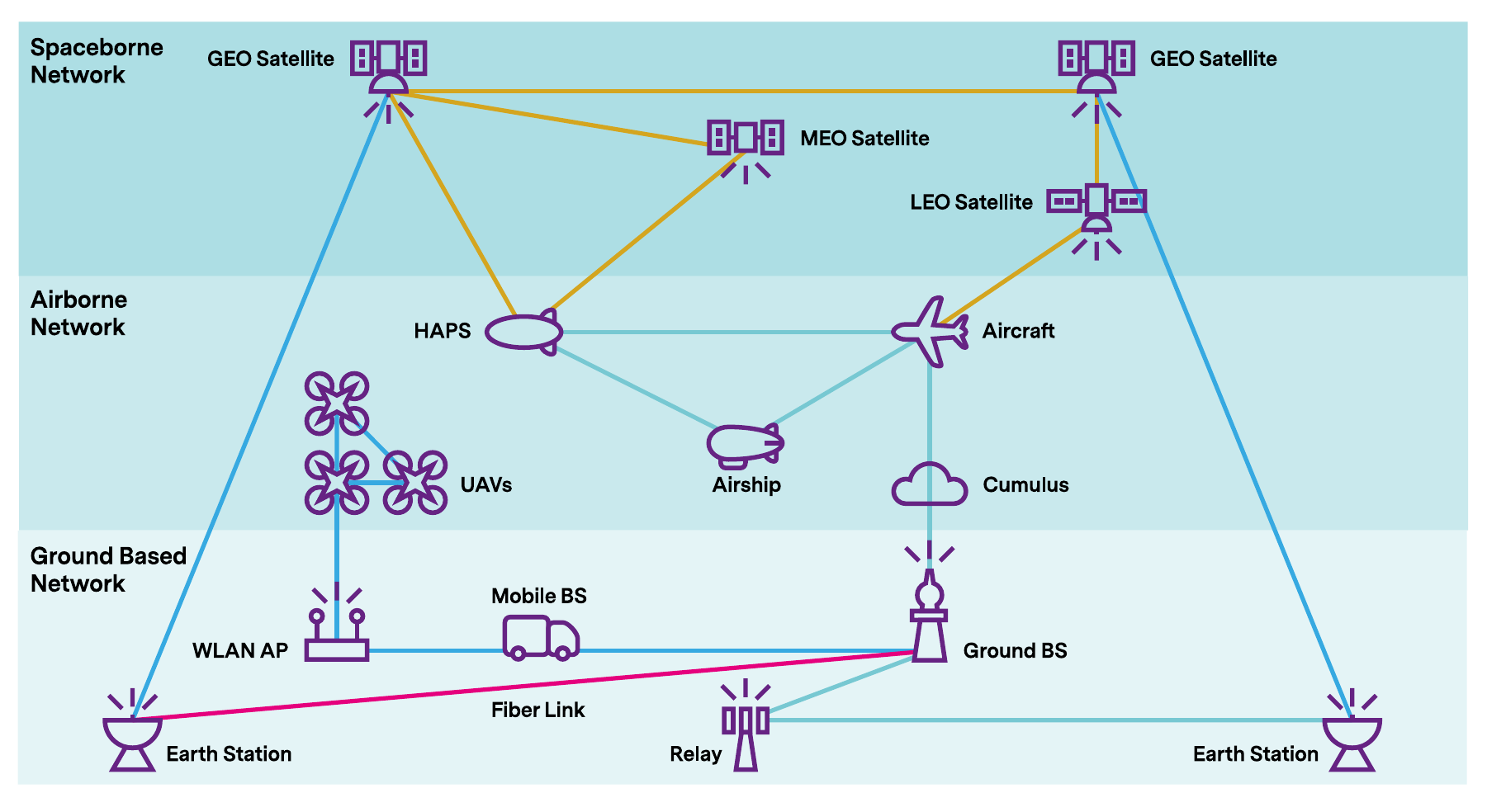}


\caption{Illustration of the ISTN architecture and its three layers. Yellow line: free space optics link; green line: mmWave link. Blue line: other microwave link).}
\label{fig:ISTN}
\end{figure*}

In parallel with the development of terrestrial mobile systems such as 5G, another major international effort in wireless communications is the development of space communications networks which enable global coverage at any time and from anywhere, such as on the sea, over the air and space, and in rural and remote areas. The concept of using various space platforms to perform data acquisition, transmission and information processing has been around for several tens of years. However, due to the limited bandwidth and large transmission delay, existing space networks alone cannot provide the sufficient capacity and guaranteed quality-of-service to meet the ever increasing demand for global wireless connectivity. Seamlessly interconnecting space networks with terrestrial networks to support truly global high-speed wireless communications will be one of the major objectives of the 6G wireless systems \cite{HuangVTM2019}. 

Fig.~\ref{fig:ISTN} shows the architecture of a typical ISTN which consists of three layers: the spaceborne network layer, the airborne network layer, and the conventional ground-based network layer. The spaceborne network consists of various orbiting satellites such as the geostationary-Earth orbit (GEO) satellites, medium-Earth orbit (MEO) and LEO satellites, and mini satellites known as CubeSats. The airborne network consists of various aerial platforms including stratospheric balloons, airships and aircrafts, UAVs, and HAPSs. The traditional ground-based networks include wireless cellular networks, satellite ground BSs, mobile satellite terminals, and many more. The integrated network can make full use of the signal propagation characteristics of large space coverage, low loss LoS transmission, etc. to achieve seamless high-speed communications with global coverage. 

There are a number of grand technological challenges to achieve the ISTN with high capacity and low cost. The two predominant bottlenecks that limit the current ISTN development are the available bandwidth for high-speed aerial backbones and the spectral efficiency for direct air-to-ground communications between the airborne and ground based networks. Aerial backbones uses high-speed links to connect the major nodes in the airborne network and between the airborne and ground based networks. They play a pivotal role in the ISTN by handling the aggregation and distribution of various data flows such as voice, video, Internet, and other data sources. Because it sits in-between the spaceborne network and ground based network, the airborne network is an indispensable and important intermediate layer in the ISTN and hence the high-speed, flexible, and all-weather mobile aerial backbones become the important infrastructure to support the services provided by the airborne network. Unfortunately, existing aerial backbones mainly use microwave links with very limited bandwidth and, hence, cannot meet the requirements of the future high capacity ISTN. Frequency reuse can be easily achieved in ground based cellular networks by limiting the coverage area of each cell within its cell boundary in order to improve the spectral efficiency, but this is not true for communications between the space and terrestrial networks. Because the satellite is far away from the ground, the coverage of a narrow communication beam will be very large when it reaches the ground, resulting in very low spectral efficiency. For the given available bandwidth, low frequency reuse will significantly affect the overall capacity of the ISTN.

The capacity of the aerial backbones is proportional to the bandwidth. The available bandwidth is limited in lower bands to a few MHz up to a few hundred MHz. One way to increase the bandwidth is to use a higher frequency band (such as mmWave or THz). However, then the signal propagation is affected by atmospheric losses, and the communication distance is greatly reduced.
Another way to increase the capacity is to utilize multiple antennas to perform spatial multiplexing, but the size and weight limits the number of antennas that can be deployed on a satellite.

Taking all the above ways into consideration, it is believed that using an mmWave communication system is the best choice for realizing the high-speed RF backbone with data rate up to 100 Gbps. Due to the over 10 GHz bandwidth in the mmWave band (such as the 71-76 and 81-86 GHz E-band), it can meet the spectrum requirements of the system. The increase in transmit power can be solved with a high power amplifier combined with an antenna array. The atmospheric attenuation is below 0.3 dB/km and the total atmospheric and cumulus loss (assuming 50 km link distance) is estimated to be only about 10 dB. Multi-channel transmission can also be implemented using line-of-sight multiple input multiple output technology (LoS-MIMO).

Although being superior to microwave systems, currently existing mmWave links still cannot meet the requirements of the high capacity ISTN due to the lack of advanced technologies. For instance, in-band full duplex has not been adopted due to the difficulty in cancelling self-interference especially in mmWave frequencies, thus only reaching 50\% of the potential data rate. Further, although LoS-MIMO has been considered to achieve spatial multiplexing, the rank of the channel is limited since the propagation distance is large and the antenna spacing is small.
Finally, yet importantly, to reach longer communication distance, sufficiently high transmit power is necessary. Even with the most powerful 40 dBm solid-state power sources in mmWave bands, achieving longer communication distance beyond 50 km is still very difficult. This calls for the employment of massive, preferably conformal and reconfigurable, antenna arrays.

Other enabling technologies to realize high-speed and low-cost ISTN include the modelling of the mmWave aeronautical transmission channels and the dynamic behaviours of spaceborne and airborne networks especially when LEO satellites are involved, three-dimensional (Space-Air-Ground) networking and optimization, and high-speed communication protocol optimization.

%% file: 4_broadcast.tex
The demands on network capabilities continue to increase, among other aspects, in terms of capacity, availability and cost. These increasing demands could be challenging to networks that only support unicast and, more precisely, when there are more users that require simultaneous service than the APs are able to separate by beamforming. When some of those users request the same data at the same time, broadcast and multicast are suitable transport mechanisms \cite{Saily_Broadcast}. They are great options for large-scale delivery, since they permit the transmission of the same content to a vast number of devices within the covered area simultaneously, and with a predefined quality-of-service.

Future 6G delivery networks need to be as flexible as possible to respond to the needs of service providers, network operators and users. Hence, broadcast and multicast mechanisms could be integrated into 6G networks as a flexible and dynamic delivery option to enable a cost-efficient and scalable delivery of content and services in situations where the APs have limited beamforming capabilities. More precisely, if the AP can only send wide beams, then it should be able to broadcast information over the coverage area of that beam.

\begin{figure}[t]
\centering
\includegraphics[width=0.45\textwidth]{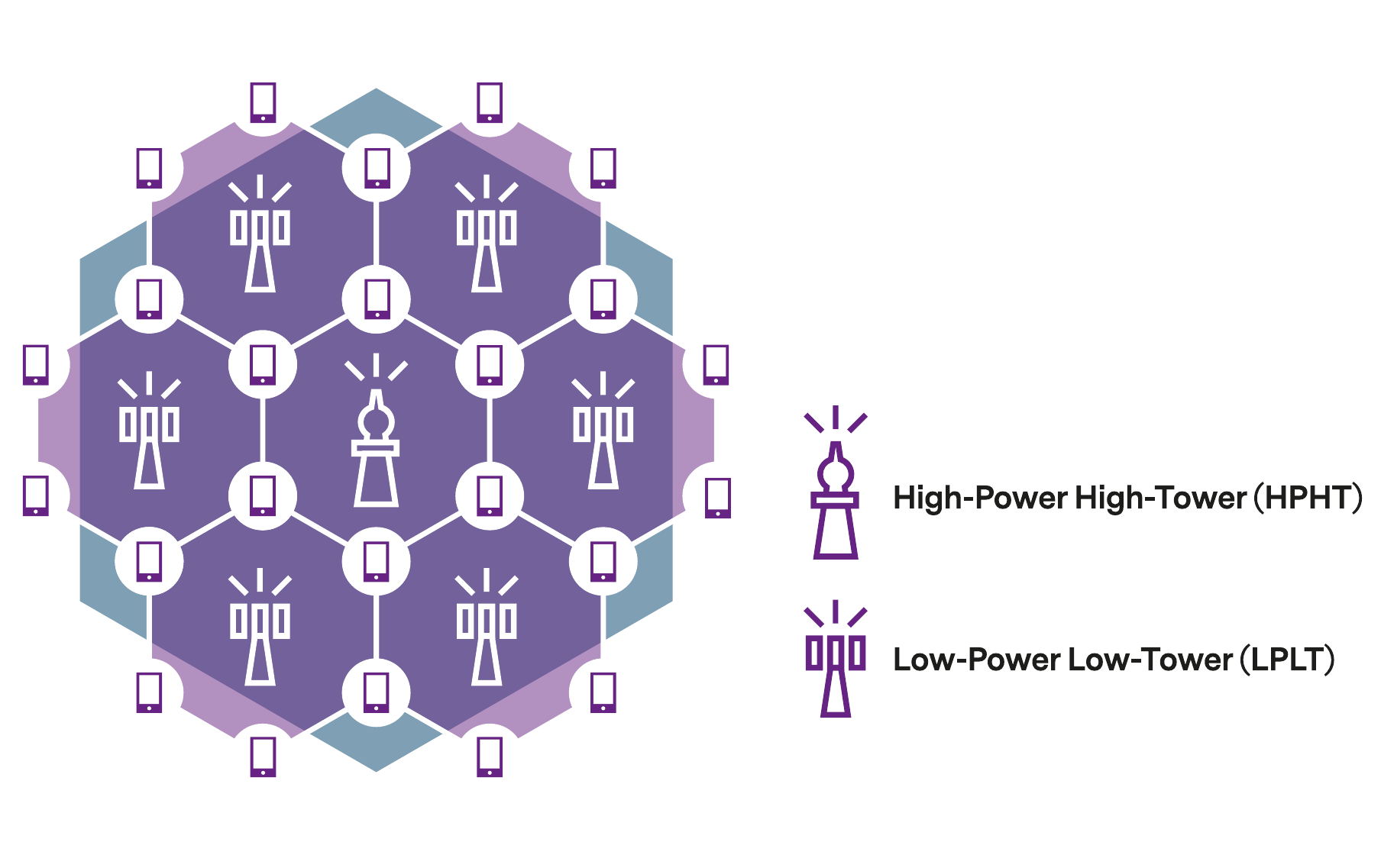}
\caption{Network convergence and different topologies involved.}
\label{Figure_2}
\end{figure}

6G also represents a great opportunity for the convergence of mobile broadband and traditional broadcast networks, usually used for TV video broadcasting  \cite{Tran_Broadcast}. Low-Power Low-Tower cellular networks with adaptive beamforming capabilities would benefit from the complementary coverage provided by High-Power High-Tower broadcast networks with fixed antenna patterns \cite{5GTerrestrialBroadcast}, as shown in Fig.~\ref{Figure_2}. A technology flexible enough to efficiently distribute content over any of these networks would allow the 6G infrastructure to better match the needs of future consumers and make more efficient use of the existing tower infrastructure. The convergence could be addressed from different perspectives, i.e., the design of a single and highly efficient radio physical layer; the use of common transport protocols across fixed and mobile networks including broadcast, multicast and unicast services; or a holistic approach that allows client applications running on handsets to better understand and, therefore, adapt to the capabilities of the underlying networks.

In Europe, it was decided to ``ensure availability at least until 2030 of the 470-694 MHz (‘sub-700 MHz’) frequency band for the terrestrial provision of broadcasting services'' \cite{EUJournal}. Some assignations are already happening in the USA, where the 600 MHz frequency band is being assigned to mobile broadband services \cite{Gomez-Barquero}. The timing is very well aligned with the release of 6G, and it is now when a full convergence could take place.

\begin{figure}[t]
\centering
\includegraphics[width=0.45\textwidth]{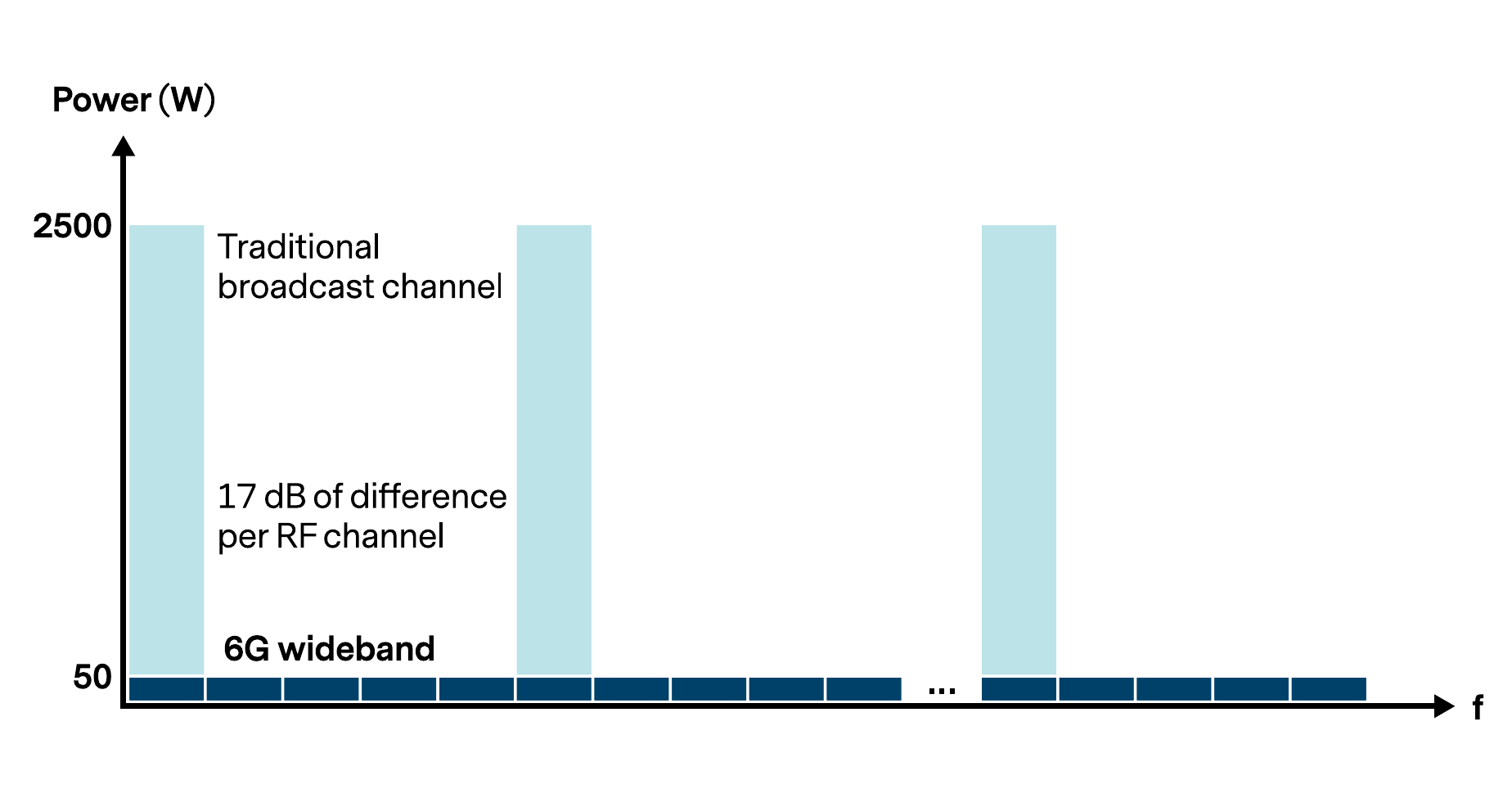}

\caption{6G wideband concept and comparison with traditional digital broadcast (Source: \cite{Stare}).}
\label{Figure_3}
\end{figure}

A potential solution would be the use of a 6G wideband broadcasting system, proposed in \cite{Stare, Wideband}. By using 6G wideband, all RF channels within a particular frequency band could be used on all high-power high-tower transmitter sites (i.e., frequency reuse-1). This is drastically different from current broadcasting networks, where usually a frequency reuse-7 is used to avoid inter-cell interference among transmitters. The entire wideband signal requires only half the transmission power of a single traditional digital RF channel, as shown in Fig.~\ref{Figure_3}. Another advantage is that a similar capacity could be obtained, thanks to the use of a much more robust modulation and coding rate combination, since the whole frequency band is employed. In terms of power, 6G wideband also permits to transmit about 17 dB less power (around 50 times) per RF channel (considering a bandwidth of 8 MHz, typical from digital terrestrial broadcasting channels), although using more RF channels per station. This leads to a total transmit power saving of around 90\% \cite{Stare}. Thanks to this higher spectrum use, the approach allows not only for a dramatic reduction in fundamental power and cost, but also about a 37-60\% increase in capacity for the same coverage as with current services.

%% file: 5_coding_modulation_waveforms_duplex.tex
\subsubsection{Channel Coding}
From 4G to 5G, the peak data rate has increased by 10-100 times, and this trend is likely to continue with 6G. The throughput of a single decoder in a 6G device will reach hundreds of Gbps. Infrastructure links are even more demanding since they aggregate user throughput in a given cell or virtual cell, which is expected to increase due to spatial multiplexing. However, it will be difficult to achieve such a high throughput, only relying on the progress of integrated circuit manufacturing technology within ten years. Solutions must be found on the algorithm side as well. Both FEC code design and corresponding encoding/decoding algorithms need to be taken into account to reduce the decoding iterations and improve the decoder's level of parallelism. Moreover, it is vital for the decoder to achieve reasonably high energy efficiency. To maintain the same energy consumption as in current devices, the energy consumption per bit needs to be reduced by 1-2 orders of magnitude. Implementation considerations such as area efficiency (in Gbps/mm$^2$), energy efficiency (in Tb/J), and absolute power consumption (W) place huge challenges on FEC code design, decoder architecture, and implementation~\cite{EPIC_project}.

FEC code design implies trade-offs between communications performance and high throughput. High communications performance typically requires complex decoding algorithms and large number of iterations to achieve near maximum-likelihood performance and large block lengths to approach the Shannon bound. On the other hand, 6G communication systems require flexibility in the codeword length and coding rate. The most commonly used coding schemes today are turbo codes, polar codes, and LDPC. Their performance and throughput have already been pushed towards the limit using 16 nm and 7 nm technology \cite{EPIC_project}. Trade-offs must be made between parallelization, pipelining, iterations, and unrolling, linking with the code design and decoder architecture.
Future performance and efficiency improvements could come from CMOS scaling, but the picture is quite complex \cite{Kestel_2018}. Indeed, trade-offs must be made to cope with issues such as power density/dark silicon, interconnect delays, variability, and reliability. Cost considerations render the picture even more complex: costs due to silicon area, design effort, test, yield and masks, explode at 7 nm and below.

\begin{figure}
    \centering
    \includegraphics[width=1\columnwidth]{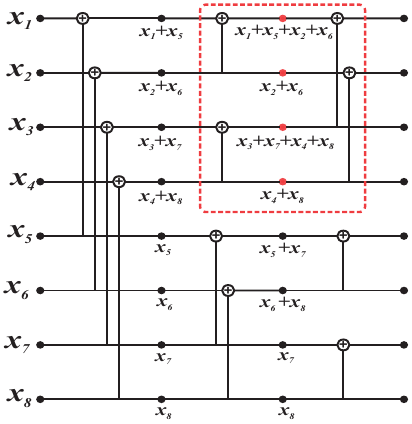}
    \caption{Partially permuted factor graph for polar code of block length 8.}
    \label{fig:partialpermute}
\end{figure}

The channel coding scheme used in 6G high-reliability scenarios
must provide a lower error floor and better “waterfall”
performance than that in 5G. Short and moderate length codes
with excellent performance need to be considered. Polar codes,
due to their error correction capabilities and lack of error floor,
might be the preferred choice in 6G. However, state-of-the art
CA-SCL decoding does not
scale up well with throughput due to the serial nature of the
algorithm. As a result, iterative algorithms like BP which are more parallelizable have become a prime candidate for channel decoding in high throughput data transmissions in 6G. However, as of now for polar codes, there exists a significant performance gap between state-of-the art CA-SCL decoding and BP. Hence, effort has been made towards improving the performance of iterative algorithms. In \cite{Elkelesh_permute}, authors propose a novel variant of BP called multi-trellis belief propagation which is based on permuted factor graphs of the polar code. Further improvements to the  algorithm are investigated in \cite{permuteList,warren_permute}. In \cite{ranasinghe2019partially}, along with a discussion on implication of various aspects of the polar code factor graph on the performance, a new variant of the multi-trellis BP decoder is proposed based on permutation of a subgraph of the original factor graph (Fig.~\ref{fig:partialpermute}). This enables the decoder to retain information about variable nodes in the subgraphs, which are not permuted, reducing the required number of iterations needed in-between the permutations. Even though the iterative algorithms are prime candidates in high throughput applications, progress has been made on reducing the latency and improving the parallelizability of  algorithms based on successive cancellation. In \cite{fastpolar} authors propose methods to calculate the information bits without traversing the complete binary tree on sub trees with specific information bit and frozen bit patterns resulting in latency improvements. Same idea is extended to successive cancellation list decoding in \cite{fastlist}. Work in \cite{gamage2019low} proposes some new algorithms to process new types of node patterns that appear within multiple levels of pruned sub-trees, and it enables the processing of certain nodes in parallel improving the throughput of the decoder. Furthermore, modified polar code constructions can be adopted to improve the performance of belief propagation by selecting the information bits such that the minimum stopping set is larger \cite{stopping_set_and_girth_polar_code}. In \cite{tradeoffcomplexity} similar approach is adopted to improve the latency in SC decoding. Here, the polar code construction is formulated as an optimization problem which minimizes the complexity under a mutual information based performance constraint. Another potential avenue of improving the throughput is the use of deep neural networks to approximate the iterative decoders which is called as deep unfolding \cite{ondeeplearning,deeplinear}. These approaches could lead to better communication performance but they will require significant advances in understanding the behavior, robustness, and generalization of neural networks.

\subsubsection{Modulation and Waveform}

Modulation is another aspect that can be revised in 6G. High-order QAM constellations have been used to improve spectral efficiency in high SNR situations. However, because of the non-linearity of hardware, the benefits obtained in higher-order QAM constellations are gradually disappearing. 
Non-uniform constellations have been adopted in the ATSC 3.0 standard for terrestrial broadcasting \cite{LZMA16}. Probabilistic shaping schemes as in \cite{bocherer2015bandwidth,PX17} provide in general even better performance and are nowadays widely used for optical fiber communication, and they can be employed for wireless backhaul channels \cite{UXK18}. Significant shaping gains can already be achieved by simple modifications of the 5G NR polar coding chain \cite{IBX18,icscan2019probabilistic, IBX20}. Similar approaches may be used to generate transmit symbol sequences with arbitrary probability distributions optimized for different channel models (see e.g. \cite{SB16, IBX19, BIX20} and the references therein).

Reducing the PAPR is another important technology direction in order to enable IoT with low-cost devices, edge coverage in THz communications, industrial-IoT applications with high reliability, etc.
There will be many different types of demanding use cases in 6G, each with its own requirements. No single waveform solution will address the requirements of all scenarios. For example, as discussed in section \ref{terahertz_technologies}, the high-frequency scenario is faced with challenges such as higher phase noise, larger propagation losses, and lower power amplifier efficiency. Single-carrier waveforms might be preferable over conventional multi-carrier waveforms to overcome these challenges \cite{OFDMvsSC}. 
For indoor hotspots, the requirements instead involve higher data rates and the need for flexible user scheduling. Waveforms based on OFDM or its variants exhibiting lower out-of-band emissions \cite{UFMC1,UFMC2,UFMC3,FBMC,EuCNC2016} will remain a good option for this scenario. 
6G needs a high level of reconfigurability to become optimized towards different use cases at different times or frequencies.

As previously mentioned, a critical impact from RF impairments is expected for THz systems. The main impairments include: IQ imbalance; oscillator phase noise; and more generally non-linearities impacting amplitude and phase. The non-linearities of analog RF front-end create new challenges in both the modeling of circuits and the design of mitigation techniques. It motivates the definition for new modulation shaping, inherently robust to these RF impairments. As an example, we consider the robustness to the phase noise. According to \cite{9013189_bicais}, when the corner frequency of the oscillator is small in comparison to the system bandwidth, the phase noise can be accurately modeled by an uncorrelated Gaussian process. Using this mathematically convenient assumption, it is demonstrated in \cite{9093997_bicais} that using a constellation defined upon a lattice in the amplitude-phase domain is particularly relevant for phase noise channels. More specifically, a constellation defined upon a lattice in the amplitude-phase domain is robust to PN and leads to a low-complexity implementation. The Polar-QAM (PQAM) constellation is introduced in \cite{9093997_bicais}. This constellation defines a set of $M$ complex points placed on $\Gamma \in \{1,2,4,\dots,M\}$ concentric circles, i.e. amplitude levels. Each of the $\Gamma$ circles contains $M/\Gamma$ signal points. Any PQAM constellation are hence entirely defined by two parameters, the modulation order $M$ and the modulation shape $\Gamma$. 

\begin{figure*}[]
\centering
\begin{tabular}{cc}
\includegraphics[width=0.45\textwidth]{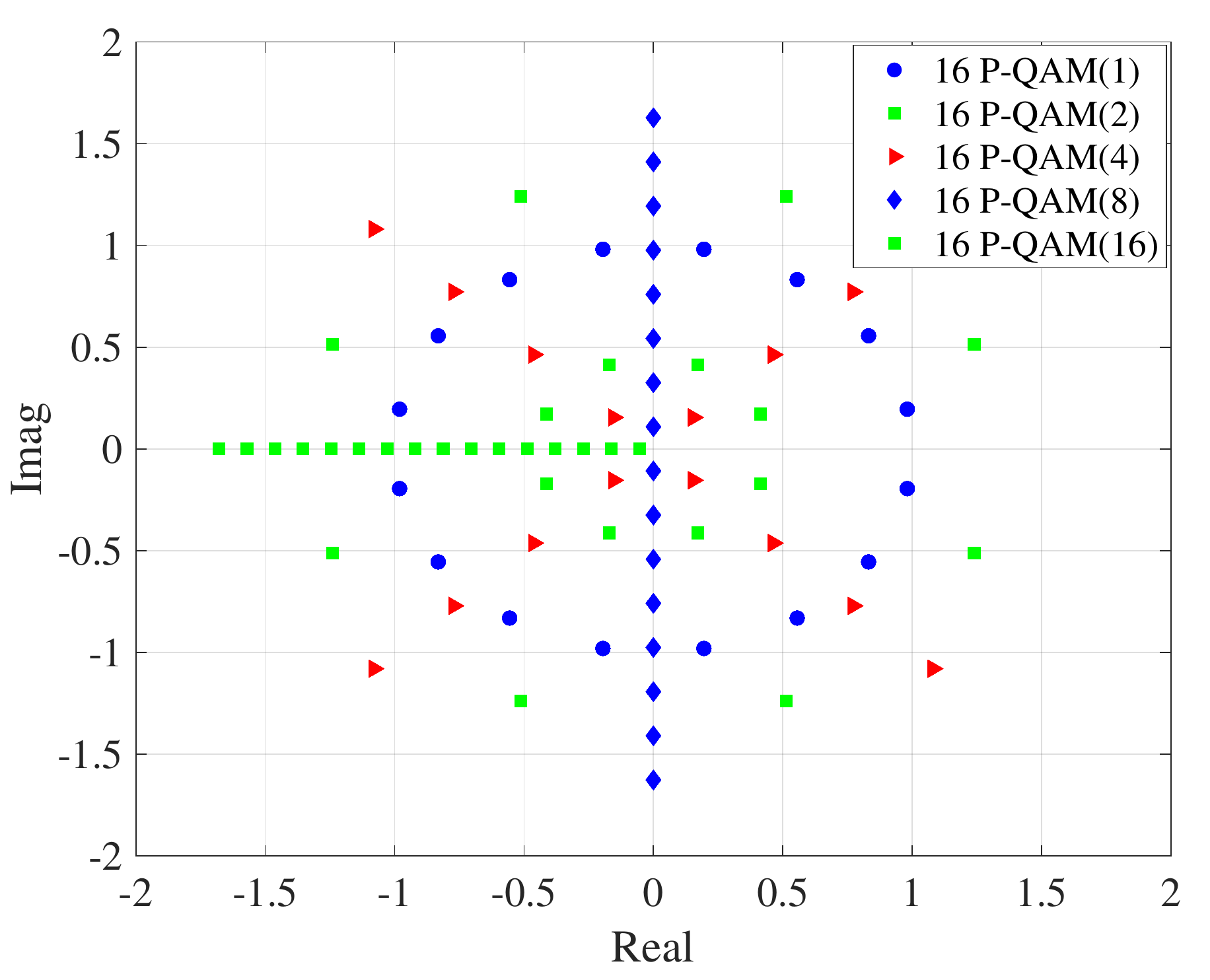} & \includegraphics[width=0.45\textwidth]{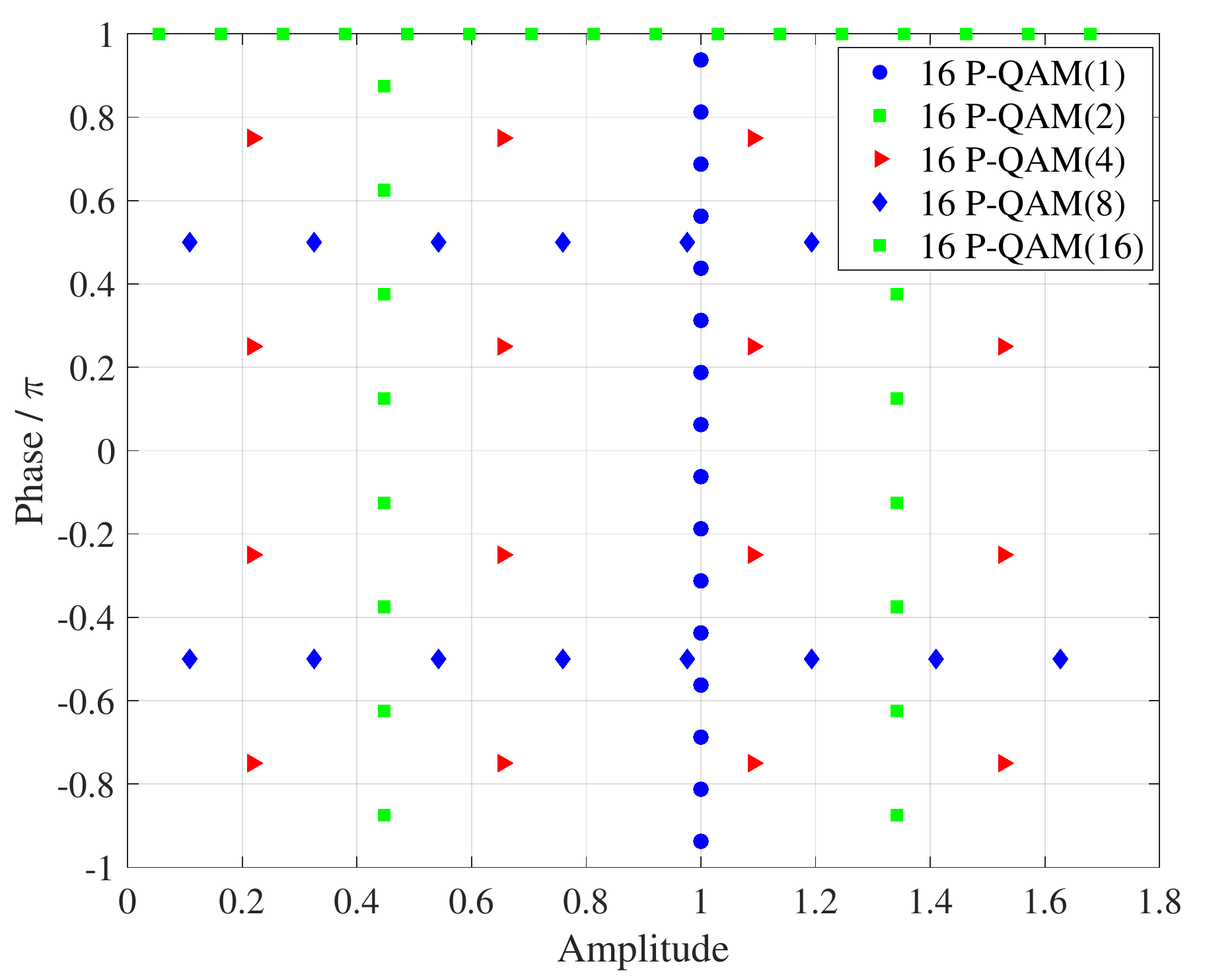} \\
(a) & (b)
\end{tabular}
\caption{Illustration of $16$-PQAM($\Gamma$) constellations in (a) the IQ plane and (b) the amplitude-phase domain.}
\label{fig:pqam_constellation}
\end{figure*}

We therefore use the notation $M$-PQAM($\Gamma$). Examples of different $16$-PQAM($\Gamma$) constellations are depicted in Fig.~\ref{fig:pqam_constellation}. Some particular cases of the PQAM fall into well known modulations: a $M$-PQAM($M/2$) describes an amplitude-shift keying while a $M$-PQAM($1$) is a phase-shift keying. 

\begin{figure}[ht]
\centering
\begin{tabular}{c}
\includegraphics[width=0.45\textwidth]{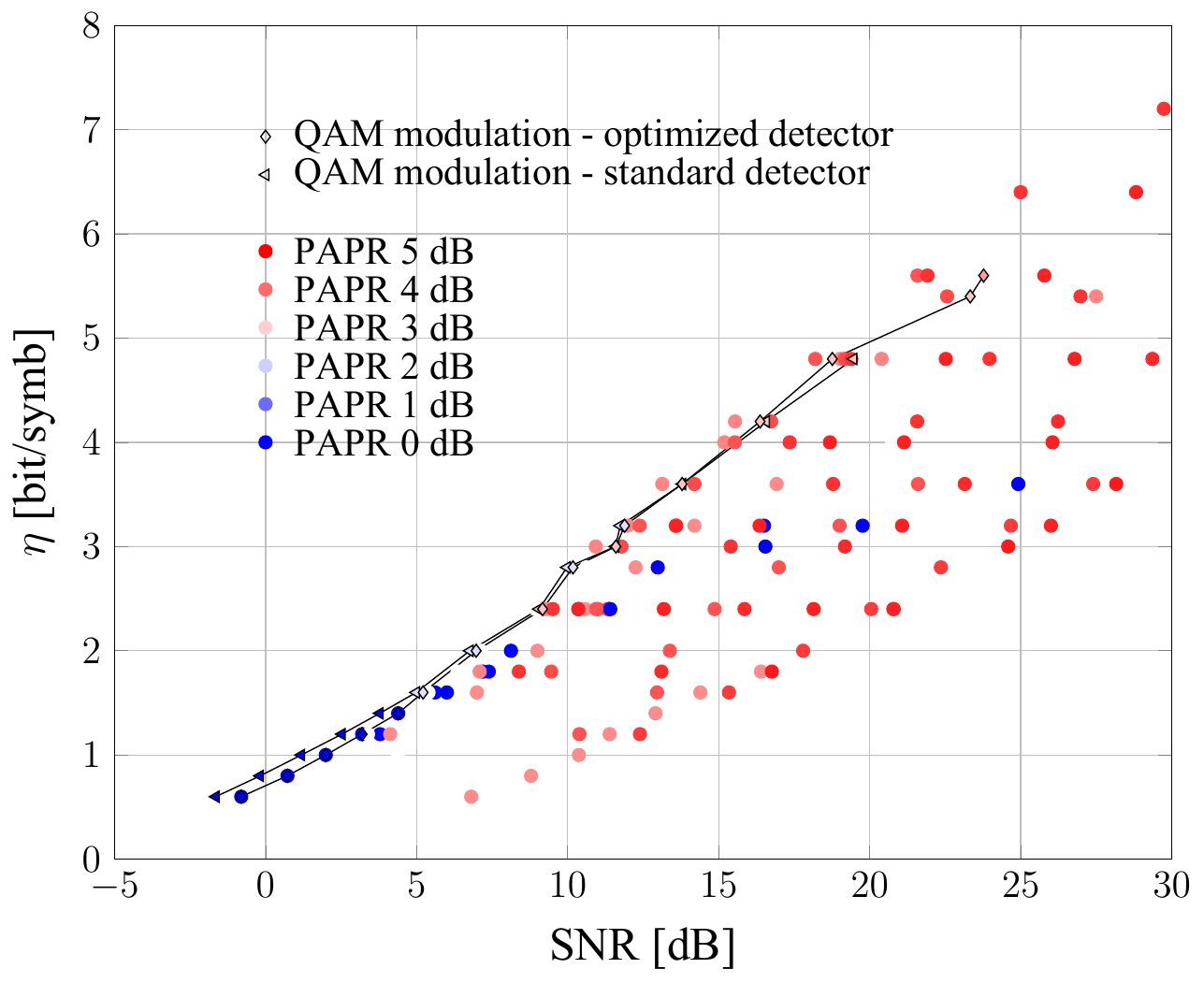} \\ \includegraphics[width=0.45\textwidth]{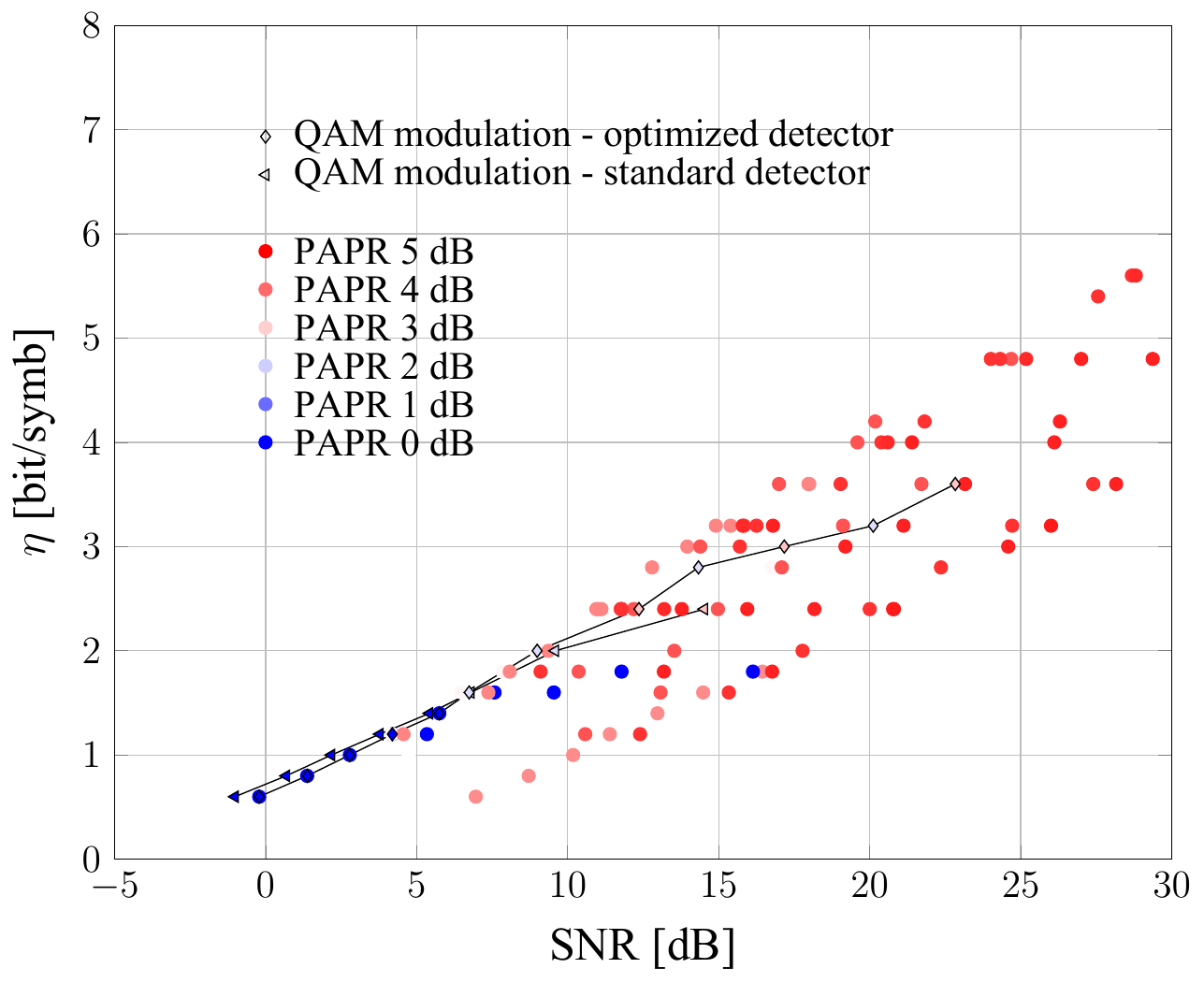}
\end{tabular}
\caption{Achievable rate in bit/symbol of the QAM and PQAM modulation schemes with a LDPC code for two phase noise configurations, medium and strong. PAPR of the constellation is also depicted through a color map.}
\label{fig:pqam}
\end{figure}

The PQAM is a structured definition of an APSK constellation. Combining this modulation scheme with a dedicated demodulation -- see for instance the polar metric detector proposed in \cite{9093997_bicais} -- leads to a simple and robust transmission technique for phase noise channels. With regard to RF power-amplifiers, directly related to the energy consumption of transmitters, the PAPR is a key performance indicator for any communication system. In the case of the PQAM, the PAPR is given by 
\begin{equation}
	3\cdot \frac{2\Gamma -1}{2\Gamma+1}. 
\end{equation}
It can be noticed that the PAPR is an increasing function of $\Gamma$ and does not depend on the modulation order $M$. We see here the trade-off to be found between the PAPR and the robustness to phase noise. Put it differently, increasing $\Gamma$ improves the robustness to phase noise, yet it increases the PAPR in the same time. To fully understand the trade-offs we are dealing with, let us present the results of numerical simulations for the Gaussian phase noise channel, with  two phase noise variances, medium and strong. We consider different settings of the PQAM combined with a state-of-the-art channel coding scheme. The coding scheme is based on the 5G-NR LDPC. It is implemented with an input packet size of $1500$ bytes and a coding rate ranging from $0.3$ to $0.9$. A Packet Error Rate (PER) of $10^{-2}$ is targeted for numerical evaluations. The phase noise variances are $\sigma_\phi^2=10^{-2}$ rad$^2$ for the medium phase noise case and ${\sigma_\phi^2=10^{-1}}$ rad$^2$ for the strong phase noise one. It should be emphasized that the ${\sigma_\phi^2=10^{-1}}$ rad$^2$ (resp. ${\sigma_\phi^2=10^{-2}}$ rad$^2$) corresponds to an oscillator spectral density floor of $-100$ dBc/Hz (resp. $-110$ dBc/Hz) for a bandwidth of $1$ GHz \cite{9013189_bicais}. Results are depicted in Fig.~\ref{fig:pqam}. We also put for comparison the performance of a QAM modulation with both the standard Euclidean detector and an optimized joint amplitude-phase detector, known to improve the demodulation performance on phase noise channels \cite{6510018_Krishnan}.

In the case of medium phase noise, the proposed Polar-QAM scheme demonstrates a minor performance loss with respect to the QAM. However, if the PAPR constraint becomes critical, the use of a Polar-QAM constellation with a low PAPR can be envisaged. The performance in the case of strong phase noise are particularly interesting to analyze. First of all, we clearly observe that the achievable information rate is severely limited when the QAM is used due to phase noise. Conversely, the PQAM allows to reach larger spectral-efficiencies. In the low SNR regime, it is possible to considered low PAPR modulation, i.e. most of the information bits are carried by the phase component. In this regime the degradation from thermal noise is dominant in comparison to phase noise. On the contrary in the high SNR regime, performance are limited by phase noise. It is clear from Fig.~\ref{fig:pqam} that it is not possible to reach high rate with low PAPR in this case. On the contrary to the QAM, the Polar-QAM provides an additional degree of freedom in the choice of the modulation scheme to find the best trade-off between the level of PAPR and the phase noise robustness. For this reason, the Polar-QAM offers an interesting modulation scheme for future coherent THz communication systems targeting high spectral-efficiency.

\subsubsection{Full-Duplex}

The current wireless systems (e.g., 4G, 5G, WiFi) use TDD or FDD methods, and the so-called half-duplex, where transmission and reception are not performed at the same time or frequency. In contrast, the full-duplex or the IBFD technology allows a device to transmit and receive simultaneously in the same frequency band. Full-duplex technology has the potential to double the spectral efficiency and significantly increase the throughput of wireless communication systems and networks. The biggest hurdle in the implementation of IBFD technology is self-interference, i.e., the interference generated by the transmit signal to the received signal, which can typically be over 100 dB higher than the receiver noise floor. Three classes of cancellation techniques are usually used to cope with self-interference: passive suppression, analog cancellation, and digital cancellation (see e.g. \cite{sabharwal2014in} and the references therein). Passive suppression involves achieving high isolation between the transmit and receive antennas before analog or digital cancellation. Analog cancellation in the RF domain is compulsory to avoid saturating the receiving chain, and nonlinear digital cancellation in the baseband is needed to further suppress self-interference, for example, down to the level of the noise floor. The analog RF cancellation can also be performed jointly with the digital cancellation using a common processor.

The full-duplex technique has a wide range of benefits, e.g., for relaying, bidirectional communication, cooperative transmission in heterogeneous networks, and cognitive radio applications. Other prospective use cases include, for instance, SWIPT, see e.g., \cite{FDSwipt1, FDSwipt2, RabieTGCNs, rabieWCLs, Galym18} and the references therein. The feasibility of full-duplex transmission has been experimentally demonstrated in small-scale wireless communications environments, and it was also considered as an enabling technique for 5G but not yet adopted by 3GPP \cite{kolodziej2019in}. However, for the full-duplex technique to be successfully employed in 6G wireless systems, there exist challenges on all layers, ranging from the antenna and circuit design (e.g., due to hardware imperfection and nonlinearity, the non-ideal frequency response of the circuits, phase noise, etc.), to the development of theoretical foundations for wireless networks with IBFD terminals. Note that IBFD becomes particularly challenging when MIMO is taken into account and is even more so with massive MIMO. Nevertheless, more antennas brings also larger degrees of freedom. Some of the antennas may be used for communication, while other may function as support antennas to cancel the self-interference, or all antennas used for both tasks jointly. The suitability of IBFD technology for 6G is an open research area, where an inter-disciplinary approach will be essential to meet the numerous challenges ahead \cite{sabharwal2014in}.


\subsubsection{Protocol-Level Interference Management}

Beamforming and sectorization have been effectively used in 4G to manage the co-user interference. In 5G, massive MIMO has been the main drive to control co-user interference together with highly adaptive beamforming, which helps to serve the spatially separated users non-orthogonally by spatial multiplexing. However, there may be situations in future communication systems when the interference management provided by massive MIMO is insufficient. For example, if the arrays are far from the users, then the spatial resolution of the antenna panels may be limited \cite{Senel2019}. In such cases, interference can be managed alternatively at the protocol level. NOMA \cite{Liu2017} and RS \cite{Clerckx2016a} schemes may be two appropriate candidates for this purpose.

The NOMA techniques are mainly based on two domains: power and code \cite{Dai2018}. Power-domain NOMA refers to the superposition of multiple messages using different transmit powers so that users with higher SNRs can decode interfering signals before decoding their own signal, while users with lower SNRs can treat interference as noise \cite{Islam2017}. Code-domain NOMA refers to the use of non-orthogonal spreading codes, which provide users with higher SNRs after despreading, at the expense of additional interference \cite{Le2018}. Both power-domain and code-domain NOMA have their advantages and disadvantages in terms of performance and implementation complexity when compared to each other. On the other hand, RS is based on dividing the users' messages into private and common parts, where each user decodes its private part and all users decode the common parts to extract the data. From the implementation point of view, RS can be viewed as a generalization of power-domain NOMA \cite{Clerckx2020}. 

As for the interference management, while massive MIMO treats any multi-user interference as noise, NOMA-based superposition coding with successive interference cancellation fully decodes and removes the multi-user interference. On the other hand, RS partially decodes the multi-user interference and partially treats it as noise providing a trade-off between fully decoding the interference and fully treating it as noise. With their modified transceiver structures, both NOMA and RS schemes may have increased complexity due to coding and receiver processing. However, there may be prospective 6G use cases where interference management at the protocol level could potentially provide sufficiently large gains to outweigh the increased implementation complexity.

Some considerations of NOMA and RS for interference management can be summarized as follows. In massive connectivity scenarios, where many devices transmit small packages intermittently, grant-free access using code-domain NOMA or similar protocols could be very competitive. In mmWave or THz communications, where the controllability of the beamforming is limited by hardware constraints (e.g., phased arrays), NOMA and RS could enable multiple users to share the same beam \cite{Zhu2019}. In VLC, where coherent adaptive beamforming may be practically impossible for multiple access, NOMA and RS could be an appropriate solution \cite{Marshoud2016a}. For massive connectivity and achieving high data rates in 6G, different implementations of NOMA techniques for cellular networks could also be considered to increase the system capacity \cite{Al-Eryani2019}. In addition to the implementation of NOMA and RS for interference management purposes, NOMA-based research that covers proposing simplified transceiver structures and hybrid multiple access schemes could help the improvement of interference management using NOMA and RS.

On the other hand, there are several practical challenges with implementing robust interference management at the protocol level. One involves the error propagation effects that appear when applying superposition coding and successive interference cancellation to finite-sized data blocks \cite{Chen2019}. Another issue is the high complexity in jointly selecting the coding rates of all messages, to enable successful decoding wherever needed, and conveying this information to the users. Implementing adaptive modulation and coding is nontrivial in fading environments with time-varying interference and beamforming, even if it is carried out on a per-user basis. With NOMA and RS, the selection of modulation/coding becomes coupled between the users, making it an extremely challenging resource allocation problem. Scheduling and ARQ retransmissions are other tasks that become more complex when the user data transmission is coupled. Hence, protocol-level interference management schemes need to be investigated in detail under practical conditions in order to determine suitable implementations for 6G use cases.

\subsubsection{THz-Band Spatial Tuning}

Besides combating the high propagation and absorption losses at high frequencies (as previously highlighted in Sec.~\ref{sec:infrastructure}), achieving any form of spatial multiplexing with ultra-massive antenna configurations is hindered by the high spatial channel correlation, caused by having a few angular propagation paths. Even though the research community is familiar with the impact of spatial correlation on MIMO links, the severity of this phenomenon at mmWave/THz frequencies and the peculiarities of THz devices result in more severe challenges and novel opportunities, respectively. This is particularly the case in doubly-massive MIMO configurations under LoS quasi-optical THz propagation, where channels tend to be of very low rank. In such scenarios, it is known that the physical separation between the antenna elements at both the transmitter and the receiver can be tuned to retain high channel ranks \cite{Torkildson6042312,do2020reconfigurable}. In particular, for a given communication range $D$, there exists an optimal separation $\Delta$ between the antennas that guarantees a sufficient number of eigenchannels to spatially multiplex multiple data streams. With highly configurable THz antenna architectures, especially plasmonic antennas, such optimal separation can be tuned in real-time \cite{hadi,hadi2}. However, if $D$ is larger than the achievable Rayleigh distance, which is a function of the physical array size, such tuning fails, and ill-conditioned channels become unavoidable. Hence, the Rayleigh distance serves as an important metric to capture THz system performance. For a number of transmit antennas $N_t$ and receive antennas $M_r$, this distance is expressed as \cite{6800118Wang}
\begin{equation}\label{eq:Rayleigh}
D_{\mathrm{Ray}} = \mathrm{max}\{M_r,N_t\} \Delta_r\Delta_t/\lambda,
\end{equation}
where $\lambda$ is the operating wavelength, and $\Delta_r$ and $\Delta_t$ are the uniform separation between antennas at the receiver and transmitter, respectively. Note that in a hybrid array-of-subarrays configuration where multiplexing is conducted at the level of subarrays, $\Delta_r$ and $\Delta_t$ can denote the corresponding separations between subarrays.

\begin{figure}[t]
\centering
\includegraphics[width=0.45\textwidth]{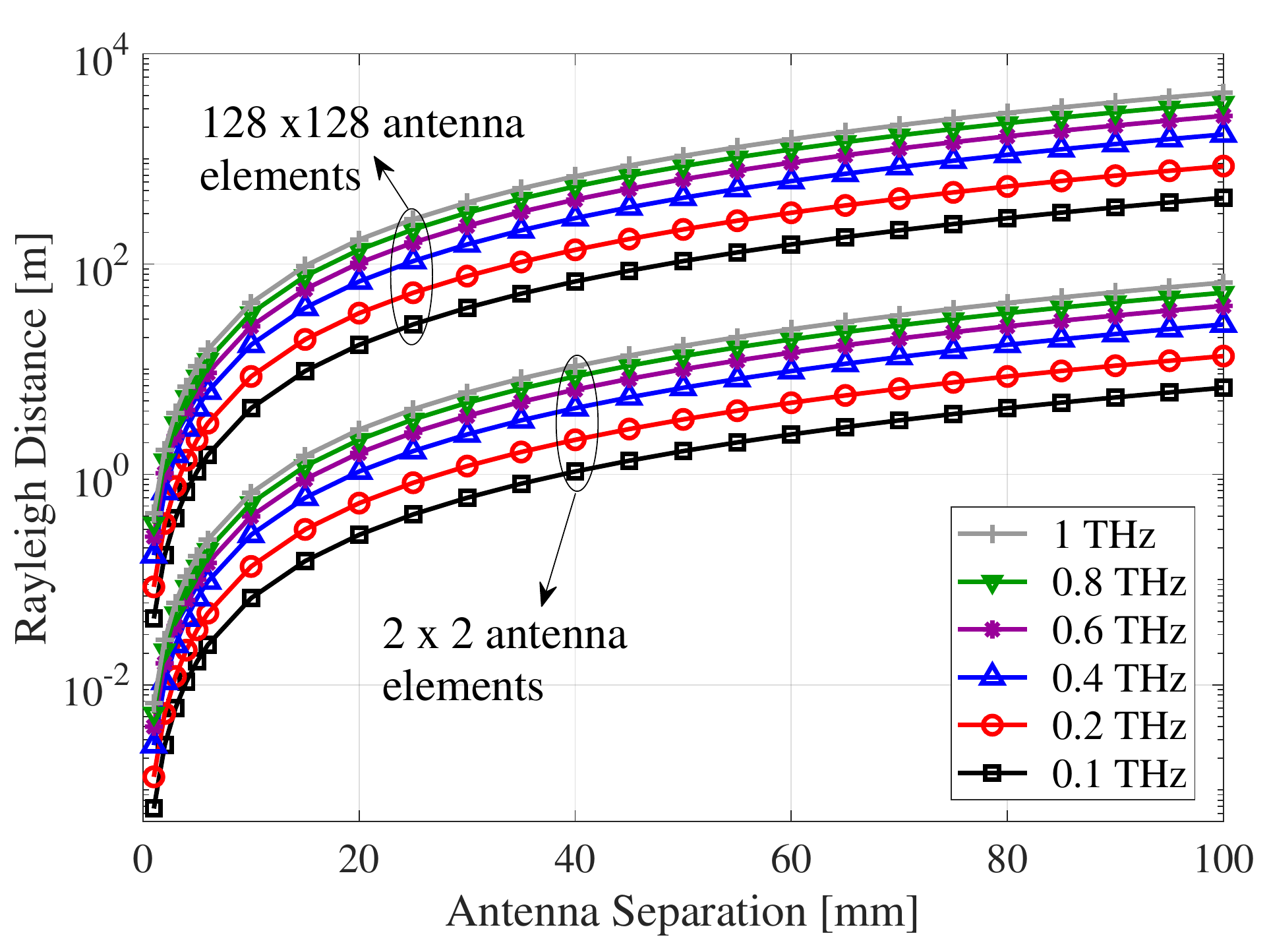}
\caption{Achievable Rayleigh distances under different THz LoS configurations.}
\label{f:Rayleigh}
\end{figure}

The achievable Rayleigh distances under different THz LoS configurations are illustrated in Fig.~\ref{f:Rayleigh}, as a function of antenna separations, array dimensions, and operating frequencies. A uniform planar array-of-subarrays configuration is assumed, and two special cases of $M_r=N_t=128\times128$ and $M_r=N_t=2\times2$ are simulated, assuming $\Delta_r=\Delta_t$. For antenna separations of a few millimeters, a large number of antennas is required to realize multiplexing-achieving distances beyond a few meters. Although for the same $\Delta$, larger arrays and higher frequencies result in larger Rayleigh distances, for the same footprint, larger antenna numbers incur a quadratic Rayleigh distance reduction in $\Delta$. While tens of mmWave-operating antennas typically occupy a few square centimeters, a huge number of THz-operating antenna elements can be embedded in few square millimeters. This adaptability in design, when combined with numerical optimization, is what we call \emph{spatial tuning}. In particular, antenna separations in plasmonic arrays can be reduced significantly, to values much below $\lambda/2$, while still evading the effects of mutual coupling. In such scenarios, to maintain an optimal $\Delta$ for multiplexing, a large number of antennas can be kept idle. Alternatively, spatial modulation \cite{hadi} or multicarrier \cite{faisal2019ultra} configurations can be applied. More importantly, antenna oversampling can be further utilized to lower the spatio-temporal frequency-domain region of support of plane waves \cite{Wang8470251}, and this can be exploited for noise shaping techniques \cite{Rap19}. While this discussion on channel characteristics is illustrative, several assumptions need to be relaxed for the exact characterization of THz channels, mainly by accounting for non-uniform array architectures \cite{7546944Wang}, spherical wave propagation \cite{Song8356240}, and wideband channels.

\begin{figure}[t]
\centering
\includegraphics[width=0.45\textwidth]{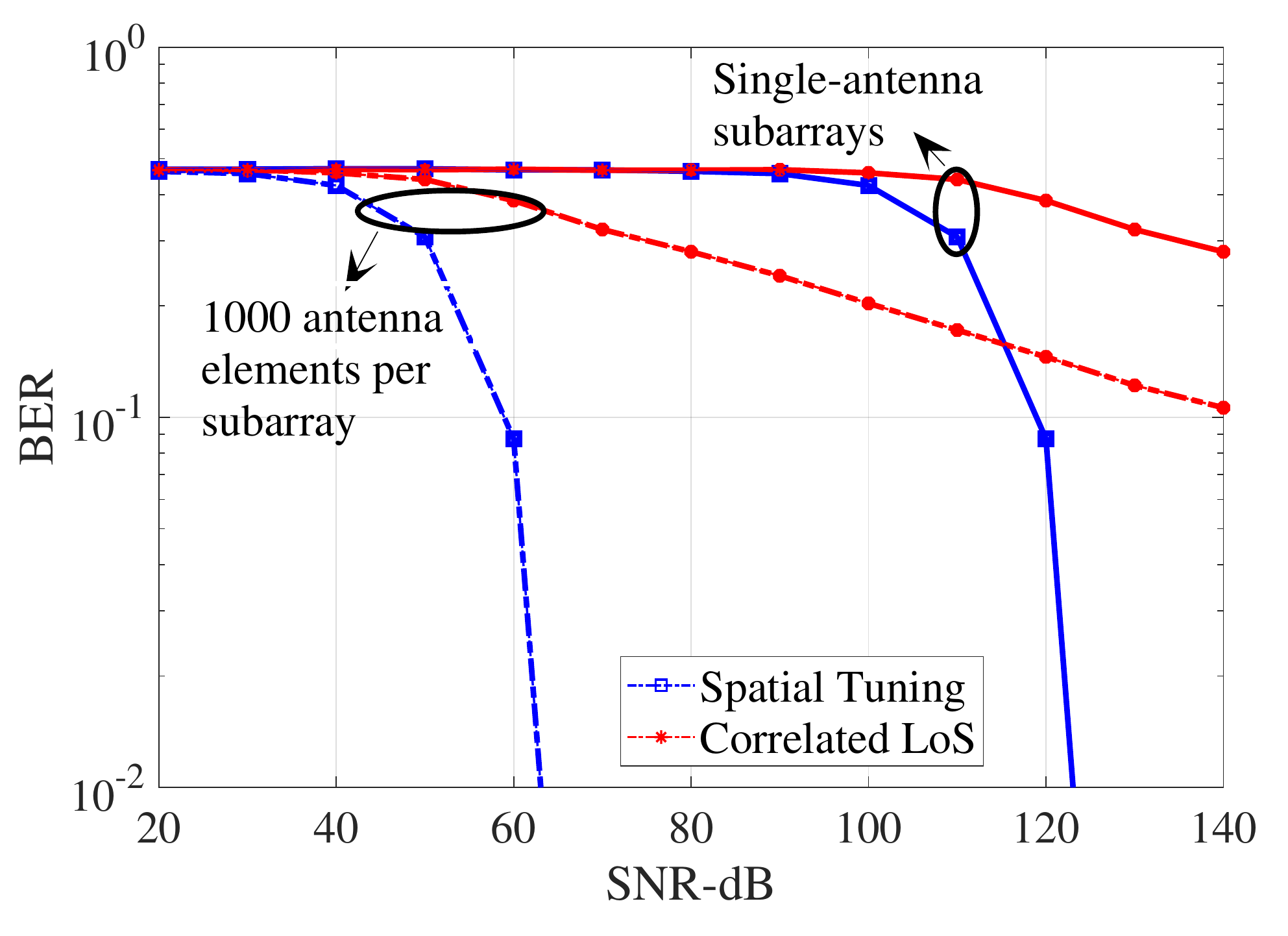}
\caption{BER performance of spatial multiplexing: $f$=1 THz, $D$=1 m,  $16\times 16$ subarrays, and 16-QAM.}
\label{f:Spatial_tuning}
\end{figure}

Efficient spatial tuning requires a sufficiently large array of antennas (a uniform graphene sheet, for example), in which antenna elements can be contiguously placed over a three-dimensinoal structure, and subarrays can be virtually formed and adapted. For a given target communication range, a specific number of antenna elements is allotted in the corresponding subarrays to achieve the required beamforming gain. Afterward, the diversity and multiplexing gains are dictated by the possible utilizations of the subarrays, limited by the number of available RF chains and the overall array dimension. Note that frequency-interleaved multicarrier utilization of antenna arrays is feasible because each antenna element can be tuned to target resonant frequency without modifying its physical dimensions. The latter can be attained via electrostatic bias or material doping at high frequencies, whereas software-defined plasmonic metamaterials can alternatively be used for sub-THz frequencies \cite{akyildiz2016realizing}.

Spatial tuning can further be extended to IRS-assisted THz NLoS environments. By controlling which element to reflect, a spatial degree of freedom is added at the IRS level, which could enhance the multiplexing gain of the NLoS system, but at the expense of a reduced total reflected power. Global solutions can be derived by jointly optimizing $\Delta_1$, $\Delta_2$, and $\Delta_3$, at the transmitting array, intermediate IRS, and receiving array, respectively. Nevertheless, true optimizations allow arbitrary precoding and combining configurations and arbitrary responses on all reflecting elements; the latter bounds the performance of lower-complexity spatial tuning techniques. Fig.~ \ref{f:Spatial_tuning} illustrates the importance of spatial tuning in allocating antenna elements for beamforming and maintaining channel orthogonality in a LoS scenario under perfect alignment; by allocating 1000 antenna elements per subarray, a 60 dB performance enhancement is noted and by maintaining channel orthogonality error floors are avoided.

%% file: 5_machine_learning.tex
Wireless technology becomes more complicated for every generation with so many sophisticated interconnected components to design and optimize. In recent research, there have been an increasing interest in utilizing ML techniques to complement the traditional model-driven algorithmic design with data-driven approaches. There are two main motivations for this \cite{Simeone2018a}: modeling or algorithmic deficiencies.

The traditional model-driven approach is optimal when the underlying models are accurate. This is often the case in the physical layer, but not always. When the system operates in complex propagation environments that have distinct but unknown channel properties, there is a modeling deficiency and, therefore, a potential benefit in tweaking the physical layer using ML.

There are also situations where the data-driven approach leads to highly complex algorithms, both when it comes to computations, run-time, and the acquisition of the necessary side information. This is called an algorithmic deficiency and it can potentially be overcome by ML techniques, which can learn how to take shortcuts in the algorithmic design. This is particularly important when we need to have effective signal processing for latency-critical applications and when jointly optimizing many blocks in a communication system.

This section provides some concrete examples and potential future research directions.

\subsubsection{End-to-End Learning}

End-to-end learning of communication systems using ML allows joint optimization of the transmitter and receiver components in a single process instead of having the artificial block structure as in conventional communication systems. In \cite{8054694_autoencoder}, autoencoder concept is used to learn the system minimizing the end-to-end message reconstruction error. Extensions of the autoencoder implementation for different system models and channel conditions can be found in \cite{8214233_hoydis, 8262721_osheaMIMO, 8422339_osheainterference, 8664650_autoencoder}. Furthermore, different approaches for end-to-end learning have been considered when the channel model is unknown or difficult to model analytically \cite{8644250_conditionalGAN, 8553233_oshea, 8792076_endtoend}. Specifically, \cite{8792076_endtoend} provides an iterative algorithm for the training of communication systems with an unknown channel model or with non-differentiable components. A unified multi-task deep neural network framework for NOMA is proposed in \cite{8952876_noma}, consisting of different components which utilize data-driven and communication-domain expertise approaches to achieve end-to-end optimization.

An example scenario of end-to-end learning is discussed below. Given the task of transmitting message $s$ out of $M$ possible messages from the transmitter to receiver over the AWGN channel using $n$ complex channel uses with a minimum error, the system is modelled as an autoencoder and implemented as a feedforward neural network. A model similar to the model proposed in \cite{8054694_autoencoder} is implemented with slight variations \cite{thesis_nuwanthika}. The transmitter consists of multiple dense layers followed by a normalization layer to guarantee the physical constraints of the transmit signal. During the model training, the AWGN channel is modelled as an additive noise layer with a fixed variance $\beta = (2RE_{b}/N_{0})^{-1}$, where $E_{b}/N_{0}$ denotes the energy per bit ($E_b$) to noise power spectral density ($N_0$) ratio. The receiver also consists of multiple dense layers and an output layer with softmax activation to output the highest probable estimated message.

\begin{figure}[ht]
\centerline{\includegraphics[width=0.5\textwidth]{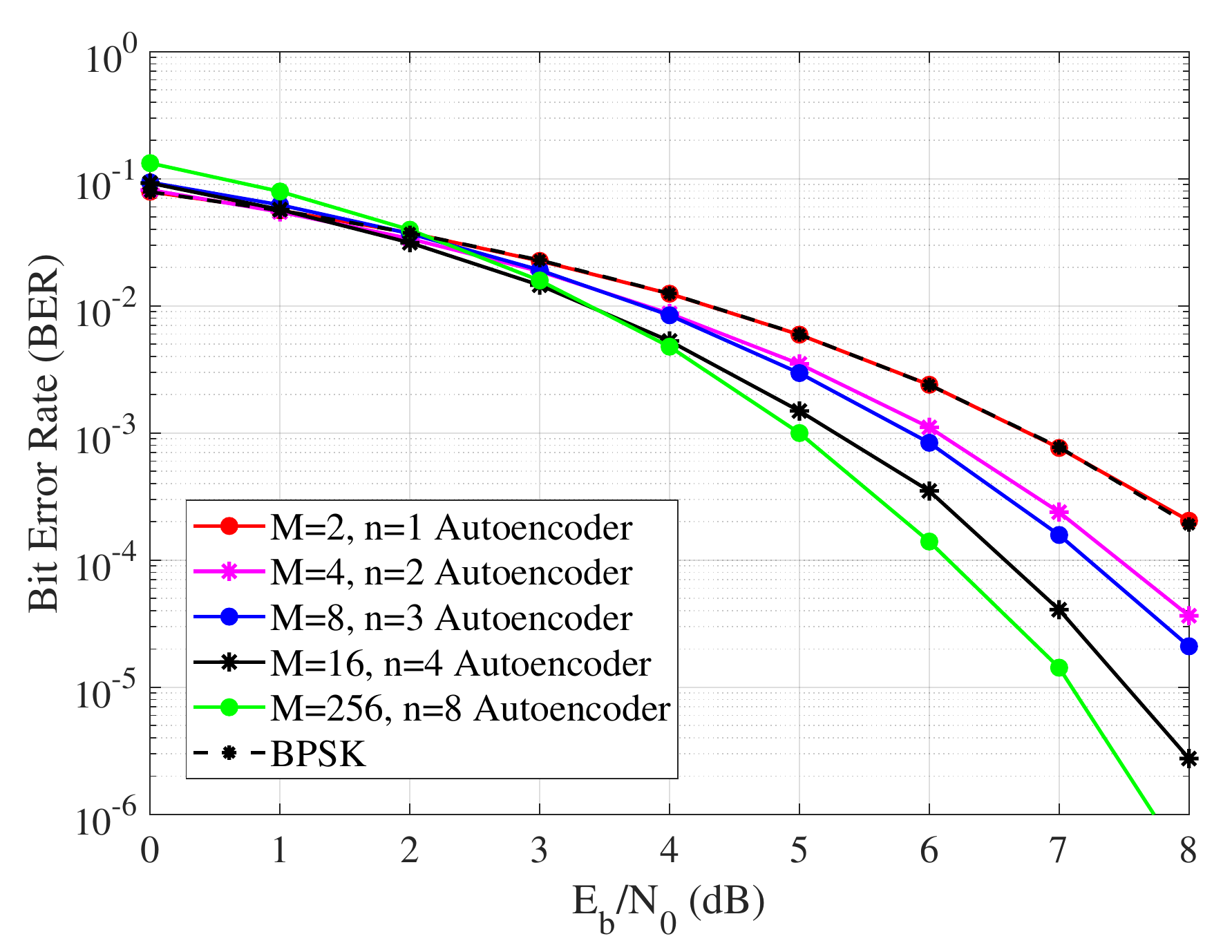}}
\caption{BER performance of the $R = 1$ bits/channel use systems compared with theoretical AWGN BPSK performance.}
\label{fig:autoencoder_bpsk}
\end{figure}

\begin{figure}[ht]
\centerline{\includegraphics[width=0.5\textwidth]{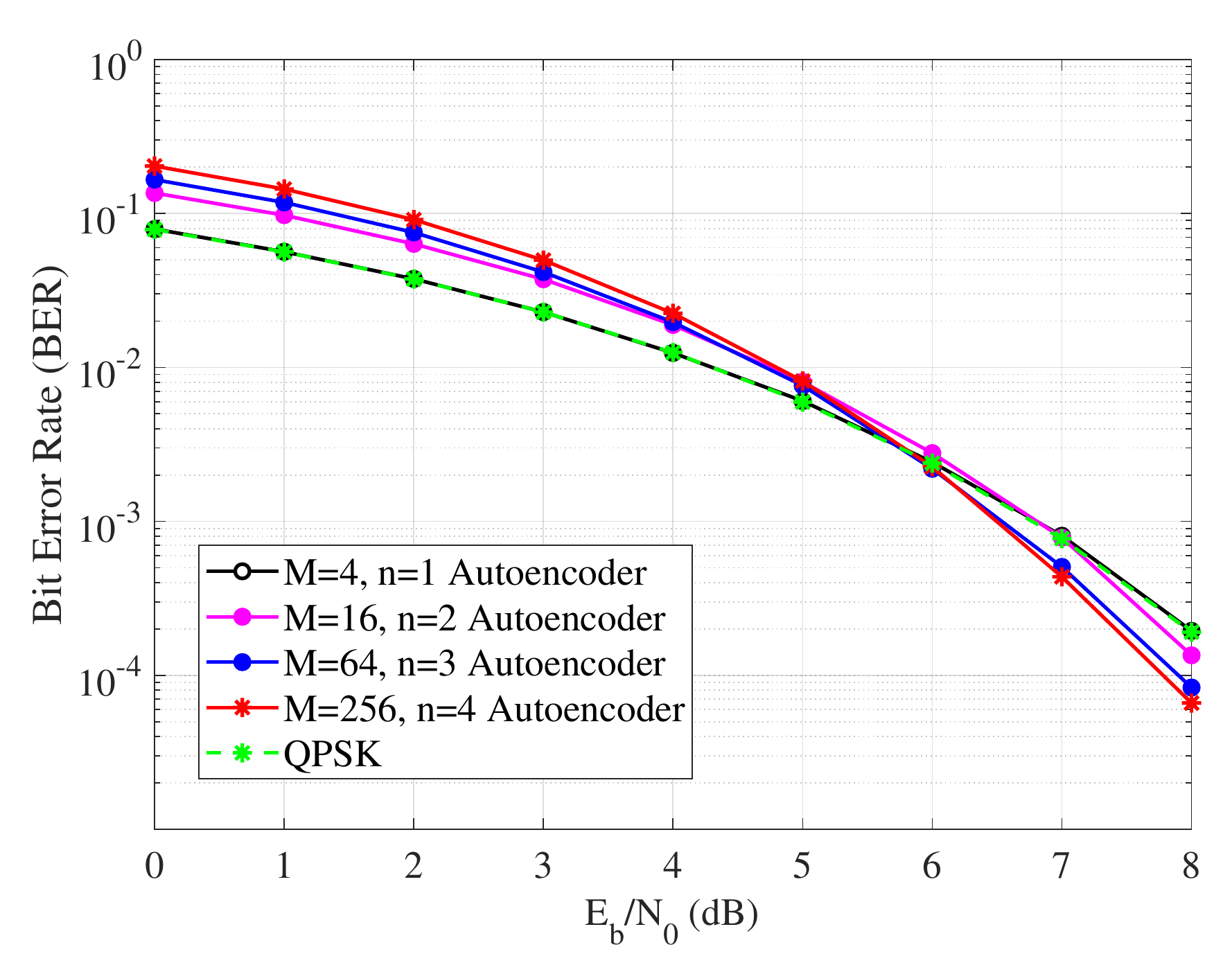}}
\caption{BER performance of the $R = 2$ bits/channel use systems compared with theoretical AWGN QPSK performance.}
\label{fig:autoencoder_qpsk}
\end{figure}

The model is trained in end-to-end manner on the set of all possible messages $s\in{\mathbb{M}}$ using the categorical cross-entropy loss function. A training set of 1,000,000 randomly generated messages with $E_{b}/N_{0} = 5$ dB is used for model training. The autoencoder learns optimum transmit symbols after the model training and the BER performance of the learnt transmit mechanism is evaluated with another set of 1,000,000 random messages across the 0 dB to 8 dB $E_{b}/N_{0}$ range. Fig. \ref{fig:autoencoder_bpsk} and Fig. \ref{fig:autoencoder_qpsk} show the BER performance of different autoencoder models in comparison with conventional BPSK and QPSK modulation schemes. It can be observed that increasing the message size $M$ improves the BER performance of the autoencoder resulting in a better BER than the conventional BPSK and QPSK schemes.

Practical aspects considering the model training implementations, methods to accompany unknown channel conditions and varying channel statistics in long-term, transmitter-receiver synchronization etc. are some of the potential future research directions.

\subsubsection{Joint Channel Estimation and Detection}


Channel estimation, equalization, and signal detection are three tasks in the physical layer that are relatively easy to carry out individually. However, their optimal joint design is substantially more complicated, making ML a potential shortcut.
In \cite{8052521_channel_est}, a learning-based joint channel estimation and signal detection algorithm is proposed for OFDM systems, which can reduce the signaling overhead and deal with nonlinear clipping. In \cite{8933050}, joint channel estimation and signal detection is carried out when both the transmitter and receiver are subject to hardware impairments with an unknown model.
Online learning-based channel estimation and equalization is proposed in \cite{8715649_channel_est}, which can jointly handle fading channels and nonlinear distortion.

Methods to overcome the challenge of offline model training, which causes performance degradation due to the discrepancies between the real channels and training channels, need to be considered. ML implementations with online training and constructing training data to match real-world channel conditions are potential future directions in this regard.

\subsubsection{Resource Management}

Resource management problems are often combinatorial in nature, which implies that optimal exhaustive-search based algorithms are impossible to utilize in practice. This provides an opportunity for ML-based algorithms to outperform existing suboptimal approaches. A generic resource allocation framework is provided in \cite{8680025}, whereas a joint beamforming, power control, and interference coordination approach is proposed in \cite{8938771_resourcemanagement}. Some examples of hybrid beamforming in mmWave MIMO systems \cite{8395149_hybrid_bf,8710287_bf,8890805_hybrid_bf, 8924932_jointantenna} utilize ML-based hybrid precoder-combiner designs, resulting in close performance and reduced complexity compared to the conventional exhaustive search-based optimization techniques. The work in \cite{8645343_powercontrol,chien2019power,8792117_powercontrol, 9022520_powercontrol} utilize different supervised learning and reinforcement learning techniques to learn and predict the optimum power allocation in the network dynamically.


Some potential research areas which are already being studied include power control, beamforming in massive MIMO and in cell-free environments, predictive scheduling and resource allocation, etc. Reinforcement learning frameworks and transfer learning techniques to learn, adapt, and optimize for varying conditions over time are expected to be useful for performing resource management tasks with minimal supervision \cite{6GFlagship_ML_WP}.

%% file: 5_caching.tex
Coded caching is originally proposed by Maddah-Ali and Niesen in~\cite{maddah2014fundamental}, as a way to increase the data rate with the help of cache memories available throughout the network. It enables a global caching gain, proportional to the total cache size of all users in the network, to be achieved in addition to the local caching gain at each user. This additional gain is achieved by multicasting carefully created codewords to various user groups, so that each codeword contains useful data for every user in the target group. Technically, if $K$ is the user count, $M$ denotes the cache size at each user and $N$ represents the number of files in the library, using CC the required data size to be transmitted over the broadcast link can be reduced by a factor of $1+t$ (equivalently, the data rate can be increased by a factor of $1+t$), where $t=\frac{KM}{N}$ is called the CC gain. 

Interestingly, the CC gain is not only achievable in multi-antenna communications, but is also additive with the spatial multiplexing gain of using multiple antennas~\cite{shariatpanahi2016multi,shariatpanahi2018physical,tolli2018multicast}.
In fact, caching non-overlapping file fragments at user terminals provides multicasting opportunities, as multiple users can be potentially served with a single multicast message. Therefore, the additive gain of CC in multi-antenna communications can be achieved through multicast beamforming of multiple parallel (partially overlapping) CC codewords to larger sets of users, while removing (or suppressing) inter-codeword interference with the help of carefully designed beamforming vectors.
The generalized multicast beamforming design enables innovative, flexible resource allocation schemes for CC. Depending on the spatial degrees of freedom and the available power resources, a varying number of multicast messages can be transmitted in parallel to distinct subsets of users~\cite{tolli2017multi}.

\begin{figure*}[t]
\begin{center}
    \includegraphics[scale=0.9]{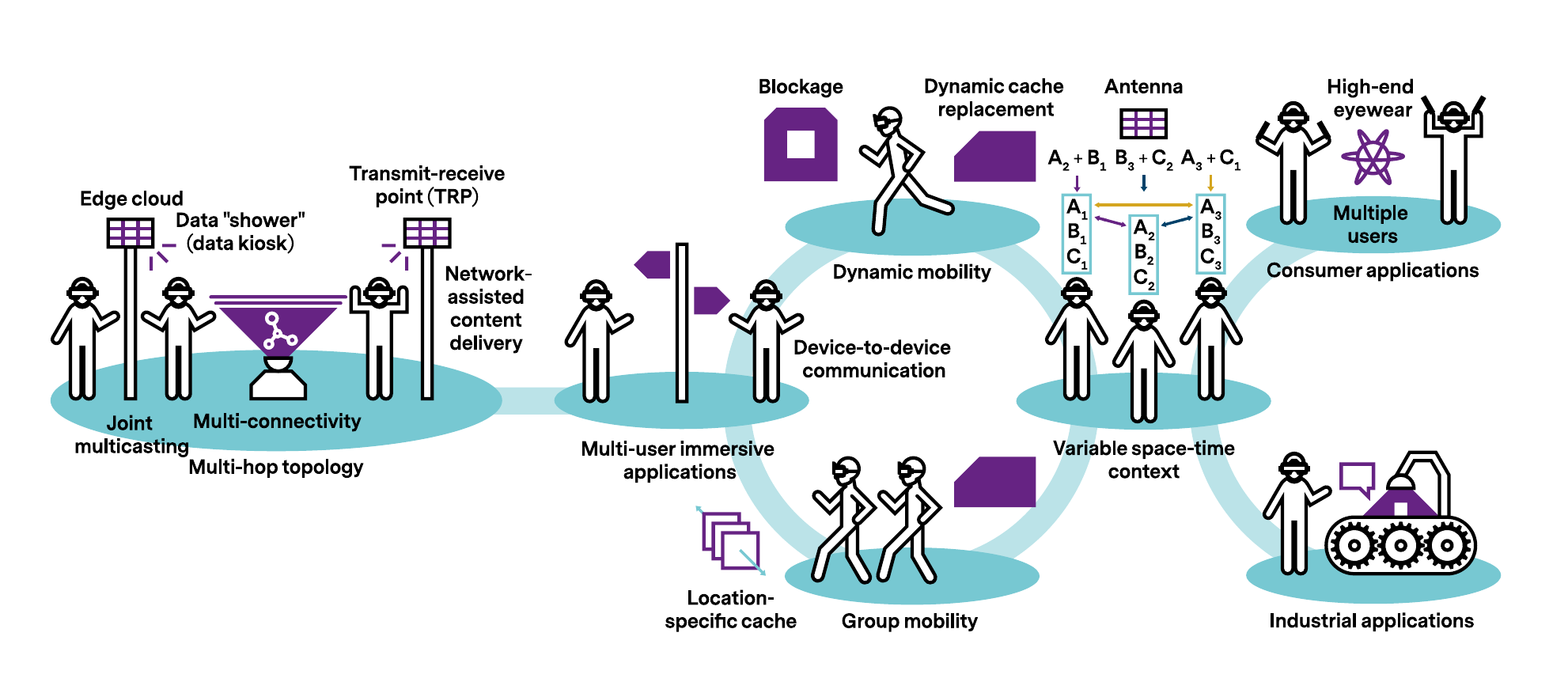}
    \caption{Immersive viewing scenario with coded caching.}
    \label{fig:immersive_vieweing}
\end{center}
\end{figure*}

The multi-antenna CC structure also provides more flexibility in reducing subpacketization, defined as the number of smaller files each file should be split into, for the CC structure to work properly. An exponentially growing subpacketization requirement is known to be a major problem in implementing original single- and multi-antenna CC schemes \cite{lampiris2018adding}. However, recently it is shown that in multi-antenna CC, especially when the spatial multiplexing gain is larger than the CC gain, linear or near-linear subpacketization growth is possible through well-defined algorithms \cite{lampiris2018adding,salehi2019codedlinear,salehi2019subpacketization}. Overall, the nice implementation possibility of CC in multi-antenna setups makes it a desirable option to be implemented in future wireless networks, where MIMO techniques are considered to be a core part.

The number of multimedia applications benefiting from CC is expected to grow in the future. One envisioned scenario assumes an extended reality or hyper-reality environment (e.g., educational, industrial, gaming, defense, social networking), as depicted in Fig.~\ref{fig:immersive_vieweing}. A large group of users is submerged in a network-based immersive application, which runs on high-end eye-wear that requires heavy multimedia traffic and is bound to guarantee a well-defined quality-of-experience level for every user in the operating theatre. 
The users are scattered across the area covered by the application and can move freely, and their streamed data is unique and highly location- and time-dependent. Notably, a large part of the rich multimedia content for rendering a certain viewpoint is common among the users. This offers the opportunity for efficient use of pooled memory resources through intelligent cache placement and multicast content delivery mechanisms. 
In such a scenario, the possibility of caching on the user devices and computation offloading onto the network edge could potentially deliver high-throughput, low-latency traffic, while ensuring its stability and reliability for a truly immersive experience. The fact that modern mobile devices are continuing to increase their storage capacity (which is one of the cheapest network resources) makes CC especially beneficial given the uniqueness of this use case, where the popularity of limited and location-dependent content becomes much higher than in any traditional network. 

\begin{figure}[ht]
\centerline{\includegraphics[width=0.5\textwidth]{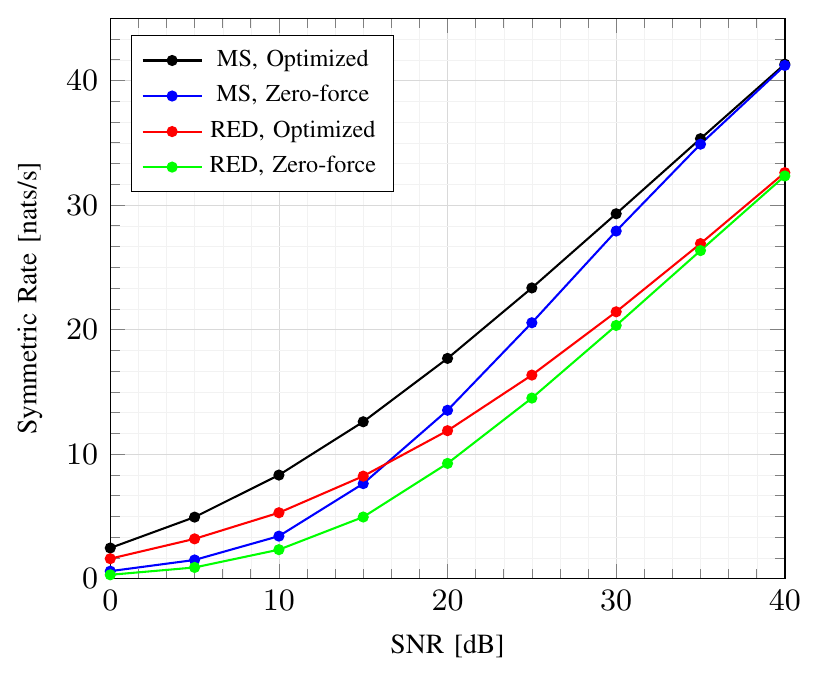}}
\caption{Rate vs SNR, $K=6$, $t=2$, $L=3$.}
\label{fig:cc_chart}
\end{figure}

Alongside its strong features, there are many practical issues with CC that must be addressed to enable its implementation in 6G. Most importantly, various parameters affecting the performance and complexity of CC schemes need to be identified, and possible trade-offs among them has to be clarified. For example, one such trade-off is depicted in Fig.~\ref{fig:cc_chart} ($L$ is the number of antennas at the transmitter side), in which the performance of the multi-antenna CC scheme in~\cite{shariatpanahi2018physical,tolli2017multi} is compared with the reduced-subpacketization scheme of~\cite{salehi2019codedlinear}, at various SNR levels and for fully-optimized (MMSE-type) and zero-force beamforming strategies. In general, the MS performs better than the RED; and optimized beamforming provides better results than zero-forcing. However, the MS scheme requires exponentially growing subpacketization, making it impractical for networks with even moderate number of users; and optimized beamforming requires non-convex optimization problems to be solved through computationally intensive methods such as successive convex approximation~\cite{tolli2017multi}. On the other hand, the difference between various strategies is dependent on the SNR regime~\cite{salehi2019beamformer}. For example, in the figure it is clear that when the SNR is below 15~dB, the RED scheme with optimized beamforming outperforms MS strategy with zero-forcing; while for SNR above 15~dB this comparison is no longer valid. Moreover, the gap between optimized and zero-force beamformers vanishes at the high-SNR regime.

The provided example only clarifies the interaction between the subpacketization, beamforming strategy and operating SNR regime. It is shown that as the CC operation is heavily dependent on the underlying multicasting implementation, its performance is degraded in case some users in the network are suffering from poor channel conditions~\cite{destounis2017alpha}. In addition, the energy efficiency of CC schemes is largely unaddressed in the literature. In fact, CC schemes require a proactive cache placement phase during which the cache memories in the network are filled with data chunks from all files in the library. This imposes a communication overhead and necessitates excess energy consumption for the required data transfer, which can considerably degrade the energy efficiency of the underlying CC scheme.


%% file: 6_full_coverage.tex
Around half of the world population, or almost 4 billion people, still lack broadband connectivity. Wireless connectivity of some form is the only realistic alternative for most people in large rural areas since building a new wired infrastructure is typically cost prohibitive.  While each new wireless generation so far has focused on giving higher data rates where the population is most concentrated, they have offered much less benefit in rural regions and sometimes resulted in even worse coverage than the previous generations, which has resulted in an increased digital divide.  This unbalanced trend still holds true today. Many emerging wireless technologies are primarily targeted at improving the wireless connectivity where it is already relatively good; at short distances to the access network and to serve densely packed users and devices with even higher capacity. While there may be some improvement also in rural areas and to remote locations, the gain is usually much lower than in urban areas. 

Any technology that aims to reduce the digital divide needs to offer as uniform performance as possible over as large area as possible.  One such technology is massive MIMO, which can improve the performance at the cell edge considerably given an altruistic max-min power allocation. While conventional high-gain sector antennas provide array gains in the cell center, the highly adaptive beamforming of massive MIMO can provide array gains wherever the user is located in the coverage area \cite[Sec.~6.1]{Marzetta2016a}. However, this comes at the price of a reduced sum-rate which may be economically unfeasible for an operator, since the long-distance users will consume a large portion of the available power for an AP in a cellular system, resulting in comparable much lower rates for nearby users compared to uniform power allocation. Long-range users may also have insufficient transmit power in the uplink.  Cell-free massive MIMO, however, with max-min power control provide a much higher energy efficiency and throughput per user in rural environments compared to a single cell cellular system. This is achieved due to the lower path-loss  and that only a few antennas need to transmit on full power in cell-free systems with distributed antennas \cite{YangMarzettaVTC18}. 

Long-range massive MIMO  can greatly increase the range and coverage of an AP and increase the capacity also at long distances, e.g., using very tall towers \cite{Taheri_Nilsson_Beek_2021_Massive-MIMO-TV}. Coherent joint transmission between adjacent APs can be utilized to enhance the performance and diversity at cell edges, effectively creating a long-range cell-free massive MIMO system. However, challenges exist in that long-distance LOS dominated  channels are more sensitive to shadowing and blocking, and frequent slow fading in general, which impose practical challenges for massive MIMO in rural areas. Unfortunate homes and places may be in a permanent deep shadow and thus always out of coverage. The near-far problem is often enhanced due to the nature of the wireless channels in rural areas, with typically decreasing spatial diversity at longer distances. Coherent joint transmissions from multiple distributed APs in a cell-free arrangement require more backhaul/fronthaul capacity and are subject to large delays. The synchronization of widely distributed transmitters is also highly challenging due to large variations in propagation delays.

LEO satellite constellations can offer almost uniform coverage and high capacity over huge rural areas, including seas and oceans, and to other remote locations using mmWaves and THz frequencies. They can also be used in conjunction with HAPS. Since these solutions view the coverage area high from above, they can fill in the coverage gaps of terrestrial long-range massive MIMO, as described above, and also offer large-capacity backhaul links for remotely located ground-based APs. High array gains are possible by using many antenna elements on the satellites and HAPS. Each user can either be served by the best available satellite/HAPS, or by coherent transmission from multiple ones. The coordination can be enabled via optical inter-satellite connections. Coherent transmission from satellites and ground-based APs is also possible, to effectively create a space-based cell-free network. Many challenges exist, however, in that a large number of LEO satellites are required to make them constantly available in rural regions, which is associated with a high deployment cost. Interference may arise between uncoordinated satellites and terrestrial systems using the same radio spectrum. The large pathloss per antenna element requires large antenna arrays and adaptive beamforming, since LEO satellites are constantly in motion. The transmit power is limited, particularly in the uplink. Coherent transmission from multiple satellites, or between space and the ground, requires higher fronthaul capacity and varying delays cause synchronization issues.

IRSs can be deployed on hills and mountains to remove coverage holes in existing networks by reflecting signals from APs or satellites towards places where the LoS path is blocked and the natural multipath propagation is insufficient. Fig.~\ref{figure:simulationIRS} illustrates such a case when the  direct path is heavily attenuated (Case 1) and where an IRS improves the spectral efficiency considerably by reflecting many coherent signal paths to the user. 
An IRS can be powered by solar panels or other renewable sources. The cost is low compared to deploying and operating additional APs since no backhaul infrastructure or connection to the power grid is needed. However, the larger the propagation distance becomes, the larger the reflecting surface area needs to be in order to counteract the large path-losses. Remote control of the IRS is challenging and might only support fixed access or low mobility. An IRS deployed in nature can be exposed to harsh climate and weather conditions and a subject of sabotage.

%% file: 7_summary.tex

\begin{table*}[t!]
\centering
\caption{Summary of Challenges, Potential 6G Solutions, and Open Research Questions}
\begin{tabular}{|p{5.5cm}|p{5.5cm}|p{5.5cm}|} \hline
\multicolumn{1}{|c|}{\textbf{Challenges}}                               & \multicolumn{1}{|c|}{\textbf{Potential 6G solutions}}                                                       & \multicolumn{1}{|c|}{\textbf{Open research questions}} \\ \hline \hline
Stable service quality in coverage area                                 & User-centric cell-free massive MIMO                                                                         & Scalable synchronization, control, and resource allocation \\ \hline
Coverage improvements                                                   & Integration of a spaceborne layer, ultra-massive MIMO from tall towers, intelligent reflecting surfaces     & Joint control of space and ground based APs, real-time control of IRS \\ \hline
Extremely wide bandwidths                                               & Sub-THz, VLC                                                                                                & Hardware development and mitigation of impairments \\ \hline
Reduced latency                                                         & Faster forward error correcting schemes, wider bandwidths                                                   & Efficient encoding and decoding algorithms \\ \hline
Efficient spectrum utilization                                          & Ultra-massive MIMO, waveform adaptation, interference cancellation                                          & Holographic radio, use-case based waveforms, full-duplex, rate-splitting \\ \hline
Efficient backhaul infrastructure                                       & Integrated access and backhauling                                                                           & Dynamic resource allocation framework using space and frequency domains \\ \hline
Smart radio environment                                                 & Intelligent reflecting surfaces                                                                             & Channel estimation, hardware development, remote control \\ \hline
Energy efficiency                                                       & Cell-free massive MIMO, suitable modulation techniques                                                      & Novel modulation methods with limited hardware complexity \\ \hline
Modeling or algorithmic deficiencies in complex and dynamic scenarios   & ML-/AI-based, model-free, data-driven learning and optimization techniques                          & End-to-end learning/joint optimization, unsupervised learning for radio resource management \\ \hline
\end{tabular} \vspace{2mm}
\label{table:summary}
\end{table*}

This paper has provided a survey of the most promising candidate technologies at the PHY and MAC layer for the realization of Tbps wireless connectivity in future 6G wireless networks. These technologies, named enablers, have been categorized into three separate classes, i.e. enablers at the spectrum level, at the infrastructure level, and at the protocol/algorithmic level. 

Our vision is that with 6G wireless networks there will be a paradigm shift in the way users are supported, moving from the network-centric view where networks are deployed to deliver extreme peak rates in special cases to a user-centric view where consistently high rates are prioritized. Such ubiquitous connectivity can be delivered by  cell-free massive MIMO and IAB, and complemented by IRSs. The sub-6 GHz spectrum will continue defining the wide-area coverage and 1 Tbps can be reached in this spectrum range by making use of infrastructure enablers for extremely high spatial multiplexing. A significant effort is needed in sub-THz and THz bands to achieve short-range connectivity with data rates in the vicinity of 1~Tbps. The extremely wide bandwidths can take us far towards the goal, but it challenging to maintain a decent spectral efficiency. Novel coding, modulation and waveforms will be needed to support extreme data rates with manageable complexity and robustness to the hardware impairments that will increase with the carrier frequency. In situations where the beamforming capabilities offered by the ultra-massive MIMO technology are insufficient to manage interference in the spatial domain, coding methods based on rate-splitting or broadcasting can be also utilized. The efficiency can also be improved by making use of caching techniques. Seamless integration between satellites and terrestrial networks will ensure that large populations outside the urban centers will be reached by high-quality broadband connectivity.

Throughout the paper the main challenges related to the use of the described technologies have been highlighted and discussed. For reader's convenience, these are briefly summarized  in Table~\ref{table:summary}. 

Our hope is that this paper will help accelerating the interest of the scientific community towards these problems, so that one day not so far into the future we will be able to score the Terabit/s goal for broadband connectivity for everyone. It will be essential both for short-range communication links and to handle the massive traffic from a large number devices.

%% file: author_bio.tex

\begin{IEEEbiography}[{\includegraphics[width=1in,height=1.25in,keepaspectratio]{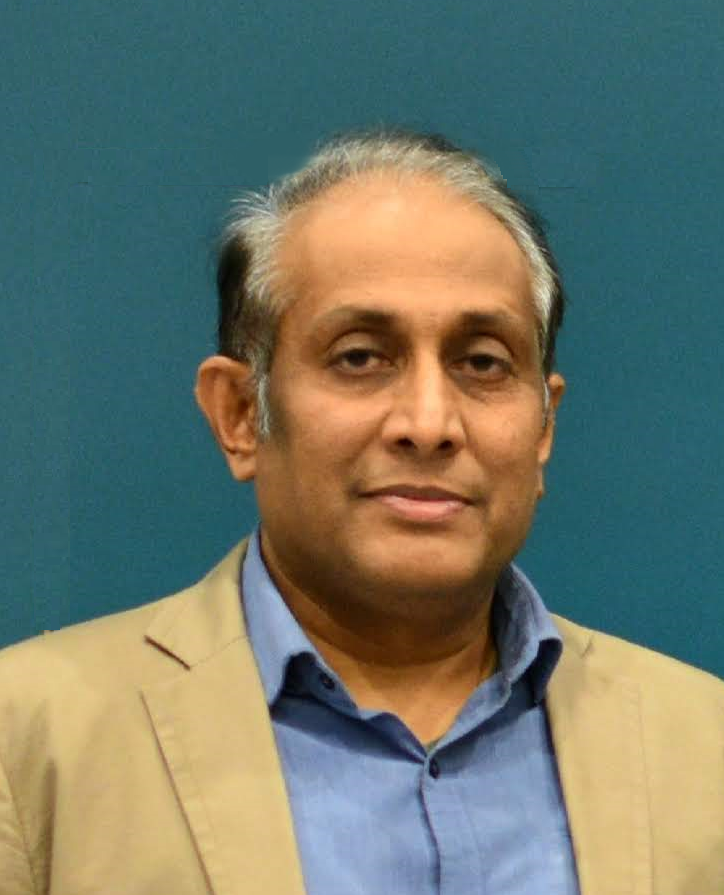}}] {Nandana Rajatheva} is currently a Professor with the Centre for Wireless Communications (CWC), University of Oulu, Finland. He is a Senior Member, IEEE and received the B.Sc. (Hons.) degree in electronics and telecommunication engineering from the University
of Moratuwa, Sri Lanka, in 1987, ranking first in the graduating class, and the M.Sc. and Ph.D. degrees from the University of Manitoba, Winnipeg, MB, Canada, in 1991 and 1995, respectively. He was a Canadian Commonwealth Scholar during the graduate studies in Manitoba. He held Professor/Associate Professor positions at the University of Moratuwa and the Asian Institute of Technology (AIT), Thailand, from 1995 to 2010. He has co-authored more than 200 refereed papers published in journals and in conference proceedings. His research interests include physical layer in beyond 5G, machine learning for PHY \& MAC, sensing for factory automation and channel coding. He is currently leading the AI-driven Air Interface design task in Hexa-X EU Project. 
\end{IEEEbiography}

\vspace{-10 mm}

\begin{IEEEbiography} [{\includegraphics[width=1in,height=1.25in,clip,keepaspectratio]{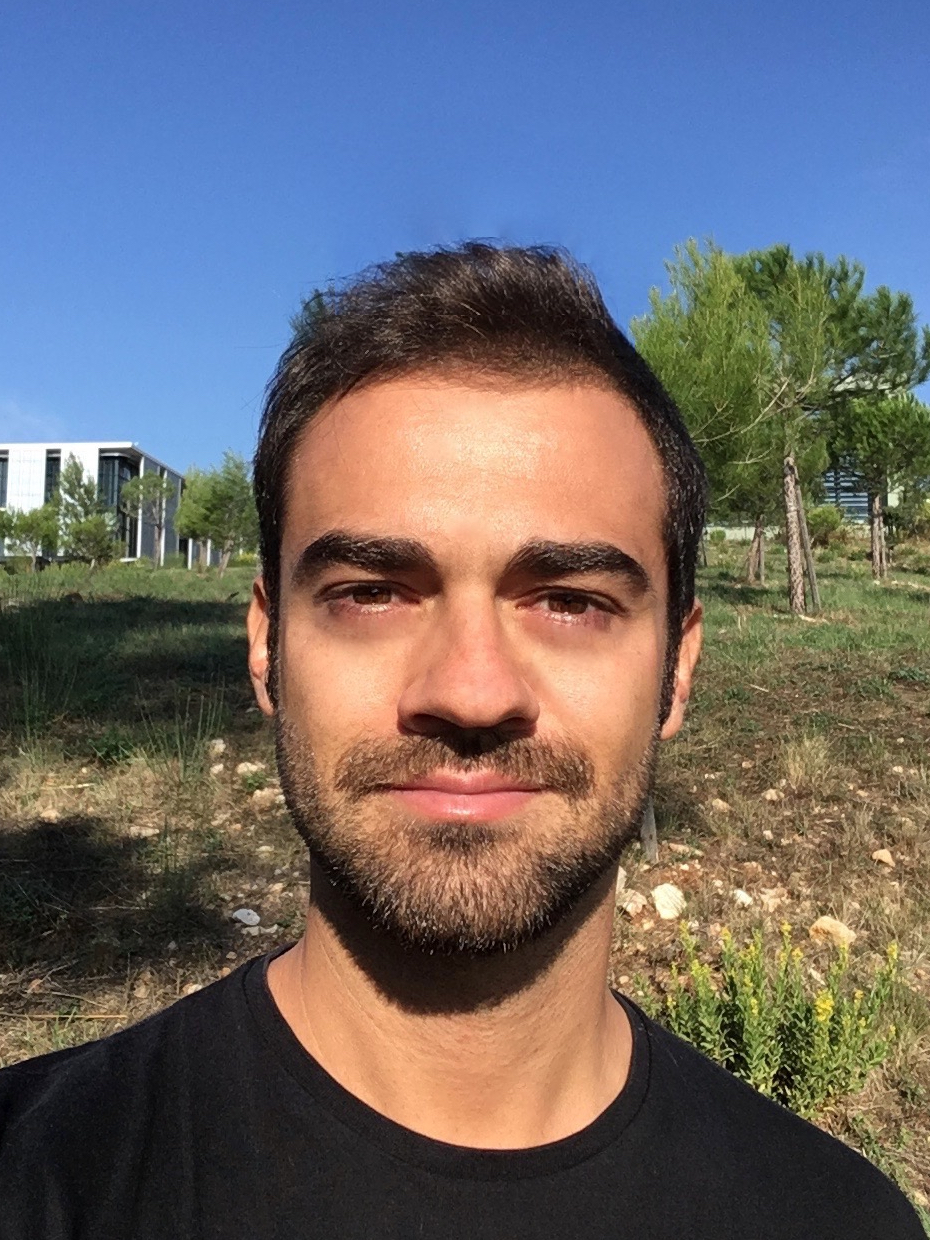}}] {Italo Atzeni} received the PhD degree (Hons.) in signal theory and communications from the Polytechnic University of Catalonia–BarcelonaTech in 2014. He is Senior Research Fellow and Adjunct Professor at the Centre for Wireless Communications, University of Oulu. He was with the Mathematical and Algorithmic Sciences Laboratory, Paris Research Center, Huawei Technologies from 2014 to 2017 and with the Communication Systems Department, EURECOM from 2017 to 2018. His primary research interests are in communication and information theory, statistical signal processing, and convex and distributed optimization theory. He received the Best Paper Award in the Wireless Communications Symposium at IEEE ICC 2019. He was recently granted the MSCA-IF for the project DELIGHT.
\end{IEEEbiography}

\vspace{-10 mm}

\begin{IEEEbiography}[{\includegraphics[width=1in,height=1.25in,clip,keepaspectratio]{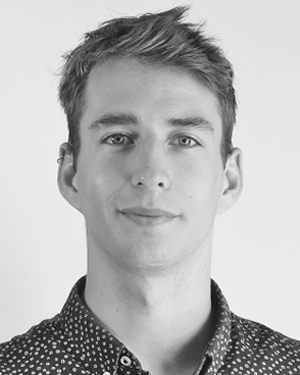}}]{Simon~Bica\"is} received the M.Sc. in telecommunications (2017), from the National Institute of Applied Sciences of Lyon (INSA Lyon), France and a PhD in 2020 from Grenoble University, France. Signal processing, wireless communications and machine learning are his current research interests. He was involved in the BRAVE  project about Beyond 5G wireless communications in the sub-TeraHertz bands. He is the main inventor of $4$ patents.
\end{IEEEbiography}

\vspace{-10 mm}

\begin{IEEEbiography}[{\includegraphics[width=1in,height=1.25in,clip,keepaspectratio]{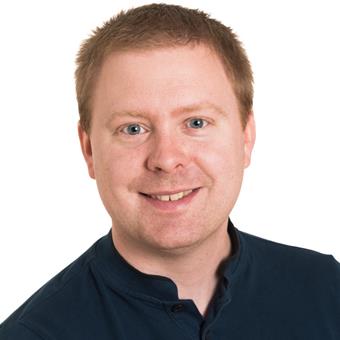}}] {Emil Bj\"ornson} is a Professor at the KTH Royal Institute of Technology, Sweden, and an Associate Professor at Link\"oping University, Sweden. He has authored the textbooks \emph{Optimal Resource Allocation in Coordinated Multi-Cell Systems} (2013), \emph{Massive MIMO Networks: Spectral, Energy, and Hardware Efficiency} (2017), and \emph{Foundations of User-Centric Cell-Free Massive MIMO} (2021). He has received the 2014 Outstanding Young Researcher Award from IEEE ComSoc EMEA, the 2016 Best Ph.D. Award from EURASIP, the 2018 IEEE Marconi Prize Paper Award in Wireless Communications, the 2019 EURASIP Early Career Award, the 2019 IEEE Communications Society Fred W. Ellersick Prize, the 2019 IEEE Signal Processing Magazine Best Column Award, the 2020 Pierre-Simon Laplace Early Career Technical Achievement Award, and the 2020 CTTC Early Achievement Award. He also co-authored papers that received Best Paper Awards at WCSP 2009, the IEEE CAMSAP 2011, the IEEE WCNC 2014, the IEEE ICC 2015, WCSP 2017, and the IEEE SAM 2014.
\end{IEEEbiography}

\vspace{-10 mm}

\begin{IEEEbiography}[{\includegraphics[width=1in,height=1.25in,clip,keepaspectratio]{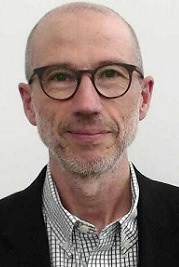}}] {Andr\'e Bourdoux} received the M.Sc. degree in electrical engineering in 1982 from the Université Catholique de Louvain-la-Neuve, Belgium. He joined IMEC in 1998 and is Principal Member of Technical Staff in the IoT Research Group of IMEC. He is a system level and signal processing expert for both the mm-wave wireless communications and radar teams. He has more than 15 years of research experience in radar systems and 15 years of research experience in broadband wireless communications. He holds several patents in these fields. He is the author and co-author of over 160 publications in books and peer-reviewed journals and conferences. His research interests are in the field of advanced architectures, signal processing and machine learning for wireless physical layer and high-resolution 3D/4D radars.
\end{IEEEbiography}

\vspace{-12 mm}

\begin{IEEEbiography}[{\includegraphics[width=1in,height=1.25in,clip,keepaspectratio]{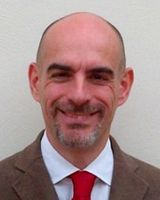}}] {Stefano Buzzi} is currently a Professor at the Department of Electrical and Information Engineering at the University of Cassino and Southern Latium. He is a former editor of the IEEE Communications Letters and of the IEEE Signal Processing Letters, while is currently serving as Associate Editor of the IEEE Transactions on Wireless Communications. He has co-authored more than 150 technical papers published in international journals and in international conference proceedings. His research interests are focused on the PHY and MAC layer of wireless networks. 
\end{IEEEbiography}

\vspace{-12 mm}

\begin{IEEEbiography}[{\includegraphics[width=1in,height=1.25in,clip,keepaspectratio]{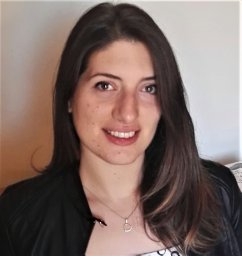}}]{Carmen~D'Andrea} was born in Caserta, Italy on 16 July 1991. She received the B.S. and M.S. degrees,  both with honors, in Telecommunications Engineering from the University of Cassino and Lazio Meridionale in 2013 and 2015, respectively.  In 2017, she was a Visiting Ph.D. student with the Wireless Communications (WiCom) Research Group in the Department of Information and Communication Technologies at Universitat Pompeu Fabra in Barcelona, Spain. In 2019, she received the Ph.D. degree with the highest marks in Electrical and Information Engineering from the University of Cassino and Lazio Meridionale where she is currently a post-doc researcher. Her research interests are focused on wireless communication and signal processing, with current emphasis on mmWave communications and massive MIMO systems, in both colocated and distributed setups.
\end{IEEEbiography}

\vspace{-12 mm}

\begin{IEEEbiography}[{\includegraphics[width=1in,height=1.25in,clip,keepaspectratio]{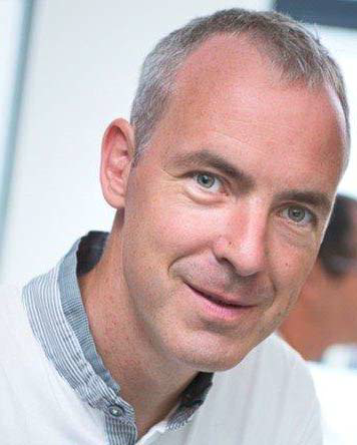}}]
{Jean-Baptiste Dor\'e} received his MS degree in 2004 from the Institut National des Sciences Appliquées (INSA) Rennes, France and his PhD in 2007. He joined NXP semiconductors as a signal processing architect. Since 2009 he has been with CEA-Leti in Grenoble, France as a research engineer and program manager. His main research topics are signal processing (waveform optimization and channel coding), hardware architecture optimizations (FPGA, ASIC), PHY and MAC layers for wireless networks. Jean-Baptiste Doré has published 50+ papers in international conference proceedings and book chapters, received 2 best papers award (ICC2017, WPNC2018). He has also been involved in standardization group (IEEE1900.7) and is the main inventor of more than 30 patents. 
\end{IEEEbiography}

\vspace{-10 mm}

\begin{IEEEbiography}[{\includegraphics[width=1in,height=1.25in,clip,keepaspectratio]{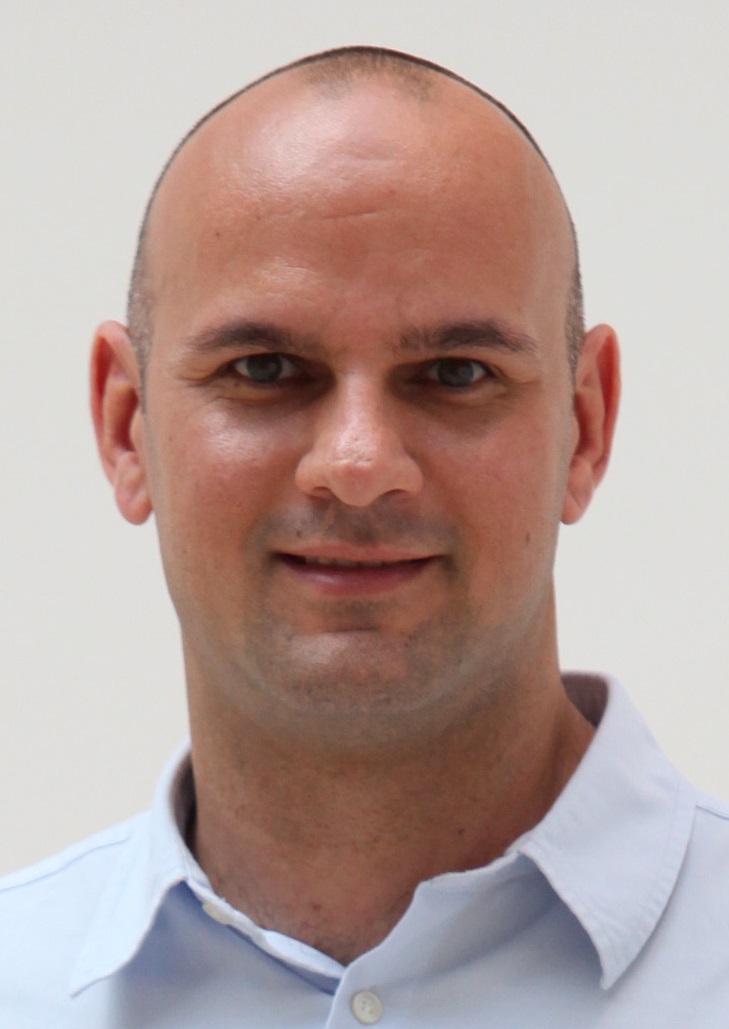}}] {Serhat Erkucuk} is a Professor in the Department of Electrical-Electronics Engineering at Kadir Has University, Turkey. He received his Ph.D. degree in engineering science from Simon Fraser University and held an NSERC postdoctoral fellowship at the University of British Columbia. He has co-authored more than 60 papers published in international journals and conference proceedings in the areas of PHY and MAC layer design of wireless communication systems, and has been granted 2 US patents. He is a Marie Curie Fellow and a recipient of Governor General's Gold Medal.
\end{IEEEbiography}

\vspace{-15 mm}

\begin{IEEEbiography} [{\includegraphics[width=1in,height=1.25in,clip,keepaspectratio]{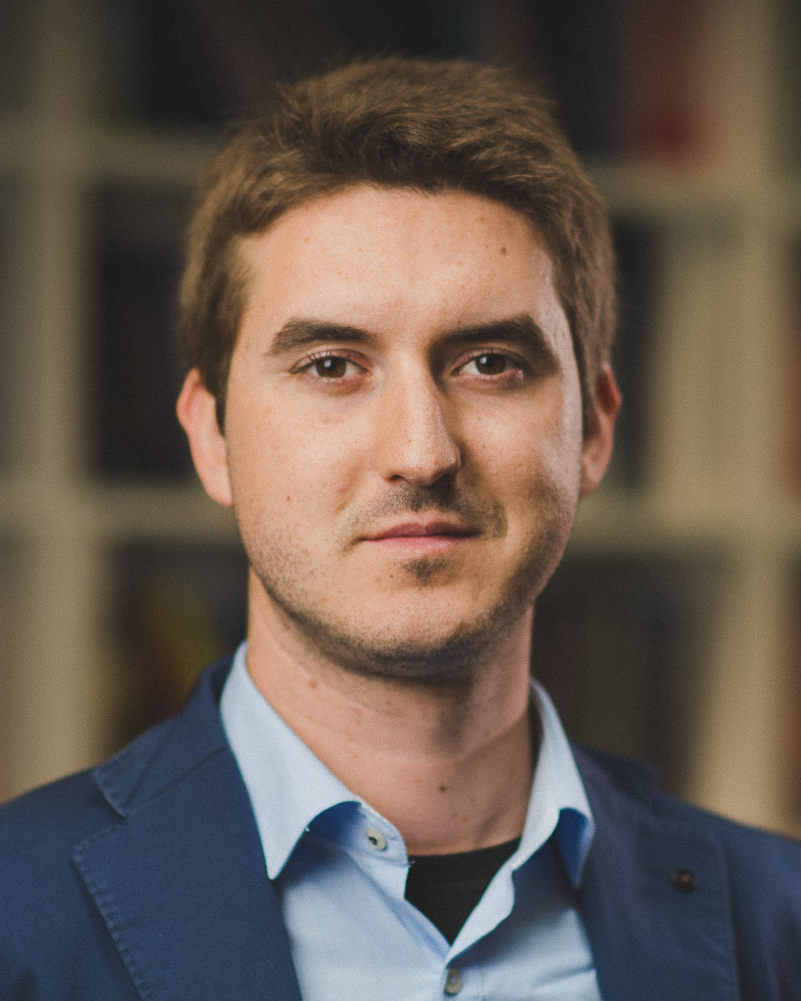}}]{Manuel Fuentes} received the Ph.D. degree in Telecommunication Engineering from Universitat Politecnica de Valencia (UPV), in 2017. In 2012-2017 and 2018-2020, he was with the the Institute of Telecommunications and Multimedia Applications (iTEAM) at UPV. He was also a guest researcher at the Vienna University of Technology, Austria, in 2016. In 2017-2018, Dr. Fuentes worked with the Samsung Electronics R{\&}D UK team as a 5G research engineer. He contributed actively to the ATSC 3.0 standardization process and participated in the IMT-2020 Evaluation Group of 5G PPP. He is co-author of 20+ international IEEE journal and conference papers. He also received the IEEE BTS Scott Helt Award to the best paper of the IEEE Transactions on Broadcasting in 2019. In September 2020, Dr. Fuentes joined Fivecomm as an R{\&}D Manager. He is currently leading a team to work in several Horizon 2020 European 5G/IoT projects. His main areas of interest include physical layer and radio procedures, 5G integration, development and demonstration, as well as Beyond-5G and 6G communications.
\end{IEEEbiography}

\vspace{-15 mm}

\begin{IEEEbiography}[{\includegraphics[width=1in,height=1.25in,clip,keepaspectratio]{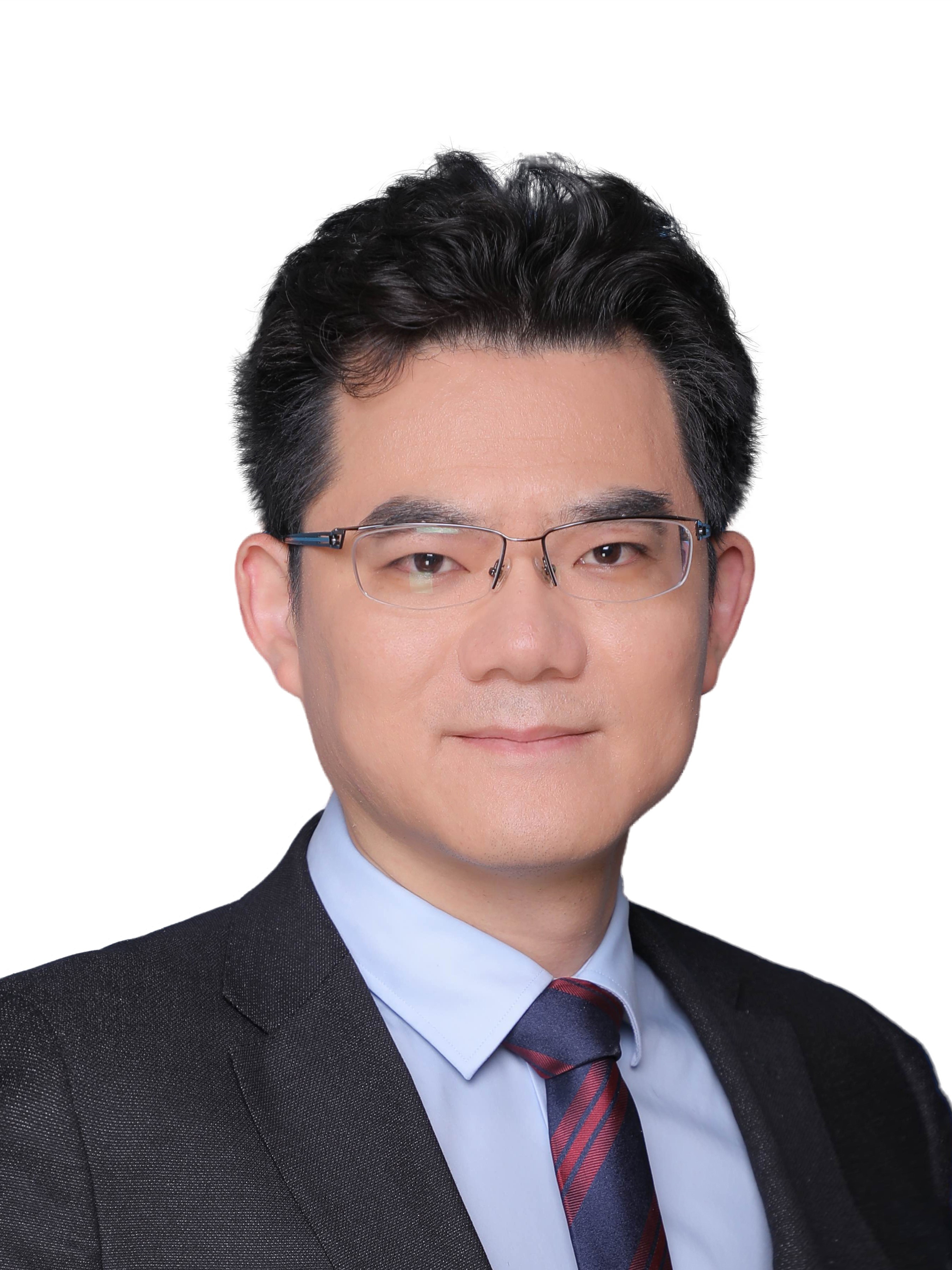}}] {Ke Guan} is a Full Professor in State Key Laboratory of Rail Traffic Control and Safety, Beijing Jiaotong University (BJTU). In 2015, he has been awarded a Humboldt Research Fellowship. He has authored/co-authored more than 240 journal and conference papers, receiving eight Best Paper Awards, including IEEE vehicular technology society 2019 Neal Shepherd memorial best propagation paper award. His current research interests include measurement and modeling of wireless propagation channels for various applications in the era of 5G and beyond. He is an Editor of the IEEE Vehicular Technology Magazine, the IEEE ACCESS, and the IET Microwave, Antenna and Propagation. He is the contact person of BJTU in 3GPP and a member of the IC1004 and CA15104 initiatives.
\end{IEEEbiography}

\vspace{-15 mm}

\begin{IEEEbiography} {Yuzhou Hu} is a senior algorithm engineer with ZTE Corporation. He has served as 3GPP delegate and researcher in the field of non-orthogonal multiple access (NOMA), 2-step random access (2SR), 5G V2X since joining ZTE in 2017. He has filed over 50 patents. He received a BSc and MSc(Summa Cum Laude) in Mathematics and Electronics Engineering from Beihang University, China.
\end{IEEEbiography}


\begin{IEEEbiography} [{\includegraphics[width=1in,height=1.25in,clip,keepaspectratio]{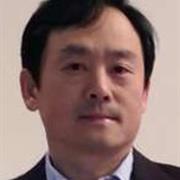}}]{Xiaojing Huang} received the B.Eng., M.Eng., and Ph.D. degrees in electronic engineering from Shanghai Jiao Tong University, Shanghai, China, in 1983, 1986, and 1989, respectively. He was a Principal Research Engineer with the Motorola Australian Research Center, Botany, NSW, Australia, from 1998 to 2003, and an Associate professor with the University of Wollongong, Wollongong, NSW, Australia, from 2004 to 2008. He had been a Principal Research Scientist with the Commonwealth Scientific and Industrial Research Organisation (CSIRO), Sydney, NSW, Australia, and the Project Leader of the CSIRO Microwave and mm-Wave Backhaul projects since 2009.  He is currently a Professor of Information and Communications Technology with the School of Electrical and Data Engineering and the Program Leader for Mobile Sensing and Communications with the Global Big Data Technologies Center, University of Technology Sydney (UTS), Sydney, NSW, Australia. 
\end{IEEEbiography}

\vspace{-15 mm}

\begin{IEEEbiography} [{\includegraphics[width=1in,height=1.25in,clip,keepaspectratio]{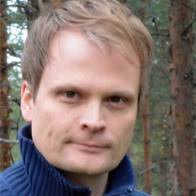}}]{Jari Hulkkonen} graduated in 1999 (M.Sc.EE) from University of Oulu, Finland. He has been working in Nokia since 1996. He started his career in GSM/EDGE research and standardization projects. Since 2006 he has been leading radio systems research in Oulu. Currently Jari is Radio Research Department Head in Nokia Bell Labs Oulu with focus on 5G New Radio evolution. He has more than 30 granted patents/patent applications in 2G-5G technologies.
\end{IEEEbiography}

\vspace{-15 mm}

\begin{IEEEbiography} [{\includegraphics[width=1in,height=1.25in,clip,keepaspectratio]{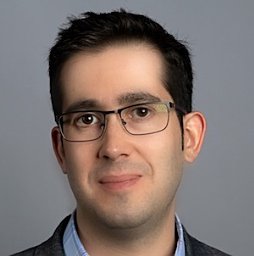}}]{Josep Miquel Jornet} received the Ph.D. degree in Electrical and Computer Engineering (ECE) from the Georgia Institute of Technology in 2013. Between 2013 and 2019, he was with the Department of Electrical Engineering at University at Buffalo. Since August 2019, he has been an Associate Professor in the Department of ECE at Northeastern University. His research interests are in terahertz communications and wireless nano-bio-communication networks. He has co-authored more than 120 peer-reviewed scientific publications, one book, and has been granted 3 US patents, and is serving as the lead PI on multiple grants from U.S. federal agencies.
\end{IEEEbiography}

\vspace{-15 mm}

\begin{IEEEbiography} [{\includegraphics[width=1in,height=1.25in,clip,keepaspectratio]{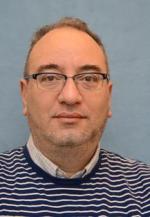}}]{Marcos Katz} is a Professor at the Centre for Wireless Communications, University of Oulu, Finland, since Dec. 2009. He received the B.S. degree in Electrical Engineering from Universidad Nacional de Tucumán, Argentina in 1987, and the M.S. and Dr. Tech. degrees in Electrical Engineering from University of Oulu, Finland, in 1995 and 2002, respectively. He worked in different positions at Nokia, Finland between 1987 and 2001. In years 2003–2005 Dr. Katz was the Principal Engineer at Samsung Electronics, Advanced Research Lab., Telecommunications R/D Center, Suwon, Korea. From 2006 to 2009 he worked as a Chief Research Scientist at VTT, the Technical Research Centre of Finland. 
\end{IEEEbiography}


\begin{IEEEbiography} [{\includegraphics[width=1in,height=1.25in,clip,keepaspectratio]{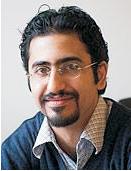}}] {Behrooz Makki} [M'19, SM'19] received his
PhD degree in Communication Engineering from Chalmers University of Technology, Gothenburg, Sweden. In 2013-2017, he was a Postdoc researcher
at Chalmers University. Currently, he works as Senior Researcher in Ericsson Research, Gothenburg, Sweden.

Behrooz is the recipient of the VR Research Link grant, Sweden, 2014, the Ericsson’s Research grant, Sweden, 2013, 2014 and 2015, the ICT SEED grant,
Sweden, 2017, as well as the Wallenbergs research grant, Sweden, 2018. Also, Behrooz is the recipient of the IEEE best reviewer
award, IEEE Transactions on Wireless Communications, 2018. Currently, he works as an Editor in IEEE Wireless Communications Letters, IEEE
Communications Letters, the journal of Communications and Information Networks, as well as the Associate Editor in Frontiers in Communications and
Networks. He was a member of European Commission projects “mm-Wave based Mobile Radio Access Network for 5G Integrated Communications”
and “ARTIST4G” as well as various national and international research collaborations. His current research interests include integrated access and
backhaul, Green communications, millimeter wave communications, finite block-length analysis and backhauling. He has co-authored 63 journal papers, 46 conference papers and 60 patent applications.
\end{IEEEbiography}

\vspace{-10 mm}

\begin{IEEEbiography}[{\includegraphics[width=1in,height=1.25in,clip,keepaspectratio]{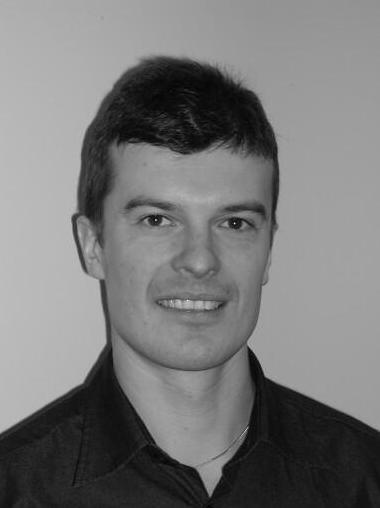}}] {Rickard Nilsson} received the Ph.D. degree from Lule{\aa} University of Technology (LTU), Sweden. With Telia Research AB, Sweden, and Stanford University, USA, he introduced a new flexible broadband access method for VDSL and contributed to its standardization. For seven years he was a senior researcher at the Telecommunications Research Center Vienna, Austria, and lectured at the Technical University. Since 2010 he is with LTU researching wireless connectivity, lecturing signal processing and communications, and cooperating with industry.  
\end{IEEEbiography}

\vspace{-10 mm}

\begin{IEEEbiography}[{\includegraphics[width=1in,height=1.25in,clip,keepaspectratio]{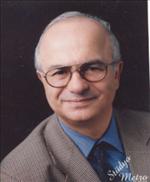}}] {Erdal Panayirci}  received the Ph.D. degree in electrical engineering from Michigan State University, Michigan, USA. He is currently a professor of electrical engineering in the Electrical and Electronics Engineering Department at Kadir Has University, Istanbul, Turkey and Visiting Research Collaborator at the Department of Electrical Engineering, Princeton University, USA. He has published extensively in leading scientific journals and international conference and co-authored the book Principles of Integrated Maritime Surveillance Systems (Kluwer Academic, 2000). His research interests are advanced signal processing techniques and their applications to wireless electrical, underwater and optical communications. Prof. Panayirci was an Editor for the IEEE transactions on communications  and served and is currently serving as a Member of IEEE Fellow Committee during 2005–2008 and 2019– 2021, respectively. He is an IEEE Life Fellow.
\end{IEEEbiography}

\vspace{-12 mm}

\begin{IEEEbiography}[{\includegraphics[width=1in,height=1.25in,clip,keepaspectratio]{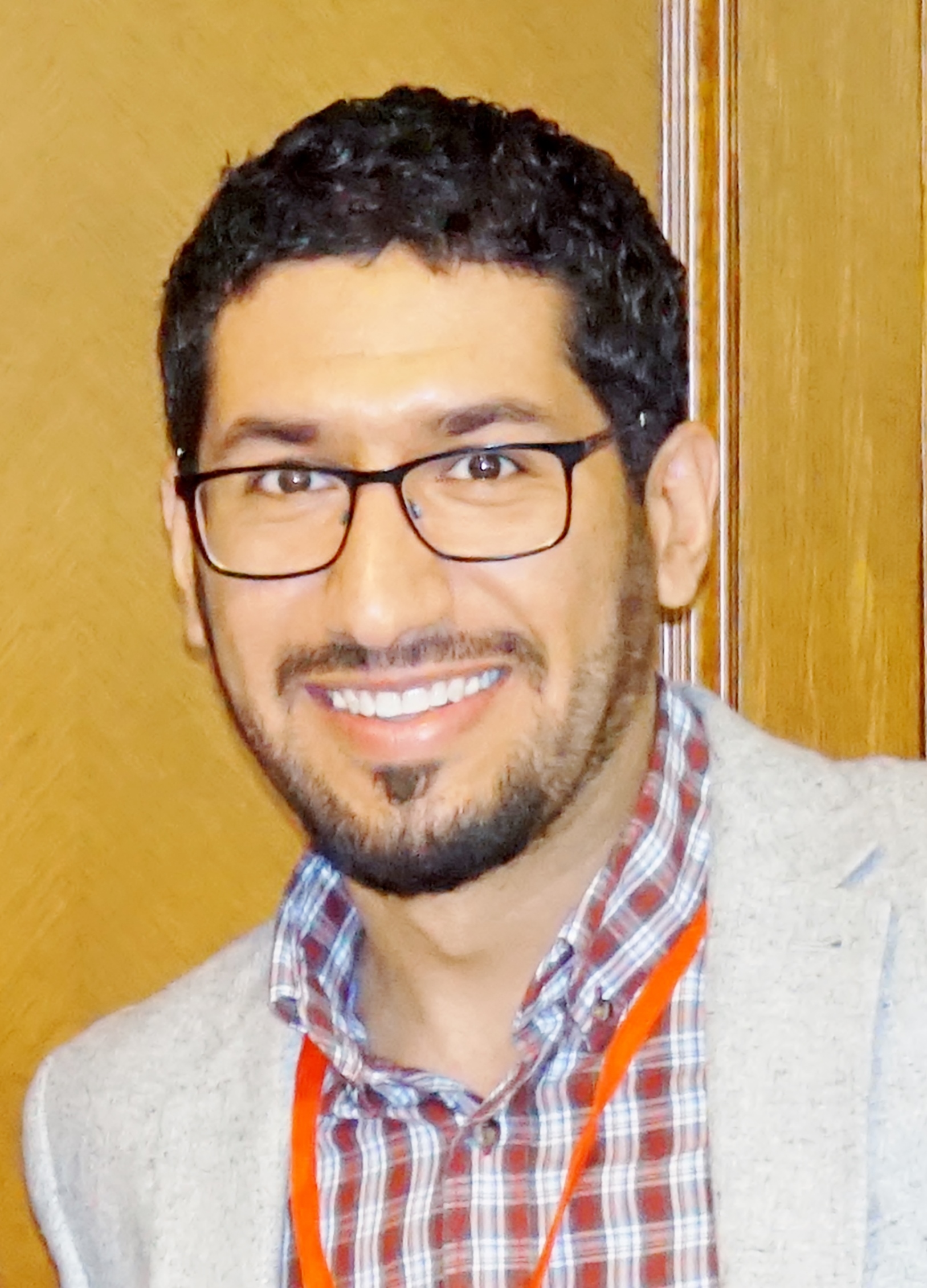}}] {Khaled Rabie} received the Ph.D. degree in Electrical and Electronic Engineering from the University of Manchester, UK, in 2015. He is currently an Assistant Professor with the department of Engineering at the Manchester Metropolitan University, UK. His primary research focuses on various aspects of the next-generation wireless communication systems. He serves as an Editor for IEEE COMMUNICATIONS LETTERS, an Associate Editor for IEEE ACCESS, and an Area Editor for PHYSICAL COMMUNICATIONS. He received the Best Paper Award at the IEEE ISPLC 2015 as well as the IEEE ACCESS Editor of the month award for August 2019. Khaled is also a Fellow of the U.K. Higher Education Academy.
\end{IEEEbiography}


\begin{IEEEbiography}[{\includegraphics[width=1in,height=1.25in,clip,keepaspectratio]{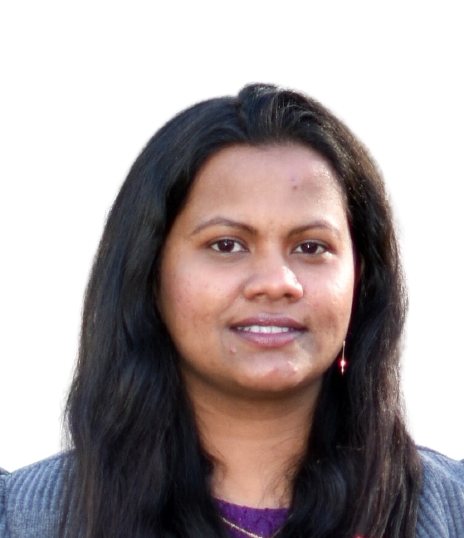}}] {Nuwanthika Rajapaksha} is currently a doctoral student at the Centre for Wireless Communications, University of Oulu, Finland. She received the B.Sc. (Hons.) degree in electronics and telecommunication engineering from University of Moratuwa, Sri Lanka, in 2014, and the M.Sc. degree in wireless communications engineering from University of Oulu, in 2019. From 2014 to 2018 she was a research engineer at Synergen Technology Labs, Sri Lanka, where she was involved in biomedical signal processing, machine learning-based algorithm development, and US FDA regulatory process for wearable medical device design. Her main research interests are signal processing and machine learning applications in PHY/MAC layers.
\end{IEEEbiography}

\vspace{-15 mm}

\begin{IEEEbiography} [{\includegraphics[width=1in,height=1.25in,clip,keepaspectratio]{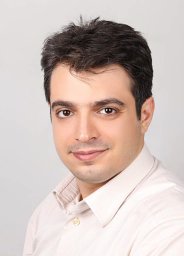}}]{MohammadJavad Salehi} received his PhD in Electrical Engineering from Sharif University of Technology, Tehran, Iran, in 2018. His career in Iran also includes 9 years of work experience in the IT industry, as a developer, manager and start-up founder. Since 2019, he is a postdoctoral researcher at University of Oulu, Finland, where he works on various aspects of practical implementation of coded caching techniques in future wireless networks. Specifically, his research now includes designing coded caching schemes with reduced subpacketization, performance analysis of multi-antenna coded caching at finite-SNR communication regime, and energy efficiency of practical coded caching setups.
\end{IEEEbiography}

\vspace{-15 mm}

\begin{IEEEbiography} [{\includegraphics[width=1in,height=1.25in,clip,keepaspectratio]{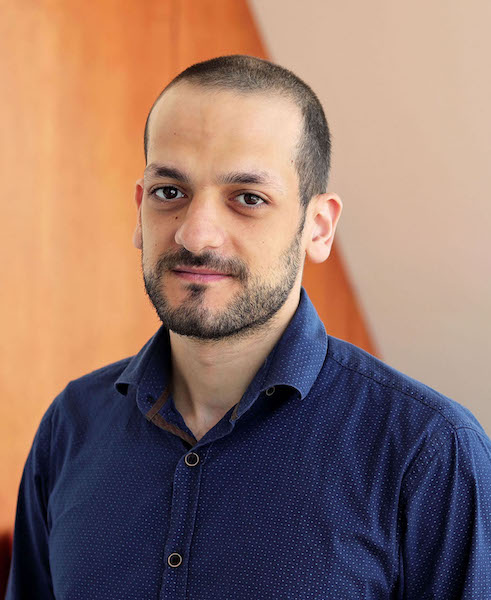}}]{Hadi Sarieddeen} received the B.E. degree (summa cum laude; first in graduating class) in computer and communications engineering from Notre Dame University-Louaize (NDU), Lebanon, in 2013, and the Ph.D. degree in electrical and computer engineering from the American University of Beirut (AUB), Lebanon, in 2018. He is currently a postdoctoral research fellow at King Abdullah University of Science and Technology (KAUST), Thuwal, Saudi Arabia. His research interests are in the areas of communication theory and signal processing for wireless communications.
\end{IEEEbiography}

\vspace{-15 mm}

\begin{IEEEbiography}[{\includegraphics[width=1in,height=1.25in,clip,keepaspectratio]{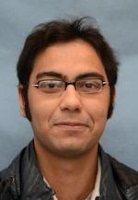}}] {Shahriar Shahabuddin} received his MSc and PhD from University of Oulu, Finland in 2012 and 2019 respectively. During Spring 2015, he worked  as a Visiting Researcher at Computer Systems Laboratory, Cornell University, USA. Shahriar received distinction in MSc and several scholarships and grants such as Nokia Foundation Scholarship, University of Oulu Scholarship Foundation Grant, Tauno Tönning Foundation Grant during his PhD. Shahriar's research interest includes VLSI signal processing, massive MIMO systems and physical layer security. Since 2017, Shahriar has been with Nokia, Finland as a SoC Specialist.
\end{IEEEbiography}


\begin{IEEEbiography}[{\includegraphics[width=1in,height=1.25in,clip,keepaspectratio]{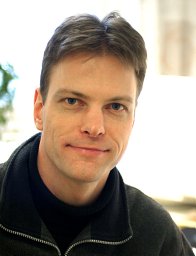}}] {Tommy Svensson} is Full Professor at Chalmers University of Technology in Gothenburg, Sweden, leading Wireless Systems. He has a PhD in Information theory from Chalmers in 2003, and has worked at Ericsson AB with core, radio access, and microwave networks. He has been deeply involved in European research towards 4G, 5G and currently towards 6G within the Hexa-X, RISE-6G and SEMANTIC EU projects, on access, backhaul/ fronthaul, C-V2X and satellite networks, with a focus on physical and MAC layer. He has coauthored 5 books, 93 journal and 129 conference papers, and 53 EU projects deliverables. He is chair of the IEEE Sweden VT/COM/IT chapter, founding editorial board member of IEEE JSAC Series on Machine Learning in Communications and Networks, been editor of IEEE TWC, WCL, guest editor of top journals, organizer of tutorials/ workshops at top IEEE conferences, and coordinator of the Communication Engineering MSc program at Chalmers.
\end{IEEEbiography}

\vspace{-15 mm}

\begin{IEEEbiography}[{\includegraphics[width=1in,height=1.25in,clip,keepaspectratio]{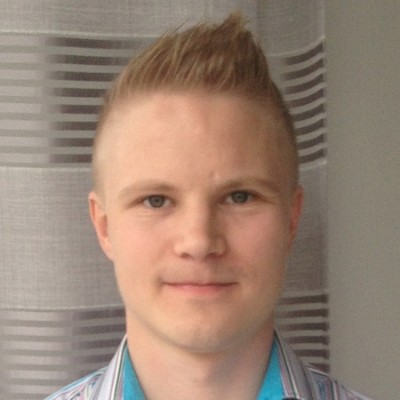}}] {Oskari Tervo} is currently working as a Senior Standardization Research Specialist in Nokia Bell Labs, Oulu, where he joined in 2018. He received a doctoral degree from Centre for Wireless Communications (CWC), University of Oulu, Finland, in 2018 with distinction. In 2014 and 2016, he was a Visiting Researcher with Kyung Hee University, Seoul, Korea, and the Interdisciplinary Centre for Security, Reliability and Trust (SnT), University of Luxembourg, Luxembourg, respectively. He has received the Best Reviewer Award from IEEE Transactions on Wireless Communications in 2017 and 2018. 
\end{IEEEbiography}

\vspace{-15 mm}

\begin{IEEEbiography} [{\includegraphics[width=1in,height=1.25in,clip,keepaspectratio]{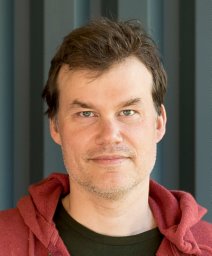}}] {Antti T\"olli} is an Associate Professor with the Centre for Wireless Communications (CWC), University of Oulu. He received the Dr.Sc. (Tech.) degree in electrical engineering from the University of Oulu, Oulu, Finland, in 2008. From 1998 to 2003, he worked at Nokia Networks as a Research Engineer and Project Manager both in Finland and Spain. In May 2014, he was granted a five year (2014-2019) Academy Research Fellow post by the Academy of Finland. During the academic year 2015-2016, he visited at EURECOM, Sophia Antipolis, France, while from August 2018 till June 2019 he was visiting at the University of California Santa Barbara, USA. 
He is currently serving as an Associate Editor for IEEE Transactions on Signal Processing. 
\end{IEEEbiography}

\vspace{-15 mm}

\begin{IEEEbiography}[{\includegraphics[width=1in,height=1.25in,clip,keepaspectratio]{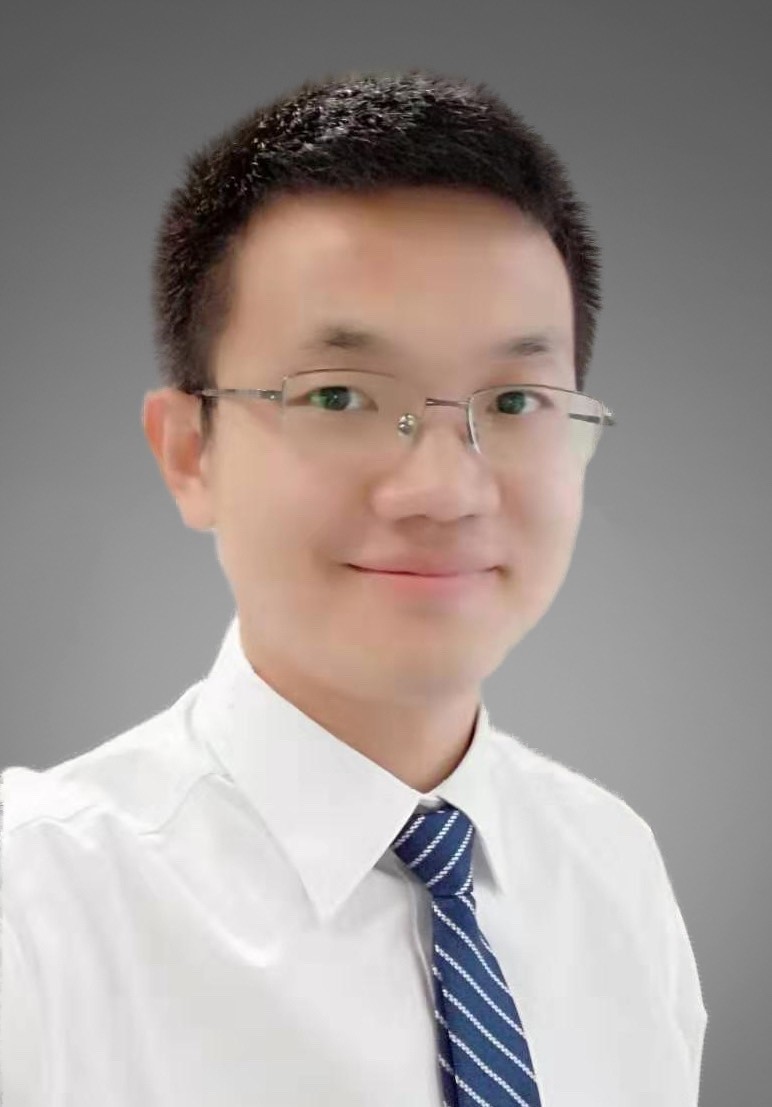}}]{Qingqing Wu} is an assistant professor in the University of Macau, Macau, China. His current research interests include IRS-enabled 6G, UAV communications, and green communications. He has published over 60 IEEE top-tier journal and conference papers, which have attracted more than 3400 Google Scholar citations. He received the Best Ph.D. Thesis Award of China Institute of Communications in 2017 and the IEEE WCSP Best Paper Award in 2015. He serves as an Associate Editor for IEEE CL, the Guest Editor for IEEE OJVT on ``6G Intelligent Communications", and the Leading Guest Editor for IEEE JSAC on  ``UAV Communications in 5G and Beyond Networks".
\end{IEEEbiography}


\begin{IEEEbiography}[{\includegraphics[width=1in,height=1.25in,clip,keepaspectratio]{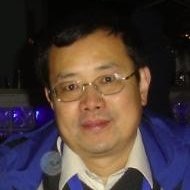}}] {Wen Xu} received the Dr.-Ing. degree from Technical University of Munich, Germany, in 1996. From 1995 to 2006, he was with Siemens AG, Munich, where he was head of the Algorithms and Standardization Lab. From 2007 to 2014, he was with Infineon Technologies AG (\textit{later} Intel Mobile Communications GmbH), Germany. In 2014, he joined Huawei Technologies Duesseldorf GmbH - Munich Research Center, where he is leading the Radio Access Technologies Dept. His research interests include signal processing, source/channel coding, and wireless communication systems. He has 100+ peer-reviewed papers published and numerous patents granted.
\end{IEEEbiography}